\documentclass{article}
\usepackage[utf8]{inputenc}
\usepackage{amsfonts}
\usepackage{geometry}
\usepackage{hyperref,amsthm,enumerate,
            imakeidx, xparse, mathtools,
            xcolor, color, colortbl, 
            amssymb, tensor, 
            soul, graphicx ,titlesec,  appendix, tikz,
            amsmath,scalerel, comment, float, multirow, multicol, caption, subcaption} 
\usepackage[most]{tcolorbox}
\usepackage{stackengine}
\usetikzlibrary{arrows,decorations.markings}
\usetikzlibrary{shapes.misc}
\usetikzlibrary{arrows.meta}
\usetikzlibrary{angles,quotes}
\usepackage{tkz-euclide}
\usetikzlibrary{intersections}
\usetikzlibrary{calc}
\usetikzlibrary{backgrounds}
\usepackage{cite}
\usepackage{booktabs}

\hypersetup{
pdfstartview = {FitH},
}
\hypersetup{
	colorlinks=true,         
	linkcolor=brown,          
	citecolor=red,        
	urlcolor=blue            
}

\newcommand{\be}{\begin{equation}}
\newcommand{\ee}{\end{equation}}

\newtheorem{definition}{Definition}
\newtheorem{theorem}{Theorem}

\newcommand{\cA}{{\cal A}}

\newcommand{\cT}{{\cal T}}
\newcommand{\cG}{{\cal G}}

\newcommand{\cV}{{\cal V}}
\newcommand{\cS}{{\cal S}}
\newcommand{\cP}{{\cal P}}

\newcommand{\cZ}{{\cal Z}}

\newcommand{\cK}{{\cal K}}

\def\g{\mathfrak g}

\def\R{\mathbb{R}}

\def\SU{\mathrm{SU}}

\def\x{\mathrm X}
\def\y{\mathrm Y}

\def\dh{[\mathrm{d h}]}

\def\dx{[\mathrm{d x}]}
\def\du{[\mathrm{d u}]}

\def\dY{[\mathrm{d Y}]}
\def\dX{[\mathrm{d X}]}

\def\rd{\mathrm{d}}

\def\rhd{\triangleright}

\def\mone{^{-1}}

\definecolor{darkblue}{rgb}{0.0, 0.0, 0.65}
\definecolor{darkred}{rgb}{0.65, 0.0, 0.0}
\definecolor{darkgreen}{rgb}{0.0, 0.65, 0.0}

\definecolor{carblue}{rgb}{0.6, 0.0, 0.85}
\definecolor{carred}{rgb}{0.85, 0.6, 0.0}
\definecolor{cargreen}{rgb}{0.0, 0.85, 0.6}

\definecolor{darkblue'}{rgb}{0.25, 0.25, 0.85}
\definecolor{darkred'}{rgb}{0.85, 0.25, 0.25}
\definecolor{darkgreen'}{rgb}{0.25, 0.85, 0.25}

\definecolor{carblue'}{rgb}{0.75, 0.35, 0.95}
\definecolor{carred'}{rgb}{0.95, 0.75, 0.35}
\definecolor{cargreen'}{rgb}{0.35, 0.95, 0.75}

\definecolor{darkorange}{rgb}{1.0, 0.5, 0.0}
\definecolor{darkbrown}{rgb}{0.6, 0.3, 0.0}
\definecolor{cargreen}{rgb}{0.0, 0.8, 0.6}
\definecolor{darkpink}{rgb}{0.9, 0.3, 0.5}
\definecolor{gold}{rgb}{0.85, 0.65, 0.1}
\definecolor{bluette}{rgb}{0.3, 0.25, 0.55}

\usepackage{authblk}

\makeatletter
\renewenvironment{abstract}{%
    \if@twocolumn
      \section*{\abstractname}%
    \else 
      \begin{center}%
        {\bfseries\sffamily\abstractname\vspace{\z@}}
      \end{center}%
      \quotation
    \fi}
    {\if@twocolumn\else\endquotation\fi}
\makeatother

\title{Group field theory on 2-groups }

\begin{document}

\author[1]{\sffamily Florian Girelli\thanks{florian.girelli@uwaterloo.ca}}
\author[2,1]{\sffamily Matteo Laudonio\thanks{matteo.laudonio@u-bordeaux.fr}}
\author[2,3]{\sffamily Adrian Tanasa\thanks{ntanasa@u-bordeaux.fr}}
\author[1, 2]{\sffamily Panagiotis Tsimiklis\thanks{ptsimiklis@uwaterloo.ca}}

\affil[1]{\small Department of Applied Mathematics, University of Waterloo, 200 University Avenue West, Waterloo, Ontario, Canada, N2L 3G1}
\affil[2]{\small Univ. Bordeaux, LABRI, 351 Cours de la Libération, 33400 Talence, France}
\affil[3]{\small DFT, H. Hulubei Nat. Inst. Nucl. Engineering, P. O. Box MG-6, 077125 Magurele, Romania}

\maketitle

\begin{abstract}
      Group field theories are quantum field theories built on groups. They can be seen as a tool to generate topological state-sums or 
      quantum gravity models. For four dimensional manifolds, different arguments have pointed towards  2-groups (such as crossed modules) as the relevant symmetry structure to probe  four dimensional topological features. Here, we introduce a group field theory built on crossed modules which generate a four dimensional topological model, as we prove that the Feynman diagram amplitudes can be related by Pachner moves. This model is presumably the dual version of the Yetter-Mackaay model.      
\end{abstract}

\tableofcontents

\section{Introduction}

\medskip

Lattice  models have been  useful tools to study many  theories. Well known examples are the Yang-Mills type lattice models  which are key to understanding the strong force. 
 Other examples include the  topological models generated by a BF-type  \cite{Dupuis:2020ndx, Baez:1999sr} or Chern-Simons action \cite{Fock:1992xy, Fock:1998nu, Alekseev:1994pa, Alekseev:1994au} which rely on such lattice techniques. In fact, since these examples are topological, the discretization is merely a regularization there is no proper loss of degrees of freedom. Note however that topological models in different spacetime dimensions might not be described by exactly the same type of gauge symmetry structures. Indeed, it is expected that with dimensions increasing we go up in the categorical ladder \cite{Crane:1994ty,Baez:1995xq, Baez:2009as}. If in 3d, (quantum) groups or Hopf algebras and their category of representations are well adapted to probe the topological excitations \cite{Witten:1988hc, Crane:1994ty, Baez:2009as, Turaev:2017uxl, Dupuis:2020ndx}, in 4d one expects that one could (should?) use instead (quantum) 2-groups and their 2-category of representations to properly describe  the 4d topological excitations \cite{Crane:1994ty, Asante:2019lki, Girelli:2021zmt, Girelli:2021khh, Bullivant:2016clk, Bullivant:2017qrv, Bullivant:2019tbp}.    

2-groups or crossed modules can be seen as a categorified version of the notion of group \cite{baez2004}. They have only been recently studied and many of the things that are known about groups are actually unknown for (quantum) 2-groups. For example, the notion of harmonic analysis is missing for 2-groups since their representations theory is not under control, except for some specific classes of 2-groups (the skeletal ones \cite{Baez:2008hz}, or also see  \cite{barrett:2004zb}). Following the categorical ladder, one can see that there are different options one can choose to define what is a quantum 2-groups \cite{Crane:1994ty}, see \cite{Majid:2012gy} for one attempt.

While many pieces of the general theoretical understanding of (quantum) 2-groups are missing, a topological model based on 2-groups exists and is  called  \textit{Yetter-MacKaay model} \cite{Yetter:1993dh, mackaay:ek}. The model is based on lattice 2-gauge theories where the gauge symmetry is actually specified by a (strict finite) 2-group or crossed module (one can also be extend to the weak case instead of strict \cite{mackaay:ek, mackaay:uo}). There is a continuum picture given by an analogue of the BF theory framework, this time defined in terms of Lie 2-groups as gauge symmetries. These are called 2-BF models (or BFCG action) \cite{Girelli:2007tt, Martins:2010ry}. In fact it can be shown that the standard  4d BF theories are themselves theories with Lie 2-group symmetries \cite{chen:2022}. It can be shown that discretization of 2-BF theories leads to the Yetter-Mackaay model built on Lie 2-groups  \cite{Girelli:2007tt}.

\medskip

Gravity can also be treated with techniques inspired by lattice gauge theory and topological models. 3d gravity is topological and is very well described using lattice gauge theory techniques \cite{Carlip:1998uc}. On the other hand in 4d, using the fact that 4d gravity can be recovered as constrained topological theory, one essentially tweaks the structures obtained from the 4d BF model to incorporate the gravitational features \cite{Barrett:1997gw, Baez:1999in, Baez:1999sr, rovelli2004}.   While 4d gravity is usually defined in terms of a standard lattice gauge picture, it was suggested that better insights could be gained if one would use 2-groups instead \cite{Crane:2003ep, Girelli:2021khh}. For example, one could have the frame field degrees of freedom explicitly present in the discrete picture \cite{Crane:2003ep, Asante:2019lki, Girelli:2021khh}. In this sense, one could expect that the construction of a quantum gravity theory also follows the categorical ladder.

Starting from a different perspective,  Boulatov introduced the notion of group field theory (GFT), which is essentially,
a quantum field theory  over some abstract space, a bunch of (quantum) groups \cite{Boulatov:1992vp},  whose Feynman diagrams are exactly the amplitudes of the Ponzano-Regge/Turaev-Viro models. 
Said otherwise these Feynman diagrams provide the natural amplitudes or quantum dynamics for quantum states using lattice gauge theory techniques. Note that the Feynman diagrams are interpreted as the 1-complex dual to the spacetime triangulation. By performing a Fourier transform, in terms of representations for example, one recovers the Ponzano-Regge/Turaev-Viro models. If instead one uses the Fourier transform in terms of dual groups \cite{Baratin:2010nn, Baratin:2010wi, Guedes:2013vi}, we recover typically a  non-commutative field theory where the non-commutative variables decorate the edges and can be interpreted as the discretized frame field. Boulatov's GFT  describes the quantization of 3d gravity.


\medskip 

GFT's can also be defined in the 4d case, either to provide the amplitude of the BF theory topological model \cite{Ooguri:1992eb}, or  quantum gravity models \cite{Freidel:2005qe, Krajewski:2011zzu, Oriti:2017ave}. They have also be used to see how spacetime notions can emerge using concepts of analogue gravity \cite{Oriti:2016acw}. Let us also mention here that GFT's can also be seen to be a particular class of tensor models
(see the review articles 
\cite{Tanasa:2012pm, Tanasa:2015uhr, Gurau:2011xp}
or the books \cite{book-razvan, book-io}), which are
 a natural generalization in dimension higher then two of the celebrated matrix models (see, for example, the review articles \cite{DiFrancesco:1993cyw, DiFrancesco:2004qj}).

As discussed before, topological models in 4d can/should  be described in terms of (Lie) 2-groups. It is then natural to ask  as a first step whether one could construct a field theory over 2-groups to recover 
the amplitudes associated to a 2-BF amplitude, namely the Yetter-Mackaay model. A second step would then consist in trying to recover gravity, if gravity can be found to be related to a Lie 2-group symmetries. 

\medskip 

In the following, we focus on the first question and construct a 2-GFT, ie a field theory built on strict (Lie) 2-groups. We show that the associated Feynman diagram amplitudes are topological invariants, that is are proportional to other diagrams, related by Pachner moves. These amplitudes are interpreted as the 2-complex dual to a 4d triangulation\footnote{The coefficient of proportionality might be infinite, as in the Ponzano-Regge model or the Turaev-Viro model for $q$ not root of unity.}. We do not prove yet the equivalence with the Yetter-Mackaay model as we would need for this the notion of Fourier transform which is still lacking.

\medskip

{ The paper is organized as follows.}
In Section \ref{Sec_2-groups}, we review all the fundamental aspects of 2-category and 2-group theory needed for the formulation of our model, with a particular emphasis on the definition of lattice gauge theories based on 2-groups.
In Section \ref{1gft}, we first provide a brief review of 3d GFT, stressing the fundamental ingredients of the model.
Section \ref{2gft} contains the  main result of the paper, namely the definition of a field theory over strict 2-groups. 
In Section \ref{topoinv}, we discuss how the amplitudes are topological invariants. The proof that they are invariants is given in the appendix as it relies on lengthy calculations.

\section{Lattice 2-gauge theory}
\label{Sec_2-groups}

In this section we introduce the key tools from lattice gauge theory that will be relevant to construct 2-group field theory.  
In this setting, holonomies generated by the gauge fields decorate the edges of the graph. For example, if $G_1$ is a group and if $u\in G_1$ is the holonomy on the oriented edge connecting vertex $i$ to vertex $j$ it transforms under local gauge transformations localized at the vertices according to 
\begin{align}
  u\rightarrow   h \mone u h', \quad h,h'\in G_1.\label{1gauge}
\end{align}
We note that the holonomies $u$ can be interpreted as the  transport of degrees of freedom $\psi$ living at the vertices. 
\begin{align}\label{1trans}
\psi'= \pi(u)\psi,    
\end{align}
where  $\pi:G_1\rightarrow GL_n$ is a representation of $G_1$ under which $\psi$ transforms.

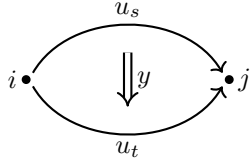
\begin{figure}[h!]
    \centering
    \begin{tikzpicture}[scale=0.9]
        \draw[fill , thick] (0,0) circle [radius=0.05];
        \draw[fill , thick] (3,0) circle [radius=0.05];
        \draw[-> , thick] (0.1,0.1) to [out=60 , in=120] (2.9,0.1);
        \draw[-> , thick] (0.1,-0.1) to [out=-60 , in=-120] (2.9,-0.1);
        \draw[-{Implies[]} , double distance=2 , thick] (1.5,0.4) -- (1.5,-0.4);
        \node[above] at (1.5,0.78) {$u_s$};
        \node[below] at (1.5,-0.78) {$u_t$};
        \node[left] at (0,0) {$i$};
        \node[right] at (3,0) {$j$};
        \node[right] at (1.5,0) {$y$};
    \end{tikzpicture}
    \caption{Plaquette decorated in terms of a 1-holonomy $u_p=u_tu_s\mone$ and a 2-holonomy $y$.}
    \label{Fig:bigone}
\end{figure}

We can extend this construction to include
decorations on the two dimensional faces of the graph. 
{ In this context,} 
the face decorations $y$, valued in a group $G_2$ can be interpreted as \textit{2-holonomies} which transport edge decorations. 
To explain the consistent way of including face decorations, consider a plaquette $p$ and let $u_p$ be the holonomy associated to the boundary of the plaquette, starting and ending at vertex $i$. We require that there is a group homomorphism $t:G_2\to G_1$ and demand that if $y\in G_2$ decorates $p$, then it satisfies $t(y) = u_p$. In particular, if we decompose the loop around $p$ into two paths from vertex $i$ to a vertex $j$ as shown in fig.\, \ref{Fig:bigone} with holonomies labelled by $u_s$ and $u_t$, we demand
\begin{align}
u_t=t(y)u_s.    
\end{align}
In analogy to \eqref{1trans}, we can think of the face decoration $y$ as transporting the path labelled by $u_s$ to the path $u_t$.

Just as the concatenation of edge holonomies is achieved by group multiplication in lattice gauge theory, we must insist on having compatibility conditions between $G_1$, $G_2$, and $t$ in order to combine faces in a consistent way. This leads us to the definition of a crossed module, AKA a strict 2-group \cite{Baez:2002jn, Baez:2004in, Baez:2005qu, Girelli:2003ev, Baez:2010ya}.

\subsection{Strict 2-groups.}\label{sec:2group}

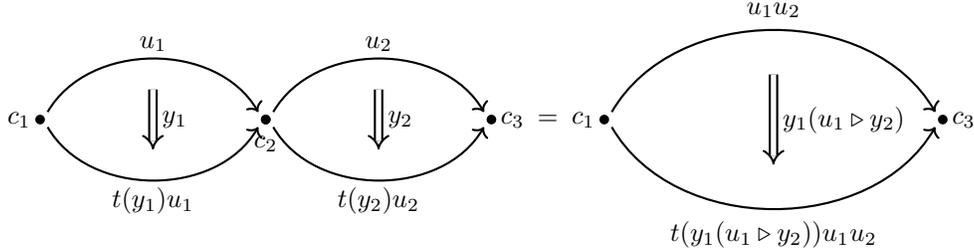
\begin{figure}
    \centering
    \begin{tikzpicture}[scale=1]
        \draw[fill , thick] (0,0) circle [radius=0.05];
        \draw[fill , thick] (3,0) circle [radius=0.05];
        \draw[-> , thick] (0.1,0.1) to [out=60 , in=120] (2.9,0.1);
        \draw[-> , thick] (0.1,-0.1) to [out=-60 , in=-120] (2.9,-0.1);
        \draw[-{Implies[]} , double distance=2 , thick] (1.5,0.4) -- (1.5,-0.4);
        \node[above] at (1.5,0.78) {$u_1$};
        \node[below] at (1.5,-0.78) {$t(y_1)u_1$};
        \node[right] at (1.5,0) {$y_1$};
        
        \draw[fill , thick] (3,0) circle [radius=0.05];
        \draw[fill , thick] (6,0) circle [radius=0.05];
        \draw[-> , thick] (3.1,0.1) to [out=60 , in=120] (5.9,0.1);
        \draw[-> , thick] (3.1,-0.1) to [out=-60 , in=-120] (5.9,-0.1);
        \draw[-{Implies[]} , double distance=2 , thick] (4.5,0.4) -- (4.5,-0.4);
        \node[above] at (4.5,0.78) {$u_2$};
        \node[below] at (4.5,-0.78) {$t(y_2)u_2$};
        \node[right] at (4.5,0) {$y_2$};

        \node[left] at (0,0) {$c_1$};
        \node[below] at (3,-0.1) {$c_2$};
        \node[right] at (6,0) {$c_3$};
        
        \node at (6.75,0) {$=$};
        
        \draw[fill , thick] (7.5,0) circle [radius=0.05];
        \draw[fill , thick] (12,0) circle [radius=0.05];
        \draw[-> , thick] (7.6,0.1) to [out=60 , in=120] (11.9,0.1);
        \draw[-> , thick] (7.6,-0.1) to [out=-60 , in=-120] (11.9,-0.1);
        \draw[-{Implies[]} , double distance=2 , thick] (9.75,0.6) -- (9.75,-0.6);
        \node[above] at (9.75,1.23) {$u_1 u_2$};
        \node[below] at (9.75,-1.23) {$t(y_1 (u_1 \rhd y_2)) u_1 u_2$};
        \node[right] at (9.75,0) {$y_1 (u_1 \rhd y_2)$};
        
        \node[left] at (7.5,0) {$c_1$};
        \node[right] at (12,0) {$c_3$};
    \end{tikzpicture}
    \caption{Horizontal composition of the 2-group elements $(h_1, g_1)$ and $(h_2, g_2)$.}
    \label{Fig:HorizontalComposition}
\end{figure}

\begin{figure}
    \centering
    \begin{tikzpicture}[scale=1]
        \draw[fill , thick] (0,0) circle [radius=0.05];
        \draw[fill , thick] (3,0) circle [radius=0.05];
        \draw[-> , thick] (0.1,0.1) to [out=80 , in=180] (1.5,1.5) to [out=0 , in=100] (2.9,0.1);
        \draw[-> , thick] (0.1,-0.1) to [out=-80 , in=-180] (1.5,-1.5) to [out=0 , in=-100] (2.9,-0.1);
        \draw[-> , thick] (0.1,0) -- (2.9,0);
        \draw[-{Implies[]} , double distance=2 , thick] (1.5,1.25) -- (1.5,0.65);
        \draw[-{Implies[]} , double distance=2 , thick] (1.5,-0.3) -- (1.5,-1.2);
        \node[above] at (1.5,1.5) {$u_1$};
        \node[above] at (1.5,0) {$t(y_1) u_1 = u_2$};
        \node[below] at (1.5,-1.5) {$t(y_1)u_1$};
        \node[right] at (1.5,1.05) {$y_1$};
        \node[right] at (1.5,-0.75) {$y_2$};
        
        \node[right] at (3,0) {$c_2$};
        \node[left] at (0,0) {$c_1$};
        
        \node at (3.75,0) {$=$};
        
        \draw[fill , thick] (4.5,0) circle [radius=0.05];
        \draw[fill , thick] (7.5,0) circle [radius=0.05];
        \draw[-> , thick] (4.6,0.1) to [out=60 , in=120] (7.4,0.1);
        \draw[-> , thick] (4.6,-0.1) to [out=-60 , in=-120] (7.4,-0.1);
        \draw[-{Implies[]} , double distance=2 , thick] (6,0.4) -- (6,-0.4);
        \node[above] at (6,0.78) {$u_1$};
        \node[below] at (6,-0.78) {$t(y_2 y_1) u_1$};
        \node[right] at (6,0) {$y_2 y_1$};
        
        \node[left] at (4.5,0) {$c_1$};
        \node[right] at (7.5,0) {$c_2$};
    \end{tikzpicture}
    \caption{Vertical composition of the 2-group elements $(y_1, u_1)$ and $(y_2, u_2)$, with $u_2 = t(y_1) u_1$.}
    \label{Fig:VerticalComposition}
\end{figure}
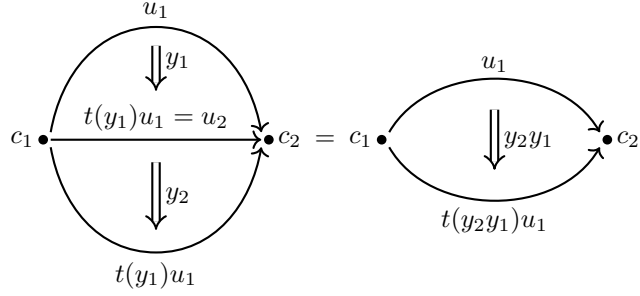

\begin{definition}
A strict 2-group or crossed module is given by
\begin{itemize}
    \item a pair of groups  $G_1$ and $G_2$;
    \item a group homomorphism $t : G_2 \to G_1$;
    \item an  action $\rhd : G_1 \times G_2 \to G_2$;
\end{itemize}
such that
\be
    \begin{aligned}
        \text{$t$ is $G_1$-equivariant:} & \quad
        t(u \rhd y) = u t(y) u\mone , \\
        \text{Peiffer identity holds:} & \quad
        t(y) \rhd y' = y y' y\mone .
    \end{aligned}
    \label{t-map_GequivariancePeifferId}
\ee
\end{definition}
We can draw elements of a 2-group as a bigon. For example, the 2-group element $(u_s,y)\in G_1\times G_2$ is shown in fig.\, \ref{Fig:bigone}. A pair of 2-group elements $(y_1, u_1)$ and $(y_2, u_2)$ are vertically composable only if 
the target of the first 2-holonomy  coincides with the source of the second 2-holonomy: $t(y_1)u_1 = u_2$.
The compositions for the 2-group elements, represented in Fig. \ref{Fig:HorizontalComposition} and \ref{Fig:VerticalComposition}, are
\begin{align}
    \text{horizontal composition:} & \quad
    (y_1, u_1) \circ (y_2, u_2) = (y_1 (u_1 \rhd y_2), u_1 u_2) , 
    \label{HorizontalComposition} \\
    \text{vertical composition:} & \quad
    (y_1, u_1) \cdot (y_2, u_2) = (y_2 y_1, u_1)
    \,,\quad
    t(y_1)u_1 = u_2 .
    \label{VerticalComposition}
\end{align}

Several remarks can now be made.
First, the vertical and horizontal products are compatible in the following sense:
\be
    (\alpha'_1 \cdot \alpha_1) \circ (\alpha'_2 \cdot \alpha_2) = (\alpha'_1 \circ \alpha'_2) \cdot (\alpha_1 \circ \alpha_2) .
    \label{InterchangeLaw}
\ee
Second, we note that the horizontal product is nothing more than the group operation in the semi-direct product $G_1\ltimes G_2$.

Finally,  one can now compose  plaquettes as in Fig. \ref{Fig:bigone} to form a closed surface bounding a polyhedron decorated by a $G_2$ element $y_{tot}$ using the vertical product (we would obtain a surface looking like a melon). 
In this case, the reader might convince themselves that due to the Bianchi identity, we must have that 
\begin{align}
    t(y_{tot})=1.
\end{align}

\begin{figure}
    \centering
    \begin{tikzpicture}[scale=0.9]
        \coordinate (x) at (0,0);
        \coordinate (y) at (3,0);
        \coordinate (X) at (6.5,0);
        \coordinate (Y) at (9.5,0);
        \coordinate (x+) at (0.1,0);
        \coordinate (x-) at (-0.1,0);
        \coordinate (y+) at (0,0.1);
        \coordinate (y-) at (0,-0.1);
        \coordinate (x') at ($ (x) + (-2,0) $);
        \coordinate (y') at ($ (y) + (2,0) $);
        
        \draw[fill] (x) circle [radius=0.03];
        \node[below left] at (x) {$c$};
        \draw[fill] (y) circle [radius=0.03];
        \node[below right] at (y) {$c''$};
        
        \draw[-> , thick] ($ (x) + (x+) + (y+) $) to [out=60 , in=120] node[above] {$u$} ($ (y) + (x-) + (y+) $);
        \draw[-> , thick , dotted] ($ (x) + (x+) + (y-) $) to [out=-60 , in=-120] node[below] {$\bar u$} ($ (y) + (x-) + (y-) $);
        \draw[-{Implies[]} , thick , double distance=2] ($ 1/2*(x) + 1/2*(y) + 4*(y+) $) -- node[right] {$y$} ($ 1/2*(x) + 1/2*(y) + 4*(y-) $);
        
        \draw[fill] (x') circle [radius=0.03];
        \node[left] at (x') {$c'$};
        \draw[fill] (y') circle [radius=0.03];
        \node[right] at (y') {$c'''$};
        
        \draw[-> , thick] ($ (x') + (x+) $) -- node[above] {$u'$} ($ (x) + (x-) $);
        \draw[-> , thick] ($ (y) + (x+) $) -- node[above] {$u''$} ($ (y') + (x-) $);
        
        \node at ($ (y') + 7.5*(x+) $) {$=$};
        
        \draw[fill] (X) circle [radius=0.03];
        \node[left] at (X) {$c'$};
        \draw[fill] (Y) circle [radius=0.03];
        \node[right] at (Y) {$c'''$};
        
        \draw[-> , thick] ($ (X) + (x+) + (y+) $) to [out=60 , in=120] node[above] {$u'uu''$} ($ (Y) + (x-) + (y+) $);
        \draw[-> , thick , dotted] ($ (X) + (x+) + (y-) $) to [out=-60 , in=-120] node[below] {$\bar u$} ($ (Y) + (x-) + (y-) $);
        \draw[-{Implies[]} , thick , double distance=2] ($ 1/2*(X) + 1/2*(Y) + 4*(y+) $) -- node[right] {$u' \rhd y$} ($ 1/2*(X) + 1/2*(Y) + 4*(y-) $);
    \end{tikzpicture}
\caption{Consider the 2-group element $(y,u)$; the wedge $y$ is rooted at the source of the link $u$, the node $c$. We can root this 2-group element  at the node $c'$, using the holonomy $u'$. This is equivalent to consider a new 1-holonomy rooted at $c'$.  To change the target of $u$, we multiply horizontally on the right by a 2-group element with trivial face decoration.  }
    \label{fig:whisker}
\end{figure}
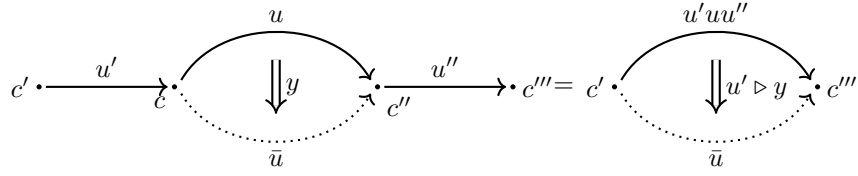

\paragraph{Whiskering.}
Recall that in establishing the geometric meaning of the face decorations and the $t$ map, we divide the holonomy of a plaquette into two paths, a source and a target. The choice of how to divide the plaquette into a source and a target should not impact the compatibility conditions we have imposed. The source or root of the bigon is changed using a process called \textit{whiskering}. 
Changing the root of the 2-holonomy from the source of the 1-holonomy to the target of it is an example of whiskering: following an arbitrary path, we can always move the root of a 2-holonomy by composing horizontally with a 2-group element which has a trivial face decoration. This is 
illustrated in Fig. \ref{fig:whisker}.  

\begin{figure}
    \centering
    \begin{tikzpicture}[scale=0.9]
        \coordinate (x) at (0,0);
        \coordinate (y) at (3,0);
        \coordinate (X) at (4,0);
        \coordinate (Y) at (7,0);
        \coordinate (x+) at (0.1,0);
        \coordinate (x-) at (-0.1,0);
        \coordinate (y+) at (0,0.1);
        \coordinate (y-) at (0,-0.1);
        \coordinate (x') at ($ (x) + (-2,0) $);
        \coordinate (y') at ($ (y) + (2,0) $);
        
        \draw[fill] 
        (x) circle [radius=0.03] node[below left] at (x) {$c$}
        (x') circle [radius=0.03] node[below left] at (x') {$c'$}
        (X) circle [radius=0.03] node[below left] at (X) {$c'$};
        
        \node at ($ 0.5*(y) + 0.5*(X) $) {$=$};
        
        \draw[-> , thick] ($ (x') + (x+) $) -- node[above] {$u'$} ($ (x) + (x-) $);
        
        \draw[-> , thick] 
        ($ (x) + (x+) + (y+) $) to [out=60 , in=120] node[above] {$u$} 
        ($ (y) + (y+) $) to [out=-70 , in=70]
        ($ (y) + (y-) $) to [out=-120 , in=-60] 
        ($ (x) + (x+) + (y-) $);
        \draw[-{Implies[]} , thick , double distance=2] ($ 1/2*(x) + 1/2*(y) + 6*(x+) $) -- node[above] {$y$} ($ 1/2*(x) + 1/2*(y) + 6*(x-) $);
        
        \draw[-> , thick] 
        ($ (X) + (x+) + (y+) $) to [out=60 , in=120] node[above] {$u'u{u'}\mone$}
        ($ (Y) + (y+) $) to [out=-70 , in=70]
        ($ (Y) + (y-) $) to [out=-120 , in=-60]
        ($ (X) + (x+) + (y-) $);
        \draw[-{Implies[]} , thick , double distance=2] ($ 1/2*(X) + 1/2*(Y) + 6*(x+) $) -- node[above] {$u' \rhd y$} ($ 1/2*(X) + 1/2*(Y) + 6*(x-) $);
    \end{tikzpicture}
\caption{Example of an adjoint whiskering changing the root of 2-holonomy. Starting from Fig. \ref{fig:whisker}, we identify $c$ and $c''$, as well as $c'''$ and $c'$.}
    \label{fig:adjwhisker}
\end{figure}
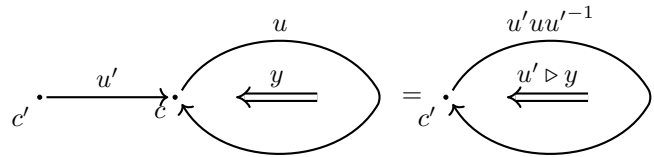

As a 
{particular}
case, 
{let}
the source of $y$ be a closed holonomy with root and target $c$, see Fig. \ref{fig:adjwhisker}. If we want to change this root to $c'$, then we need to use an adjoint whiskering, since we need to transport both the source of $u$ and its target to $c'$.  Indeed, 
from the definition of the $t$-map,
we have 
\begin{align}\label{adjwhisk}
    1=t(y)u= u't(y) u  u'{}^{-1} = t(u'\rhd y) u' u  u'{}^{-1}
\end{align}

\medskip 

There are two families of 2-groups that stand out.

\paragraph{Skeletal/trivial 2-groups.}
A skeletal/trivial 2-group $\cG$ is a 2-group with a trivial $t$-map, $t(y) = 1$ for all $y \in G_2$, which implies that the group $G_2$ is abelian. Typical examples are:  
the Poincar\'e 2-group with $G_1=SO(3,1)$ and $G_2=\R^4$, or the adjoint 2-group with $G_2= \g_1$, the Lie algebra of $G_1$ which is viewed as an abelian group.

\paragraph{Inner/identity 2-group.}
The identity 2-group $\cG$ is a 2-group with $G_2=G_1$ such that the  $t$-map is given by the identity, $t(y) = y$ for all $y \in G_2$. The action is then given by the adjoint action.

\subsection{Lattice 2-gauge transformations.}\label{sec:2gauge}
The crossed module structure provides a natural framework to decorate a 2-complex with group decorations.

\begin{figure}
   \centering
    \begin{tikzpicture}
        \coordinate (a) at (0,0);
        \coordinate (b) at (0,2);
        \coordinate (c) at (1.5,2);
        \coordinate (d) at (1.5,0);
        
        \coordinate (incr) at (4.5,0);
        
        \coordinate (a') at ($ (a) + (incr) $);
        \coordinate (b') at ($ (b) + (incr) $);
        \coordinate (c') at ($ (c) + (incr) $);
        \coordinate (d') at ($ (d) + (incr) $);
        
        \coordinate (x) at (0.1,0);
        \coordinate (y) at (0,0.1);:
        
        \draw[fill] (a) circle [radius=0.025];
        \draw[fill] (b) circle [radius=0.025];
        \draw[fill] (c) circle [radius=0.025];
        \draw[fill] (d) circle [radius=0.025];
        
        \draw[fill] (a') circle [radius=0.025];
        \draw[fill] (b') circle [radius=0.025];
        \draw[fill] (c') circle [radius=0.025];
        \draw[fill] (d') circle [radius=0.025];
        
        \draw[-> , thick] ($ (a) + (y) $) -- node[left] {$h_1$} ($ (b) - (y) $);
        \draw[-> , thick] ($ (b) + (x) $) -- node[above] {$u'$} ($ (c) - (x) $);
        \draw[<- , thick] ($ (c) - (y) $) -- node[right] {$h_2$} ($ (d) + (y) $);
        \draw[<- , thick] ($ (d) - (x) $) -- node[below] {$u$} ($ (a) + (x) $);
        
        \draw[-{Implies} , thick , double distance=2] (2.5,1) -- (3.5,1);
        
        \draw[-> , thick] ($ (a') + (y) $) -- node[left] {$h_1$} ($ (b') - (y) $);
        \draw[-> , thick] ($ (b') + (x) $) -- node[above] {$u'$} ($ (c') - (x) $);
        \draw[<- , thick] ($ (c') - (y) $) -- node[right] {$h_2$} ($ (d') + (y) $);
        \draw[<- , thick] ($ (d') - (x) $) -- node[below] {$u$} ($ (a') + (x) $);
        
        \draw[-{Implies} , thick , double distance=2] ($ (b') +4.5*(x) - 6*(y) $) -- node[above right] {$x$} ($ (d') - 5*(x) + 6*(y) $);
    \end{tikzpicture}
  \caption{Generalizing the standard gauge transformation to include a 2-gauge transformation.}
    \label{fig:gauge1}
\end{figure}
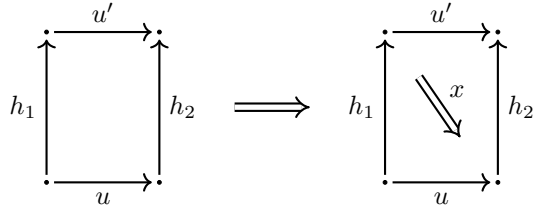

As  in the usual gauge theory framework, $1$-gauge transformations, parametrized by group elements $h_i\in G_1$, act as in \eqref{1gauge}.  
Note that the gauge transformations at both ends of a holonomy can be understood as stating that the closed 1-holonomy  in the loop generated by the gauge fiber and the 1-holonomy is flat, see Fig. \ref{fig:gauge1}.
{One has:}
\begin{equation}\label{1-g}
    u'=h_1 \mone u h_2 \Leftrightarrow u'h_2 \mone u\mone h_{1} =1
\end{equation}

\medskip 

A  2-gauge transformation, parametrized by $x\in G_2$, acts on both 1- and 2-holonomies. Considering the 1-gauge transformation of the 1-holonomy, we can generalize the gauge transformation \eqref{1-g}, by adding a face decoration in the loop, as illustrated in Fig. \ref{fig:gauge1}. In this case, we have the transformation 
\begin{equation}\label{1-gbis}
    u'=t(x)h_1\mone u h_2,
\end{equation}
where $x$ is rooted at the source of $u'$. 

\begin{figure}
    \centering
   \begin{tikzpicture}[scale=0.9]
        \coordinate (A) at (0,3,0);
        \coordinate (B) at (0,-3,0);
        
        \coordinate (a) at (1.3,0,1.25);
        \coordinate (b) at (-1.3,0,1.25);
        \coordinate (c) at (1.3,0,-1.25);
        \coordinate (d) at (-1.3,0,-1.25);
        
        \coordinate (x) at (0.1,0,0);
        \coordinate (y) at (0,0.1,0);
        \coordinate (z) at (0,0,0.1);
        
        \draw[fill] (A) circle [radius=0.025];
        \draw[fill] (B) circle [radius=0.025];
        
        \draw[- , thick] ($ (A) - (y) + (x) + (z) $) to [out=-60 , in=90] (a);
        \draw[- , thick] (a) to [out=-90 , in=60] ($ (B) + (y) + (x) + (z) $);
        
        \draw[- , thick] ($ (A) - (y) - (x) + (z) $) to [out=-150 , in=95] (b);
        \draw[- , thick] (b) to [out=-80 , in=150] ($ (B) + (y) - (x) + (z) $);
        
        \draw[- , thick] ($ (A) - (y) + (x) - (z) $) to [out=-30 , in=95] (c);
        \draw[- , thick] (c) to [out=-80 , in=30] ($ (B) + (y) + (x) - (z) $);
        
        \draw[- , thick , dotted] ($ (A) - (y) - (x) - (z) $) to [out=-120 , in=90] (d);
        \draw[- , thick , dotted] (d) to [out=-90 , in=120] ($ (B) + (y) - (x) - (z) $);
        
        \draw[-{Implies} , double distance=2] ($ 3/4*(b) + 1/4*(a) $) -- node[above] {$y$} ($ 1/4*(b) + 3/4*(a) $);
        
        \draw[-{Implies} , double distance=2 , dotted] ($ 3/4*(d) + 1/4*(c) $) -- node[above left] {$y'$} ($ 1/4*(d) + 3/4*(c) $);
        
        \draw[-{Implies} , double distance=2] ($ 3/4*(a) + 1/4*(c) $) -- node[below=0.15] {$x_1$} ($ 1/4*(a) + 3/4*(c) $);
        
        \draw[-{Implies} , double distance=2 , dotted] ($ 3/4*(b) + 1/4*(d) $) -- node[above=0.15] {$x_2$} ($ 1/4*(b) + 3/4*(d) $);
        
        \node[right] at (3,0,0) {$y' = x_1 y x_2\mone$};
    \end{tikzpicture}
    \caption{2-gauge transformations on the 2-holonomy. This is the 2-analogue of the closed holonomy on the LHS of Fig. \ref{fig:gauge1}. }
    \label{fig:gauge2}
\end{figure}
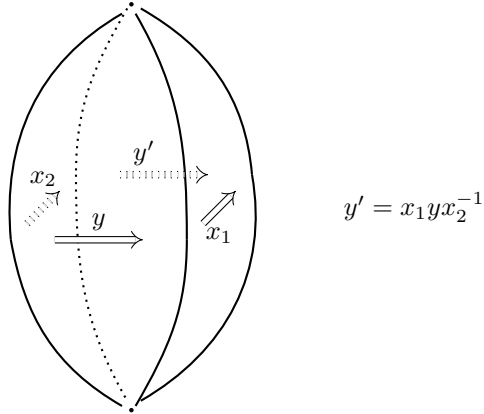

Similarly to the 1-gauge transformations, we can transform both the source and the target of the 2-holonomy. Using the product of the 2-holonomies, namely the vertical product, we can obtain the 2-gauge transformation on the 2-holonomy by imposing that the closed 2-holonomy, melon like, built from the face contribution in the  fiber and the lattice is flat, see Fig \ref{fig:gauge2}.  
\begin{equation}\label{2-gbis}
    y'=x_1 y x_2\mone \Leftrightarrow y'x_2 y\mone x_1\mone =1
\end{equation}

\medskip 

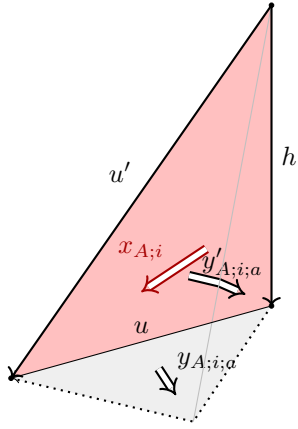
\begin{figure}
     \centering
    \begin{tikzpicture}
        \coordinate (c) at (0,4,0);
        \coordinate (c') at (0,0,0);
        \coordinate (c'') at (-2.5,0,2.5);
        \coordinate (v) at (0.5,0,4);
        
        \path[fill=lightgray , nearly transparent] (c') -- (c'') -- (v);
        \path[fill=red , nearly transparent] (c) -- (c') -- (c'');
        
        \draw[-{Implies} , thick , double distance=2 , darkred] ($ 1/4*(c) + 1/2*(c') + 1/4*(c'') $) -- node[above left] {$x_{A;i}$} ($ 1/6*(c) + 1/3*(c') + 1/2*(c'') $);
        \draw[-{Implies} , thick , double distance=2] ($ 1/3*(c') + 1/3*(c'') + 1/3*(v) $) -- node[above right] {$y_{A;i;a}$} ($ 1/6*(c') + 1/6*(c'') + 2/3*(v) $);
        \draw[-{Implies} , thick , double distance=2] ($ 1/3*(c) + 1/6*(c'') + 1/2*(v) $) -- ($ 1/3*(c) + 2/3*(v) $) -- node[above] {$y'_{A;i;a}$} ($ 1/6*(c) + 1/2*(c') + 1/3*(v) $);
        
        \draw[- , thin , lightgray] (c) -- (v);
        
        \draw[fill] (c) circle [radius=0.03];
        \draw[fill] (c') circle [radius=0.03];
        \draw[fill] (c'') circle [radius=0.03];
        
        \draw[-> , thick] (c) -- node[right] {$h$} (c');
        \draw[-> , thick] (c) -- node[above left] {$u'$} (c'');
        \draw[->] (c') -- node[above] {$u$} (c'');
        \draw[- , thick , dotted] (c') -- (v) -- (c'');
        
        \node[right] at (3,2.5,3) {$ (y',u') = (x,h)^{-1 \, _V} \circ \big(y,u\big)\rightarrow
        \left\{\,
        \begin{aligned}
            &
            u' = t(x) h u \\
            &
            y' = (h \rhd y) x\mone
        \end{aligned}
        \right.
        $};
    \end{tikzpicture}
    \caption{A combination of 1- and 2-gauge transformations. We note that while we have a left (horizontal) transformation, the $y$ is transformed on the right, see \eqref{2-GaugeTransformation0}. This is due to the presence of the $t$-map. When the  $t$-map is trivial, $y$ is abelian so left or right transformation does not matter for the $y$ transformation.  }
    \label{fig:gauge3}
\end{figure}

Simultaneous 2- and 1-gauge transformations give rise to the 
{so-called}
``tin can" \cite{Baez:2002jn, Girelli:2007tt}. In particular, we can focus on the ``half squashed" tin can where we only perform a gauge transformation at 
{the}
source of the 1-holonomy and at the source of the 2-holonomy, see Fig \ref{fig:gauge3}.  

It is interesting to note that the above pair of constraints \eqref{1-gbis}, \eqref{2-gbis} encoding the 2-gauge transformations   can be written as an horizontal composition in the following way,
\be
    (y',u') = (x,h)^{-1 \, _V} \circ \big(y,u\big)= \big(x\mone, t(x)h\big)\circ \big(y,u\big)=\big(x\mone \, ([t(x)h]\rhd y), t(x)h u \big) = \big((h\rhd y)\, x\mone, t(x)h u \big),
    \label{2-GaugeTransformation0}
\ee
where $(y,u)^{-1 \, _V}=(y\mone, t(y)u)$.   

\section{Review of  (1-)group field theory}\label{1gft}

We  review 
{in this section}
the construction of the 
Boulatov group field theory (GFT) 
\cite{Boulatov:1992vp}. The aim is to construct a field theory which allows to recover,  through its Feynman diagrams,   the Ponzano-Regge (PR)
model \cite{Ponzano_Regge_1969} when dealing with the Lie group SU(2). Feynman diagrams themselves are dual to (possibly degenerate) triangulations of a $3$-manifold and the corresponding Feynman amplitudes are the Ponzano-Regge state sum for that triangulation.

The Ponzano-Regge model is often expressed in the triangulation picture, in terms of representations of
{the} group $SU(2)$,
 through the 6j symbol \cite{Ponzano_Regge_1969}{,}
\begin{align}\label{PR1}
    \cZ(\triangle)=\sum_{\{j\}}\prod_\tau \{6j\}_\tau(j_i), 
\end{align}
where $j_i$ are half integers encoding the representations of SU(2), decorating the six edges of the tetrahedron $\tau$ in the triangulation $\triangle$.  

We can also express Ponzano partition function in terms of the dual complex, decorated by SU(2) holonomies,   
\begin{align}\label{PR2}
    \cZ(\triangle^*)=\int\dh \prod_e\delta(h_e), 
\end{align}
where $e$ are the edges of the triangulation, while $h_e$ is the SU(2) holonomy decorating the boundary of the face dual to $e$. 

There is yet a third formulation of the PR model which is also defined on the triangulation, in terms of non-commutative variables \cite{Baratin:2010wi, Baratin:2010nn}. 
\begin{align}\label{PR3}
    \cZ(\triangle)=\int\dx \prod_t\delta(x^t_1+x^t_2+x^t_3), 
\end{align}
where $x_i^t\in\R^3$ are vectors associated to the edges of the triangle $t$ in the triangulation.

These three partition functions are related through different types of Fourier transforms. One can go from \eqref{PR2} to \eqref{PR1} through the standard Fourier transform on groups in terms of representation theory,  using the Peter-Weyl theorem. 
{Moreover, one} 
can relate \eqref{PR2} to \eqref{PR3} through a generalized Fourier transform, which is most rigorously described in terms of dual Hopf algebras \cite{Majid:1996kd}.  

In this review section, we 
construct the GFT such that the associated Feynman diagram amplitude are  
\eqref{PR2}.
 We 
 emphasize the geometry behind 
 the construction, 
 as this will provide useful hints on how to generalize this {notion} to the $2$-group case.

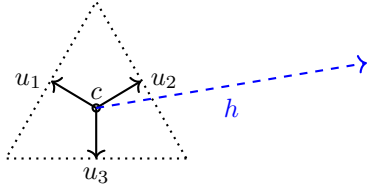
\begin{figure}
    \centering
    \begin{tikzpicture}[scale=1.2]
        \draw[- , thick , dotted] (1,0) -- (-1,0) -- (0,1.74) -- (1,0);
        
        \draw[<- , thick] (-0.5,0.86) -- (0,0.56);
        \draw[<- , thick] (0.5,0.86) -- (0,0.56);
        \draw[<- , thick] (0,0) -- (0,0.56);
        
        \draw[thick] (0,0.56) circle [radius=0.04];
        \node[above] at (0,0.56) {$c$};
        
        \node[left] at (-0.5,0.86) {$u_1$};
        \node[right] at (0.5,0.86) {$u_2$};
        \node[below] at (0,0) {$u_3$};
    
        \draw[-> , thick , blue , dashed] (0,0.56) -- (3,1.06); 
        \node[blue] at (1.5,0.56) {$h$};
    \end{tikzpicture}
    \caption{From a 2d perspective, the three holonomies $u_i$ are dual to three edges of a triangle.}
    \label{fig:1gft1}
\end{figure}

\paragraph{Field.} 
Let us specify $G=\SU(2)$ for simplicity. In the standard Boulatov GFT construction, the field  $\phi$ is a real function $\phi: G\times G\times G \to \R$.   
The field $\phi(u_1, u_2, u_3)$ is invariant under the global gauge transformation
\be\label{gauge1}
u_i \,\,\, \to \,\,\, u'=h u_i, \quad    \phi(u_1, u_2, u_3) \,\,\, \to \,\,\, \phi(hu_1, hu_2, hu_3). 
\ee\label{symm}
To implement this invariance, we use gauge averaging, through the projector $\cP$,
\be
    \big(\cP\phi\big)(u_1,u_2,u_3) = \int \dh \, \phi(hu_1, hu_2, hu_3).
\ee
{Moreover,}
the fields 
are the eigenvectors of  the projector $\cP${:}
\be
    \phi(u_1, u_2, u_3) = \int \dh \, \phi(hu_1, hu_2, hu_3) = \big(\cP \phi\big)(u_1, u_2, u_3).
\ee
The variables $u_i$ can then be seen as group variables assigned to three line segments (links) which share a common point, that is they are rooted at the same node. These 
{variables} 
$u_i$ play the role of the holonomies in the dual complex. If these three holonomies are in a two-dimensional surface, then the gauge transformation $h$ can be seen as decorating an edge with a component perpendicular to the surface as illustrated in Fig. \ref{fig:1gft1}. This holonomy 
becomes a component of the holonomy appearing in \eqref{PR2}.

On top of gauge invariance, we can also consider a symmetry under permutation of the arguments $u_i$ in the field $\phi$ \cite{Boulatov:1992vp, Freidel:2005qe, Krajewski:2011zzu}. If we note $\sigma$ the permutation in permutation group $S_3$, the field would be then
\begin{align}
    \phi(u_1,u_2,u_3) \equiv (-1)^{|\sigma|} \phi(u_{\sigma(1)},u_{\sigma(2)}, u_{\sigma(3)} ),  
\end{align}
where $|\sigma|$ is the parity of the permutation. 
In the following, we will not emphasize this symmetry and keep it un-explicit to not burden the presentation. 

The action for the scalar field is given in terms of a propagator and an interaction term. 

\paragraph{Propagator.} 
If the group elements $u_i$ are interpreted as momentum variables\footnote{We then have a curved momentum space, unlike standard QFT.} 
the propagator implements a 
conservation of momentum
\be 
    \int \du^3 \,
    \phi(u_1, u_2, u_3) \,
    \phi(u_1, u_2, u_3)
    =
    \int \du^6 \, 
    \cK(u_i) \, 
    \phi(u_1, u_2, u_3) \,
    \phi(u_4, u_5, u_6) ,
\ee
with 
\be
    \cK(u_1,u_2,u_3; u_4,u_5,u_6) = \delta(u_1u_4\mone) \delta(u_2u_5\mone) \delta(u_3u_6\mone).
    \label{PropagatorAmplitude_3d}
\ee
Geometrically, this conservation of momenta can be seen as a pairwise identification of holonomy variables $u_i$, decorating the $2d$ graph dual of the triangle.

\begin{figure}
     \centering
        \centering
        \begin{tikzpicture}[scale=1.5 , rotate around y=-45]
        \coordinate (v1) at (1.5,0,0);
        \coordinate (v2) at (-1.5,0,0);
        \coordinate (v3) at (0,2.5,0);
        
        \coordinate (w1) at (1.5,0,-1.5);
        \coordinate (w2) at (-1.5,0,-1.5);
        \coordinate (w3) at (0,2.5,-1.5);
        
        \coordinate (c) at ($ 1/3*(v1) + 1/3*(v2) + 1/3*(v3) $);
        \coordinate (ct) at ($ (c) + (0,0,2.5) $);
        \coordinate (c') at ($ 1/3*(w1) + 1/3*(w2) + 1/3*(w3) $);
        \coordinate (ct') at ($ (c') - (0,0,2.5) $);
        \coordinate (c'') at ($ 0.85*(c') + 0.15*(ct') $);
        
        \draw[-> , thick] ($ 0.95*(v1) + 0.05*(v2) $) -- ($ 0.05*(v1) + 0.95*(v2) $);
        \draw[-> , thick] ($ 0.95*(v2) + 0.05*(v3) $) -- ($ 0.05*(v2) + 0.95*(v3) $);
        \draw[-> , thick] ($ 0.95*(v3) + 0.05*(v1) $) -- ($ 0.05*(v3) + 0.95*(v1) $);
        \draw[<- , thick] ($ 0.95*(w1) + 0.05*(w2) $) -- ($ 0.05*(w1) + 0.95*(w2) $);
        \draw[<- , thick] ($ 0.95*(w2) + 0.05*(w3) $) -- ($ 0.05*(w2) + 0.95*(w3) $);
        \draw[<- , thick] ($ 0.95*(w3) + 0.05*(w1) $) -- ($ 0.05*(w3) + 0.95*(w1) $);
        
        \draw[fill , darkblue] (c) circle [radius=0.03];
        \draw[fill , darkblue] (c') circle [radius=0.03]; 
        
        \draw[-> , thick , darkblue] (c') -- (0,-0.6,-1.5);
        \draw[-> , thick , darkblue] (c') -- (-1.5,1.5,-1.5);
        \draw[-> , thick , darkblue] (c') -- (1.5,1.5,-1.5);
        
        \draw[-> , thick , darkblue] (c) -- (0,-0.6,0);
        \draw[-> , thick , darkblue] (c) -- (-1.5,1.5,0);
        \draw[-> , thick , darkblue] (c) -- (1.5,1.5,0);
        
        \draw[fill , darkred] 
        (ct) circle [radius=0.03] node[above left , scale=0.9] {$c_1$}
        (ct') circle [radius=0.03] node[above , scale=0.9] {$c_2$};
        
        \draw[-> , thick , darkred]
        ($ 0.975*(ct) + 0.025*(c) $) -- node[above , scale=0.9] {$h_{c_1}$} ($ 0.975*(c) + 0.025*(ct) $);
        \draw[- , thick , darkred]
        ($ 0.975*(ct') + 0.025*(c') $) -- node[above , scale=0.9] {$h_{c_2}$} (c'');
        \draw[-> , thick , darkred , dotted]
        (c'') -- ($ 0.975*(c') + 0.025*(ct') $);
    \end{tikzpicture} 
    \caption{The propagator 
leads to the
identification of the pair of triangles. This means that the holonomies $h_{c_i}$ going from the center  $c_i$ to the triangle are fused. 
    }
    \label{recoveringholonomy0}
\end{figure}
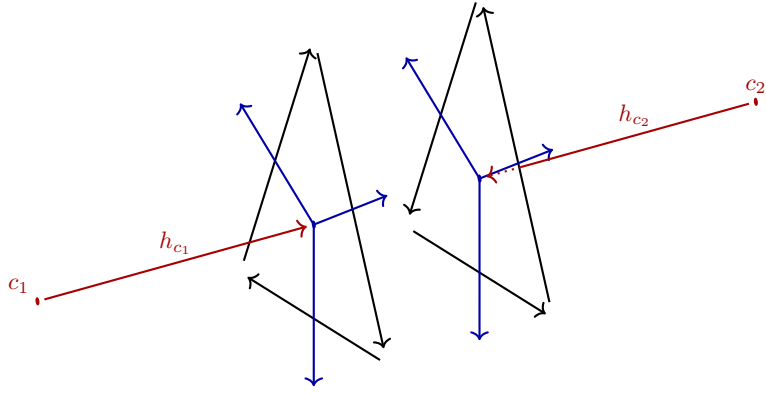

The identification of the ``boundary" triangle data leads therefore to gluing the  $h$'s associated to each field, see Fig. \ref{recoveringholonomy0}. 

\medskip

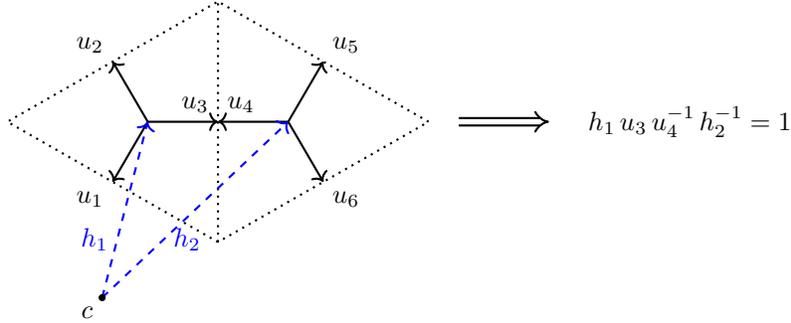
\begin{figure}
    \centering
    \begin{tikzpicture}[scale=0.8]
    \coordinate (c) at (0,1,5);
    
    \coordinate (v1) at (0,0,0);
    \coordinate (v2) at (0,4,0);
    \coordinate (v31) at (3.5,2,0);
    \coordinate (v32) at (-3.5,2,0);
    
    \coordinate (z) at ($ 1/2*(v1) + 1/2*(v2) $);
    \coordinate (x1) at ($ 1/2*(v1) + 1/2*(v32) $);
    \coordinate (x2) at ($ 1/2*(v2) + 1/2*(v32) $);
    \coordinate (y1) at ($ 1/2*(v1) + 1/2*(v31) $);
    \coordinate (y2) at ($ 1/2*(v2) + 1/2*(v31) $);
    
    \coordinate (c1) at ($ 1/3*(v1) + 1/3*(v2) + 1/3*(v32)$);
    \coordinate (c2) at ($ 1/3*(v1) + 1/3*(v2) + 1/3*(v31)$);
    
    \draw[- , thick , dotted] (v1) -- (v2) -- (v31) -- cycle;
    \draw[- , thick , dotted] (v1) -- (v32) -- (v2);
    
    \draw[-> , thick] (c1) -- (z);
    \draw[-> , thick] (c1) -- (x1);
    \draw[-> , thick] (c1) -- (x2);
    
    \draw[-> , thick] (c2) -- (z);
    \draw[-> , thick] (c2) -- (y1);
    \draw[-> , thick] (c2) -- (y2);
    
    \draw[-> , thick , dashed , blue] (c) -- (c1);
    \draw[-> , thick , dashed , blue] (c) -- (c2);
    
    \draw[fill] (c) circle [radius=0.05];
    \node[below left] at (c) {$c$};
    
    \node[above left] at (z) {$u_3$};
    \node[above left] at (x2) {$u_2$};
    \node[below left] at (x1) {$u_1$};
    
    \node[above right] at (z) {$u_4$};
    \node[above right] at (y2) {$u_5$};
    \node[below right] at (y1) {$u_6$};
    
    \node[left , blue] at ($ 1/3*(c1) + 2/3*(c) $) {$h_1$};
    \node[right , blue] at ($ 1/3*(c2) + 2/3*(c) $) {$h_2$};
    
    \draw[-{Implies[]} , double distance=2 , thick] ($ (v31) + (0.5,0,0) $) -- ($ (z) + (5.5,0,0) $);
    
    \node[right] at ($ (z) + (6,0,0) $) {$h_1 \, u_3 \, u_4\mone \, h_2\mone = 1$};
    \end{tikzpicture}
    \caption{The interaction term generates flat holonomies, spanned by the wedge dual to an edge.   }
    \label{fig:vertex}
\end{figure}

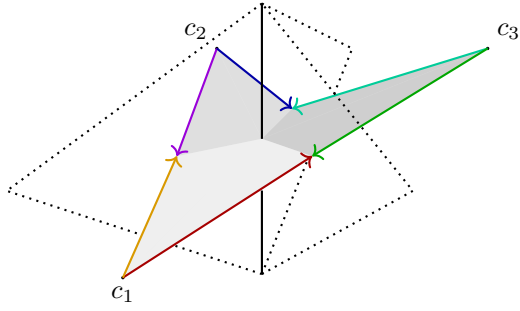
\begin{figure}
    \centering
    \begin{tikzpicture}[scale=0.6, rotate around y=0]
        \coordinate (a) at (-4.5,3,3);
        \coordinate (b) at (0,6,0);
        \coordinate (c) at (0,0,0);
        \coordinate (d) at (4.5,3,3);
        \coordinate (e) at (0,3,-5.2);
        \coordinate (bc) at ($ 1/2*(b) + 1/2*(c) $);
        \coordinate (ct) at ($ 1/3*(a) + 1/3*(b) + 1/3*(c) $);
        \coordinate (ct') at ($ 1/3*(d) + 1/3*(b) + 1/3*(c) $);
        \coordinate (ct'') at ($ 1/3*(e) + 1/3*(b) + 1/3*(c) $);
        \coordinate (C1) at ($ (bc) + (0,0,8) $);
        \coordinate (C2) at ($ (bc) + (-3,0,-5.2) $);
        \coordinate (C3) at ($ (bc) + (3,0,-5.2) $);
        \coordinate (C) at ($ (C1) + (0,7,0) $);
        
        \draw[fill] (C1) circle [radius=0.025] node[below] {$c_1$};
        \draw[fill] (C2) circle [radius=0.025] node[above left] {$c_2$};
        \draw[fill] (C3) circle [radius=0.025] node[above right] {$c_3$};
        
        \path[nearly transparent , fill=gray] (C2) -- (ct'') -- (bc);
        \path[nearly transparent , fill=gray] (C2) -- (ct) -- (bc);
        
        \draw[-> , thick , carblue] (C2) -- (ct);
        \draw[-> , thick , darkblue] (C2) -- (ct'');
        
        \path[nearly transparent , fill=darkgray] (C3) -- (ct') -- (bc);
        \path[nearly transparent , fill=darkgray] (C3) -- (ct'') -- (bc);
        
        \draw[-> , thick , cargreen] (C3) -- (ct'');
        \draw[-> , thick , darkgreen] (C3) -- (ct');
        
        \path[nearly transparent , fill=lightgray] (C1) -- (ct') -- (bc);
        \path[nearly transparent , fill=lightgray] (C1) -- (ct) -- (bc);
        
        \draw[-> , thick , darkred] (C1) -- (ct');
        \draw[-> , thick , carred] (C1) -- (ct);

        \draw[- , thick , dotted] (c) -- ($ 0.35*(a) + 0.65*(c) $);
        \draw[- , thick , dotted] (b) -- (a) -- ($ 0.48*(a) + 0.52*(c) $);
        \draw[- , thick , dotted] (c) -- (d) -- (b);
        \draw[- , thick , dotted] (c) -- ($ 0.51*(e) + 0.49*(c) $);
        \draw[- , thick , dotted] (b) -- (e) -- ($ 0.79*(e) + 0.21*(c) $);
        \draw[- , thick , fill=black] (b) -- (bc);
        \draw[- , thick , fill=black] ($ (c) + (0,1.85,0) $) -- (c);
    \end{tikzpicture}
    \caption{We illustrate how the we can glue together the boundary of three internal wedges to form the holonomy $h_e= h_{c_1c_2}h_{c_2c_3}h_{c_3c_1}$ dual to the (solid) edge $e$. A similar picture will appear when discussing the 2-group case, where the half wedges will themselves carry some decoration.}
    \label{recoveringholonomy1}
\end{figure}

\paragraph{Interaction term.} 
The 
interaction term consists of building the wedges - the faces of the dual complex -
in order to recover the graph dual to a tetrahedron. This eventually leads to constructing the $h_e$ appearing in 
\eqref{PR2} which form the wedge's boundary.
This is illustrated in Fig. \ref{fig:vertex}. 
In order to recover the 6j symbol combinatorics, we consider a quartic interaction. We have then four
variables 
$h$'s 
which 
generate six wedges, which are dual to the edges of the tetrahedron.  
{One writes}
\be
    \int {\du^6} \, 
    (\cP\phi)_{123}\, 
    (\cP\phi)_{345} \,
    (\cP\phi)_{561} \,
    (\cP\phi)_{642} 
    =
    \int \du^{12} \dh^4 \,
    \cV \,
    \phi_{123} \,
    \phi_{456} \,
    \phi_{789} \,
    \phi_{10 \, 11 \, 12} ,
\ee
with 
\be
    \cV = 
    \delta(h_1 u_1u_9\mone h_3\mone) \, 
    \delta(h_1 u_2u_{12}\mone h_4\mone) \,
    \delta(h_1 u_3u_4\mone h_2\mone) \,
    \delta(h_2 u_5u_{11}\mone h_4\mone) \,
    \delta(h_2 u_6u_7\mone h_3\mone) \, 
    \delta(h_3 u_8u_{10}\mone h_4\mone) .
    \label{VertexAmplitude_3d_Gauge}
\ee 
The propagator term fuses the $h$'s contribution by identifying the momenta of the fields. The quartic interaction makes sure we have the combinatorics of a  tetrahedron. An edge in the triangulation can be shared by several tetrahedra. Dually, this amounts to gluing the internal wedges together. 
Unlike the $2$-group case, 
we only have data on the boundary of the internal wedges (see Fig. \ref{recoveringholonomy1}). 
At the end of the day, we recover the topological invariant partition function \eqref{PR2}.  

\paragraph{Topological invariance.}
The GFT model we introduced defines a topological model. This is reflected by the fact the partition function \eqref{PR2} is invariant under 
Pachner moves up to some (potentially infinite) constant of proportionality.  We will note $V_G$ the volume of the group $G$.
In three dimensions, there exist  two Pachner moves:
\begin{itemize}
    \item[$\mathrm{P_{1,4}}$:]
    relates the amplitude of one 3-simplex $\cV_1$ to the amplitude of the combination of four 3-simplices $\cV_4$
    \be
        \cA_{\cV_4} = V_G^4  \, \cA_{\cV_1} \,,
    \ee
    where we removed by hand  some potential divergences coming from some $\delta(1)$ contribution.
    \item[$\mathrm{P_{2,3}}$:]
    relates the amplitude of the combination of two 3-simplices $\cV_2$ to the generating function of the combination of three 3-simplices $\cV_3$
    \be
        \cA_{\cV_3} = V_G^2  \, \cA_{\cV_2} \,,
    \ee
    where we again removed by hand  some potential divergences coming from some $\delta(1)$ contributions.
\end{itemize}

Note that the divergences we removed by hand coming from $\delta(1)$ contributions is a well-known issue and  moving to the quantum group case with $q$ the deformation parameter being of root of unity,  allows to remove these divergences in a more proper way.

\section{2-group field theory: definition and action}\label{2gft}

In this section we first 
introduce the different notations and restrictions on the choice of 2-groups which will simplify the construction. Then we will define the action in terms of a kinetic term and an interaction term.

\subsection{Conventions and assumptions}

\paragraph{2-group choice.}
We 
work with the strict 2-group/crossed module  $\cG=(G_1,G_2, t, \rhd)$. In this section, the crossed module $\cG$ 
can be 
a finite crossed module, {\it i.e.} $G_i$ are finite groups, or a crossed module defined in terms of  Lie groups. In the latter case, we 
consider only the class of crossed module where the $G_i$ are unimodular and such that the Haar measure on $G_2$ is invariant under the action of $G_1$. While the construction can be probably extended to a more general situation for Lie groups, it would complicate the presentation to deal with the more general case.

\paragraph{Notations. } 
In the following,  
$4$-simplices $\sigma_\mu$ will be indexed by $\mu=1...$ ,   the five boundary tetrahedra are labelled $\tau_{\mu; A}$ with $A=1,..,5$,  triangles $t_{\mu; A; i}$ with  $i=1,..4$ and  edges $e_{\mu;A;i;a}$  with $a=1,2,3$ . 
When there is no ambiguity, we will suppress some indices.

In the dual 2-complex,  links are labelled by their dual counterparts. The duality can be in 3d or 4d according to whether we are on the boundary or the bulk. A boundary holonomy would be labelled $u_{A;i}$ as being dual to the triangle $i$ in the tetrahedron $A$. If we deal with the boundary of different $4$-simplices we 
would also include  the Greek index. A bulk holonomy, dual to a tetrahedron $A$ of the 4-simplex $\mu$, would be labelled $h_{\mu;A}$. 

On the other hand, a boundary  wedge dual to the edge shared by triangles $i$ and $j$ of tetrahedron $A$  would be labelled by $\y_{A;i,j}$. If we deal with the boundary of different $4$-simplices we 
would also include  the Greek index. A bulk wedge, dual to a triangle $i$ of tetrahedron $A$ in the 4-simplex $\mu$ would be labelled by $x_{\mu;A;i}$. Later on we will also introduce the (fused) wedge shared by tetrahedra $A$ and $B$ in the 4-simplex $\mu$ which will be labelled by $X_{\mu;A,B}$.

Let us 
fix a $4$-simplex.
Let $c$  be a node which is the center of a 4-simplex, $c_A$ be the center of a tetrahedron $\tau_A$ and $c_{A;i}$ the center of the $i^{th}$ triangle of the boundary of $\tau_A$. The holonomy $u_{A;i}$ has for  source  $c_A$ and target  $c_{A;i}$, for $i=1,..,4$. 
Once again, we will add the index $\mu$ if we consider different 4-simplices.



\subsection{Action for a group field theory on 2-groups}

In this subsection, we introduce the field and its geometric interpretation and we build an action which will generate topological invariant amplitudes.
In the 3+1 case, in the standard GFT context the field is interpreted as the 1-complex dual to a tetrahedron. In the 2-group case, the field will be interpreted as the 2-complex dual to a tetrahedron.
It encodes the kinematical picture. The dynamics will be given in terms of the  interaction term which will provide the rules of how to construct the 2-complex dual to a 4d triangulation.

\subsubsection{Field.} 
Since we focus on the 2-complex,  in addition to the 4 variables associated to the 1-simplices or links, we also include six (internal) wedges 
which are subtended by pairs of links.

\smallskip

We introduce the variables $\y_{i,j}$ with $j>i$, $i,j=1..4$, associated to the wedges which are subtended by the holonomies $u_i$ and $u_j$. These wedges variables are rooted at the center of the tetrahedron. The set of variables $\y_{i,j}$ and $u_i$ undergo 2-gauge transformations as illustrated in Fig. \ref{fig:gauge30} which is a generalization of Fig. \ref{fig:gauge3}. 
 \begin{align}\label{Y2gauge}
           & u_i' = t(x_i) h u_i \,\,,\quad i=1,2 \qquad
            \y_{12}' = (h \rhd \y_{12})\, x_2\mone x_1
        \end{align}

\begin{definition}
The geometric (real scalar) field is a function of four copies of the group $G_1$ and six copies of the group $G_2$
\be
    \Phi(\y_{1,2},\y_{1,3},\y_{1,4},\y_{2,3},\y_{2,4}\y_{3,4};u_1,u_2,u_3,u_4).
    \label{FusedInternalWedges}
\ee
Given the 2-gauge transformation \eqref{Y2gauge}
 the projected geometric field $(\cP \, \Phi)$ is the  2-gauge averaged  field
\begin{align}
    (\cP \, \Phi)(\y_{i,j};u_i) 
    & =
    \int \dh \dx^4 \, \Phi\big((h\rhd \y_{i,j}) \, x_j\mone x_i;\, t(x_i)hu_i\big) 
    \,.
    \label{GaugeProjField}
\end{align}
\end{definition}
If we were using a finite crossed modules, we would replace all the integrals by sums. 
\begin{figure}
    \centering
    \begin{tikzpicture}
        \coordinate (c) at (0,5,0);
        \coordinate (c') at (0,0,0);
        \coordinate (c'') at (-1.5,0,3.5);
        \coordinate (c''') at (3,0,3.5);
        \coordinate (v) at (1.25,0,5);
        
        \coordinate (x) at (0.1,0,0);
        \coordinate (y) at (0,0.1,0);
        \coordinate (z) at (0,0,0.1);
        
        \path[fill=lightgray , nearly transparent] (c') -- (c'') -- (v) -- (c''');
        \path[fill=red , nearly transparent] (c) -- (c') -- (c'');
        \path[fill=green , nearly transparent] (c) -- (c') -- (c''');
        
        \draw[-{Implies} , thick , double distance=2 , darkred] ($ 1/4*(c) + 1/2*(c') + 1/4*(c'') $) -- node[above left] {$x_2$} ($ 1/6*(c) + 1/3*(c') + 1/2*(c'') $);
        \draw[-{Implies} , thick , double distance=2 , darkgreen] ($ 1/4*(c) + 1/2*(c') + 1/4*(c''') $) -- node[right] {$x_1$} ($ 1/6*(c) + 1/3*(c') + 1/2*(c''') $);
        \draw[-{Implies} , thick , double distance=2] ($ 1/3*(c') + 1/3*(c'') + 1/3*(v) $) -- ($ 1/3*(c') + 2/3*(v) $) -- node[above] {$Y_{12}$} ($ 1/3*(c') + 1/3*(c''') + 1/3*(v) $);
        \draw[-{Implies} , thick , double distance=2] ($ 1/3*(c) + 1/6*(c'') + 1/2*(v) $) -- ($ 1/3*(c) + 2/3*(v) $) -- node[above] {$Y_{12}'$} ($ 1/3*(c) + 1/6*(c''') + 1/2*(v) $);
        
        \draw[- , thin , lightgray] (c) -- (v);
        
        \draw[fill] (c) circle [radius=0.03];
        \draw[fill] (c') circle [radius=0.03];
        \draw[fill] (c'') circle [radius=0.03];
        \draw[fill] (c''') circle [radius=0.03];
        
        \draw[-> , thick] ($ (c) - (y) $) -- node[left] {$h$} ($ (c') + (y) $);
        
        \draw[-> , thick] ($ (c) - (y) + 0.5*(z) - 0.25*(x) $) -- node[above left] {$u_2'$} ($ (c'') + (y) - 0.5*(z) + 0.25*(x) $);
        \draw[->] ($ (c') + (z) - (x) $) -- node[above] {$u_2$} ($ (c'') - (z) + 0.5*(x) $);
        \draw[-> , thick] ($ (c) - (y) + 0.5*(z) + 0.25*(x) $) -- node[above right] {$u_1'$} ($ (c''') + (y) - 0.5*(z) - 0.25*(x) $);
        \draw[->] ($ (c') + (z) + (x) $) -- node[above] {$u_1$} ($ (c''') - (z) - 0.5*(x) $);
        \draw[- , thick , dotted] (c'') -- (v) -- (c''');
        
        \node[right] at (3.5,2.5,3) {$
        \left\{\,
        \begin{aligned}
            &
            u_i' = t(x_i) h u_i \,\,,\quad i=1,2 \\
            &
            \y_{12}' = (h \rhd \y_{12})\, x_2\mone x_1
        \end{aligned}
        \right.
        $};
\end{tikzpicture} 
\caption{The top node (source of $h$) is the center of the would be 4-simplex, while the bottom one (target of $h$) is the center of the tetrahedron. The internal wedge $\y_{12}$ is transforming under $x_1$, $x_2$, and $h$.}\label{fig:gauge30}
\end{figure}
We note that  $\cP$ is clearly a projector due to the invariance of the different Haar measures, up to possible diverging contribution coming from the infinite volume\footnote{This is similar to the construction of the 1-GFT with a non-compact group.} of $\cG$. 

From a 4d perspective, the $(x,h)$'s variables are \textit{bulk variables}, whereas the $(\y,u)$'s variables are the \textit{boundary variables}. 

Finally, just like for the 1-GFT, we can also demand the field to be invariant under permutations $\sigma\in S_4$, up to the parity of the permutation.  Since the $\y$ are bounded by a pair of $u_i$'s, permuting the $u$'s  changes accordingly the indices for the $\y$.      
\be
\Phi(\y_{i,j};u_i)\equiv (-1)^{|\sigma|} \Phi(\y_{\sigma(1),\sigma(2)},\y_{\sigma(1),\sigma(3)},\y_{\sigma(1),\sigma(4)},\y_{\sigma(2),\sigma(3)},\y_{\sigma(2),\sigma(4)}\y_{\sigma(3),\sigma(4)};u_{\sigma(1)},u_{\sigma(2)},u_{\sigma(3),u_\sigma(4)}),
\ee 
where $|\sigma|$ is the parity of the permutation. 
In the following, we will typically assume this symmetry without making it explicit in order to not burden the notations.

The action of 2-GFT is given by the contribution of a kinematic and an interaction term
\be\label{2GFT_Action}
    \cS = \cS_{\cK} + \cS_{\cV} \,.
\ee

\subsubsection{Propagator.}

The kinetic term is given by the  product of a pair of fields.
\be
    \cS_{\cK} = \int \du^4 [\mathrm{d}\y]^{6} \, \Phi(\y_{i,j},u_{i}) \,\Phi(\y_{i,j},u_{i}) \,.
\ee
It can be written as an integral operator
\be
    \cS_{\cK} = \int \du^8 \dY^{12} \, \cK \,
    \phi(\y_{A;i,j},u_{A;i}) \phi(\y_{B;i,j},u_{B;i}) \,, \quad A\neq B.
    \label{KineticTerm_DualComplex}
\ee
The integration kernel
\be
    \cK = \prod_{i=1 \, j>i}^4 \, 
    \delta_{G_2}\big(\y_{A;i,j}\mone \, \y_{B;i,j}) \,
    \delta_{G_1}(u_{A;i}\mone \, u_{B;i})\, , \quad A\neq B
    \label{PropagatorAmplitude_DualComplex}
\ee
is called integration kernel of the \textit{propagator amplitude}. 
It can also be seen as a conservation of momenta as in the 1-GFT case. 

\medskip 

\subsubsection{Interaction}
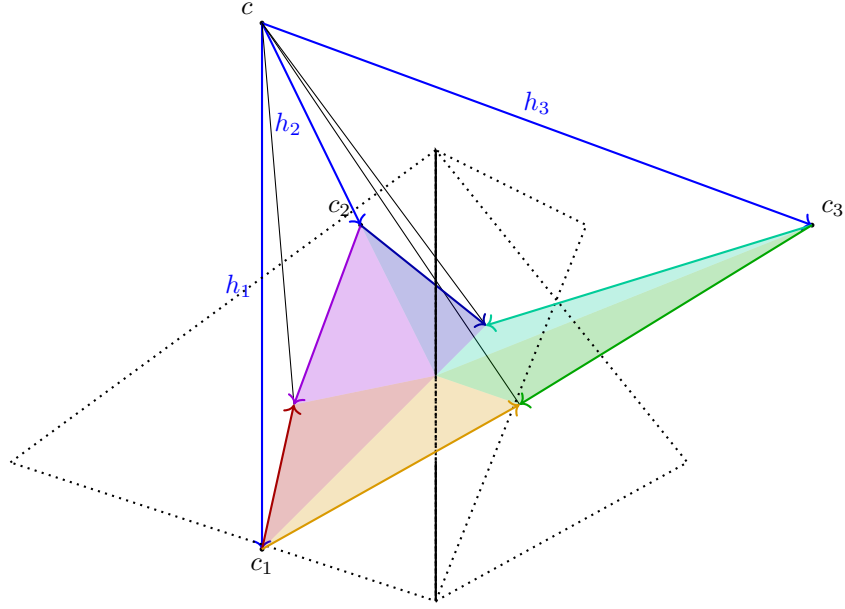
\begin{figure}
\centering
    \begin{tikzpicture}[rotate around y=0]
        \coordinate (a) at (-4.5,3,3);
        \coordinate (b) at (0,6,0);
        \coordinate (c) at (0,0,0);
        \coordinate (d) at (4.5,3,3);
        \coordinate (e) at (0,3,-5.2);
        \coordinate (bc) at ($ 1/2*(b) + 1/2*(c) $);
        \coordinate (ct) at ($ 1/3*(a) + 1/3*(b) + 1/3*(c) $);
        \coordinate (ct') at ($ 1/3*(d) + 1/3*(b) + 1/3*(c) $);
        \coordinate (ct'') at ($ 1/3*(e) + 1/3*(b) + 1/3*(c) $);
        \coordinate (C1) at ($ (bc) + (0,0,6) $);
        \coordinate (C2) at ($ (bc) + (-3,0,-5.2) $);
        \coordinate (C3) at ($ (bc) + (3,0,-5.2) $);
        \coordinate (C) at ($ (C1) + (0,7,0) $);
        
        \draw[fill] (C) circle [radius=0.025] node[above left] {$c$};
        \draw[fill] (C1) circle [radius=0.025] node[below] {$c_1$};
        \draw[fill] (C2) circle [radius=0.025] node[above left] {$c_2$};
        \draw[fill] (C3) circle [radius=0.025] node[above right] {$c_3$};
        
        \draw[-> , thick, blue] (C) -- node[left] {$h_1$} (C1);
        \draw[-> , thick, blue] (C) -- node[left] {$h_2$} (C2);
        \draw[-> , thick, blue] (C) -- node[above] {$h_3$} (C3);
        \draw[-] (C) -- (ct);
        \draw[-] (C) -- (ct');
        \draw[-] (C) -- (ct'');
        
        \path[nearly transparent , fill=darkred] (C1) -- (ct) -- (bc);
        \path[nearly transparent , fill=carred] (C1) -- (ct') -- (bc);
        \path[nearly transparent , fill=darkblue] (C2) -- (ct'') -- (bc);
        \path[nearly transparent , fill=carblue] (C2) -- (ct) -- (bc);
        \path[nearly transparent , fill=darkgreen] (C3) -- (ct') -- (bc);
        \path[nearly transparent , fill=cargreen] (C3) -- (ct'') -- (bc);

        \draw[-> , thick , carblue] (C2) -- (ct);
        \draw[-> , thick , darkblue] (C2) -- (ct'');
        
        \draw[- , thick , dotted] (a) -- (b) -- (c) -- cycle;
        \draw[- , thick , dotted] (c) -- (e) -- (b) -- cycle;
        
        \draw[-> , thick , cargreen] (C3) -- (ct'');
        \draw[-> , thick , darkgreen] (C3) -- (ct');
        
        \draw[- , thick , dotted] (c) -- (d) -- (b) -- cycle;
         \draw[- , thick , fill=black] (b) -- (bc);
        \draw[- , thick , fill=black] ($ (c) + (0,1.85,0) $) -- (c);
        
        \draw[-> , thick , darkred] (C1) -- (ct);
        \draw[-> , thick , carred] (C1) -- (ct');
    \end{tikzpicture}
    \caption{A portion of the vertex term. The dotted lines are the combination of three faces shared by three tetrahedra and sharing a common edge. The solid blue lines are three of the five  links $h_i$ that have their source at $c$, the center of the 4-simplex and their target at $c_A$, the center of the tetrahedron $A$. In different colors the six half wedges, that combine pairwise to give the fused internal wedges $\y$. 
    The combination of the six half wedges is the total wedge dual the the central edge.
    }
    \label{Fig_Vertex 0}
\end{figure}

\begin{figure}
    \centering
        \begin{tikzpicture}[scale=0.55]
        \coordinate (v11) at (0,-5);
        \coordinate (v12) at (2,-5);
        \coordinate (v13) at (1,-6.73);
        \coordinate (v142) at (1,-3.27);
        \coordinate (v143) at (-1,-6.73);
        \coordinate (v144) at (3,-6.73);
        
        \coordinate (c11) at ($1/3*(v11) + 1/3*(v12) + 1/3*(v13)$);
        \coordinate (c12) at ($1/3*(v11) + 1/3*(v12) + 1/3*(v142)$);
        \coordinate (c13) at ($1/3*(v11) + 1/3*(v13) + 1/3*(v143)$);
        \coordinate (c14) at ($1/3*(v12) + 1/3*(v13) + 1/3*(v144)$);
        
        \node[above left , scale=1.2] at (v11) {$\tau_1$};
        
        \draw[- , thick] (v11) -- (v12) -- (v13) -- cycle;
        \draw[- , thick] (v12) -- (v144) -- (v13);
        \draw[- , thick] (v13) -- (v143) -- (v11);
        \draw[- , thick] (v11) -- (v142) -- (v12);
        
        \node[violet] at (c11) {$t_{1;4}$};
        \node[blue] at (c12) {$t_{1;2}$};
        \node[green] at (c13) {$t_{1;3}$};
        \node[red] at (c14) {$t_{1;1}$};
        
        \draw[decoration={markings,mark=at position 0.6 with {\arrow[scale=1.5,thick,>=stealth]{>}}},postaction={decorate}] (v12) -- (v11);
        \draw[decoration={markings,mark=at position 0.6 with {\arrow[scale=1.5,thick,>=stealth]{>}}},postaction={decorate}] (v12) -- (v13);
        \draw[decoration={markings,mark=at position 0.6 with {\arrow[scale=1.5,thick,>=stealth]{>}}},postaction={decorate}] (v12) -- (v142);
        \draw[decoration={markings,mark=at position 0.6 with {\arrow[scale=1.5,thick,>=stealth]{>}}},postaction={decorate}] (v12) -- (v144);
        \draw[decoration={markings,mark=at position 0.6 with {\arrow[scale=1.5,thick,>=stealth]{>}}},postaction={decorate}] (v13) -- (v144);
        \draw[decoration={markings,mark=at position 0.6 with {\arrow[scale=1.5,thick,>=stealth]{>}}},postaction={decorate}] (v11) -- (v142);
        \draw[decoration={markings,mark=at position 0.6 with {\arrow[scale=1.5,thick,>=stealth]{>}}},postaction={decorate}] (v13) -- (v143);
        \draw[decoration={markings,mark=at position 0.6 with {\arrow[scale=1.5,thick,>=stealth]{>}}},postaction={decorate}] (v11) -- (v143);
        \draw[decoration={markings,mark=at position 0.6 with {\arrow[scale=1.5,thick,>=stealth]{>}}},postaction={decorate}] (v11) -- (v13);
        
        \coordinate (v21) at (5,-9);
        \coordinate (v22) at (7,-9);
        \coordinate (v23) at (6,-7.23);
        \coordinate (v242) at (8,-7.23);
        \coordinate (v243) at (6,-10.73);
        \coordinate (v244) at (4,-7.23);
        
        \coordinate (c21) at ($1/3*(v21) + 1/3*(v22) + 1/3*(v23)$);
        \coordinate (c22) at ($1/3*(v22) + 1/3*(v23) + 1/3*(v242)$);
        \coordinate (c23) at ($1/3*(v21) + 1/3*(v22) + 1/3*(v243)$);
        \coordinate (c24) at ($1/3*(v21) + 1/3*(v23) + 1/3*(v244)$);
        
        \node[below right , scale=1.2] at (v22) {$\tau_2$};
        
        \draw[- , thick] (v21) -- (v22) -- (v23) -- cycle;
        \draw[- , thick] (v22) -- (v242) -- (v23);
        \draw[- , thick] (v22) -- (v243) -- (v21);
        \draw[- , thick] (v21) -- (v244) -- (v23);
        
        \node[teal] at (c21) {$t_{2;3}$};
        \node[magenta] at (c22) {$t_{2;1}$};
        \node[brown] at (c23) {$t_{2;2}$};
        \node[red] at (c24) {$t_{2;4}$};
        
        \draw[decoration={markings,mark=at position 0.6 with {\arrow[scale=1.5,thick,>=stealth]{>}}},postaction={decorate}] (v23) -- (v244);
        \draw[decoration={markings,mark=at position 0.6 with {\arrow[scale=1.5,thick,>=stealth]{>}}},postaction={decorate}] (v23) -- (v242);
        \draw[decoration={markings,mark=at position 0.6 with {\arrow[scale=1.5,thick,>=stealth]{>}}},postaction={decorate}] (v23) -- (v21);
        \draw[decoration={markings,mark=at position 0.6 with {\arrow[scale=1.5,thick,>=stealth]{>}}},postaction={decorate}] (v23) -- (v22);
        \draw[decoration={markings,mark=at position 0.6 with {\arrow[scale=1.5,thick,>=stealth]{>}}},postaction={decorate}] (v21) -- (v22);
        \draw[decoration={markings,mark=at position 0.6 with {\arrow[scale=1.5,thick,>=stealth]{>}}},postaction={decorate}] (v21) -- (v244);
        \draw[decoration={markings,mark=at position 0.6 with {\arrow[scale=1.5,thick,>=stealth]{>}}},postaction={decorate}] (v22) -- (v242);
        \draw[decoration={markings,mark=at position 0.6 with {\arrow[scale=1.5,thick,>=stealth]{>}}},postaction={decorate}] (v21) -- (v243);
        \draw[decoration={markings,mark=at position 0.6 with {\arrow[scale=1.5,thick,>=stealth]{>}}},postaction={decorate}] (v22) -- (v243);

        \coordinate (v31) at (0,0);
        \coordinate (v32) at (2,0);
        \coordinate (v33) at (1,1.73);
        \coordinate (v342) at (3,1.73);
        \coordinate (v343) at (1,-1.73);
        \coordinate (v344) at (-1,1.73);
        
        \coordinate (c31) at ($1/3*(v31) + 1/3*(v32) + 1/3*(v33)$);
        \coordinate (c32) at ($1/3*(v32) + 1/3*(v33) + 1/3*(v342)$);
        \coordinate (c33) at ($1/3*(v31) + 1/3*(v32) + 1/3*(v343)$);
        \coordinate (c34) at ($1/3*(v31) + 1/3*(v33) + 1/3*(v344)$);
        
        \node[below left , scale=1.2] at (v31) {$\tau_3$};
        
        \draw[- , thick] (v31) -- (v32) -- (v33) -- cycle;
        \draw[- , thick] (v32) -- (v342) -- (v33);
        \draw[- , thick] (v32) -- (v343) -- (v31);
        \draw[- , thick] (v31) -- (v344) -- (v33);
        
        \node[olive] at (c31) {$t_{3;2}$};
        \node[magenta] at (c32) {$t_{3;4}$};
        \node[blue] at (c33) {$t_{3;3}$};
        \node[orange] at (c34) {$t_{3;1}$};
        
        \draw[decoration={markings,mark=at position 0.6 with {\arrow[scale=1.5,thick,>=stealth]{>}}},postaction={decorate}] (v32) -- (v342);
        \draw[decoration={markings,mark=at position 0.6 with {\arrow[scale=1.5,thick,>=stealth]{>}}},postaction={decorate}] (v32) -- (v343);
        \draw[decoration={markings,mark=at position 0.6 with {\arrow[scale=1.5,thick,>=stealth]{>}}},postaction={decorate}] (v32) -- (v31);
        \draw[decoration={markings,mark=at position 0.6 with {\arrow[scale=1.5,thick,>=stealth]{>}}},postaction={decorate}] (v32) -- (v33);
        \draw[decoration={markings,mark=at position 0.6 with {\arrow[scale=1.5,thick,>=stealth]{>}}},postaction={decorate}] (v33) -- (v342);
        \draw[decoration={markings,mark=at position 0.6 with {\arrow[scale=1.5,thick,>=stealth]{>}}},postaction={decorate}] (v33) -- (v344);
        \draw[decoration={markings,mark=at position 0.6 with {\arrow[scale=1.5,thick,>=stealth]{>}}},postaction={decorate}] (v31) -- (v344);
        \draw[decoration={markings,mark=at position 0.6 with {\arrow[scale=1.5,thick,>=stealth]{>}}},postaction={decorate}] (v31) -- (v343);
        \draw[decoration={markings,mark=at position 0.6 with {\arrow[scale=1.5,thick,>=stealth]{>}}},postaction={decorate}] (v31) -- (v33);

        \coordinate (v41) at (-5,-9);
        \coordinate (v42) at (-3,-9);
        \coordinate (v43) at (-4,-7.23);
        \coordinate (v442) at (-2,-7.23);
        \coordinate (v443) at (-4,-10.73);
        \coordinate (v444) at (-6,-7.23);
        
        \coordinate (c41) at ($1/3*(v41) + 1/3*(v42) + 1/3*(v43)$);
        \coordinate (c42) at ($1/3*(v42) + 1/3*(v43) + 1/3*(v442)$);
        \coordinate (c43) at ($1/3*(v41) + 1/3*(v42) + 1/3*(v443)$);
        \coordinate (c44) at ($1/3*(v41) + 1/3*(v43) + 1/3*(v444)$);
        
        \node[below left , scale=1.2] at (v41) {$\tau_4$};
        
        \draw[- , thick] (v41) -- (v42) -- (v43) -- cycle;
        \draw[- , thick] (v42) -- (v442) -- (v43);
        \draw[- , thick] (v42) -- (v443) -- (v41);
        \draw[- , thick] (v41) -- (v444) -- (v43);
        
        \node[cyan] at (c41) {$t_{4;1}$};
        \node[green] at (c42) {$t_{4;2}$};
        \node[brown] at (c43) {$t_{4;3}$};
        \node[orange] at (c44) {$t_{4;4}$};
        
        \draw[decoration={markings,mark=at position 0.6 with {\arrow[scale=1.5,thick,>=stealth]{>}}},postaction={decorate}] (v43) -- (v442);
        \draw[decoration={markings,mark=at position 0.6 with {\arrow[scale=1.5,thick,>=stealth]{>}}},postaction={decorate}] (v42) -- (v442);
        \draw[decoration={markings,mark=at position 0.6 with {\arrow[scale=1.5,thick,>=stealth]{>}}},postaction={decorate}] (v43) -- (v42);
        \draw[decoration={markings,mark=at position 0.6 with {\arrow[scale=1.5,thick,>=stealth]{>}}},postaction={decorate}] (v43) -- (v41);
        \draw[decoration={markings,mark=at position 0.6 with {\arrow[scale=1.5,thick,>=stealth]{>}}},postaction={decorate}] (v42) -- (v41);
        \draw[decoration={markings,mark=at position 0.6 with {\arrow[scale=1.5,thick,>=stealth]{>}}},postaction={decorate}] (v43) -- (v444);
        \draw[decoration={markings,mark=at position 0.6 with {\arrow[scale=1.5,thick,>=stealth]{>}}},postaction={decorate}] (v41) -- (v444);
        \draw[decoration={markings,mark=at position 0.6 with {\arrow[scale=1.5,thick,>=stealth]{>}}},postaction={decorate}] (v41) -- (v443);
        \draw[decoration={markings,mark=at position 0.6 with {\arrow[scale=1.5,thick,>=stealth]{>}}},postaction={decorate}] (v42) -- (v443);

        \coordinate (v51) at (0,-11);
        \coordinate (v52) at (2,-11);
        \coordinate (v53) at (1,-12.73);
        \coordinate (v542) at (1,-9.27);
        \coordinate (v543) at (-1,-12.73);
        \coordinate (v544) at (3,-12.73);
        
        \coordinate (c51) at ($1/3*(v51) + 1/3*(v52) + 1/3*(v53)$);
        \coordinate (c52) at ($1/3*(v51) + 1/3*(v52) + 1/3*(v542)$);
        \coordinate (c53) at ($1/3*(v51) + 1/3*(v53) + 1/3*(v543)$);
        \coordinate (c54) at ($1/3*(v52) + 1/3*(v53) + 1/3*(v544)$);
        
        \node[above left , scale=1.2] at (v51) {$\tau_5$};
        
        \draw[- , thick] (v51) -- (v52) -- (v53) -- cycle;
        \draw[- , thick] (v52) -- (v544) -- (v53);
        \draw[- , thick] (v53) -- (v543) -- (v51);
        \draw[- , thick] (v51) -- (v542) -- (v52);
        
        \node[olive] at (c51) {$t_{5;3}$};
        \node[violet] at (c52) {$t_{5;1}$};
        \node[cyan] at (c53) {$t_{5;4}$};
        \node[teal] at (c54) {$t_{5;2}$};
        
        \draw[decoration={markings,mark=at position 0.6 with {\arrow[scale=1.5,thick,>=stealth]{>}}},postaction={decorate}] (v52) -- (v542);
        \draw[decoration={markings,mark=at position 0.6 with {\arrow[scale=1.5,thick,>=stealth]{>}}},postaction={decorate}] (v52) -- (v51);
        \draw[decoration={markings,mark=at position 0.6 with {\arrow[scale=1.5,thick,>=stealth]{>}}},postaction={decorate}] (v52) -- (v53);
        \draw[decoration={markings,mark=at position 0.6 with {\arrow[scale=1.5,thick,>=stealth]{>}}},postaction={decorate}] (v52) -- (v544);
        \draw[decoration={markings,mark=at position 0.6 with {\arrow[scale=1.5,thick,>=stealth]{>}}},postaction={decorate}] (v51) -- (v542);
        \draw[decoration={markings,mark=at position 0.6 with {\arrow[scale=1.5,thick,>=stealth]{>}}},postaction={decorate}] (v51) -- (v53);
        \draw[decoration={markings,mark=at position 0.6 with {\arrow[scale=1.5,thick,>=stealth]{>}}},postaction={decorate}] (v544) -- (v53);
        \draw[decoration={markings,mark=at position 0.6 with {\arrow[scale=1.5,thick,>=stealth]{>}}},postaction={decorate}] (v543) -- (v53);
        \draw[decoration={markings,mark=at position 0.6 with {\arrow[scale=1.5,thick,>=stealth]{>}}},postaction={decorate}] (v51) -- (v543);

        \draw[- , double distance=2pt , thick , dotted , blue] (v142) -- (v343);
        \draw[- , double distance=2pt , thick , dotted , green] (v143) -- (v442);
        \draw[- , double distance=2pt , thick , dotted , red] (v144) -- (v244);
        \draw[- , double distance=2pt , thick , dotted , magenta] (v242) to [out=30 , in=30] (v342);
        \draw[- , double distance=2pt , thick , dotted , orange] (v444) to [out=150 , in=150] (v344);
        \draw[- , double distance=2pt , thick , dotted , brown] (v443) to [out=-90 , in=180] (1,-15) to [out=0 , in=-90] (v243);
        \draw[- , double distance=2pt , thick , dotted , violet] (v542) -- (v13);
        \draw[- , double distance=2pt , thick , dotted , teal] (v544) to [out=-30 , in=-150] (v21);
        \draw[- , double distance=2pt , thick , dotted , cyan] (v543) to [out=-150 , in=-30] (v42);
        
        \draw[- , double distance=2pt , thick , dotted , olive] (v33) to [out=90 , in=90] ($ (v33) + (9,0) $) -- (10,-15) to [out=-90 , in=-90] (1,-15.5);
        \draw[- , double distance=2pt , thick , dotted , olive] (1,-14.5) -- (v53);

        \draw[fill] (v12) circle [radius=0.1];
        \draw[thick , fill=white] (v11) circle [radius=0.1];
        \draw[thick , fill=gray] (v13) circle [radius=0.1];
        \draw[thick , fill=lime] (v142) circle [radius=0.1];
        \draw[thick , fill=lime] (v143) circle [radius=0.1];
        \draw[thick , fill=lime] (v144) circle [radius=0.1];
        
        \draw[fill] (v23) circle [radius=0.1];
        \draw[thick , fill=gray] (v21) circle [radius=0.1];
        \draw[thick , fill=yellow] (v22) circle [radius=0.1];
        \draw[thick , fill=lime] (v242) circle [radius=0.1];
        \draw[thick , fill=lime] (v243) circle [radius=0.1];
        \draw[thick , fill=lime] (v244) circle [radius=0.1];
        
        \draw[fill] (v32) circle [radius=0.1];
        \draw[thick , fill=white] (v31) circle [radius=0.1];
        \draw[thick , fill=yellow] (v33) circle [radius=0.1];
        \draw[thick , fill=lime] (v342) circle [radius=0.1];
        \draw[thick , fill=lime] (v343) circle [radius=0.1];
        \draw[thick , fill=lime] (v344) circle [radius=0.1];
        
        \draw[thick , fill=white] (v43) circle [radius=0.1];
        \draw[thick , fill=yellow] (v41) circle [radius=0.1];
        \draw[thick , fill=gray] (v42) circle [radius=0.1];
        \draw[thick , fill=lime] (v442) circle [radius=0.1];
        \draw[thick , fill=lime] (v443) circle [radius=0.1];
        \draw[thick , fill=lime] (v444) circle [radius=0.1];
        
        \draw[fill] (v52) circle [radius=0.1];
        \draw[thick , fill=white] (v51) circle [radius=0.1];
        \draw[thick , fill=yellow] (v53) circle [radius=0.1];
        \draw[thick , fill=gray] (v542) circle [radius=0.1];
        \draw[thick , fill=gray] (v543) circle [radius=0.1];
        \draw[thick , fill=gray] (v544) circle [radius=0.1];

        \draw[fill] ($(v343) + (-4,1)$) circle [radius=0.1];
        \node[right] at ($(v343) + (-4,1)$) {$:v$};
        
        \draw[fill=white] ($(v343) + (-4,0)$) circle [radius=0.1];
        \node[right] at ($(v343) + (-4,0)$) {$:v'$};
        
        \draw[fill=gray] ($(v343) + (-4,-1)$) circle [radius=0.1];
        \node[right] at ($(v343) + (-4,-1)$) {$:v''$};
        
        \draw[thick , fill=yellow] ($(v343) + (-4,-2)$) circle [radius=0.1];
        \node[right] at ($(v343) + (-4,-2)$) {$:v'''$};
        
        \draw[thick , fill=lime] ($(v343) + (-4,-3)$) circle [radius=0.1];
        \node[right] at ($(v343) + (-4,-3)$) {$:v''''$};
    \end{tikzpicture}
    \caption{4-simplex boundary construction: five tetrahedra share five vertices. Each of the four faces of each tetrahedron is identified with one of the faces of the other four tetrahedra. We use the same color and a double dotted line for the identified faces.}
    \label{Fig_4-simplex}
\end{figure}
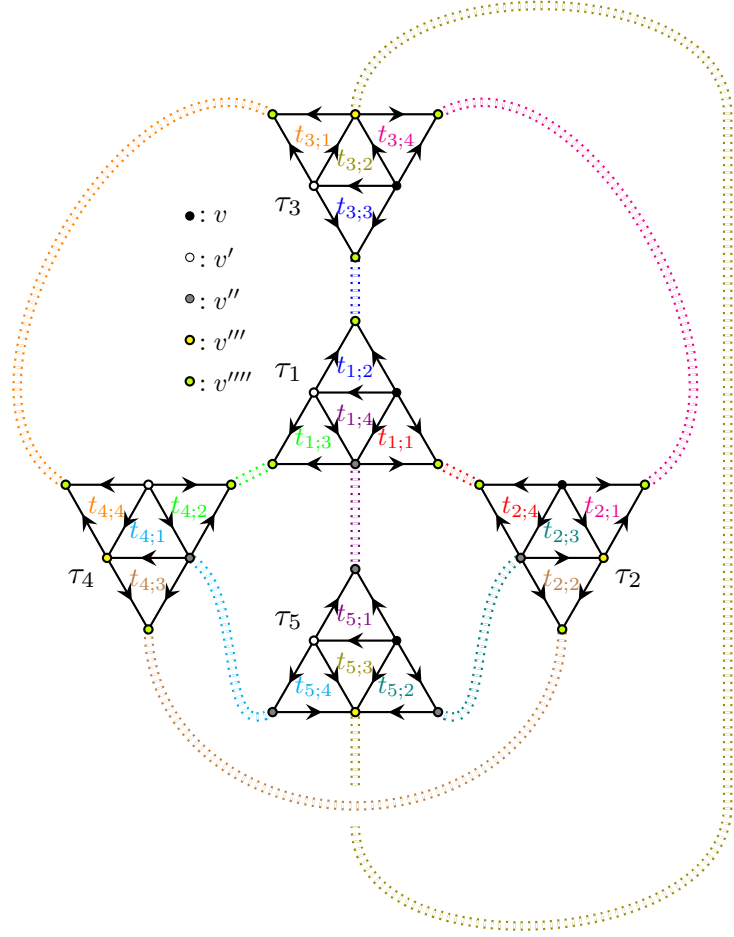

The interaction term is given by the proper identification of five  fields, such that they respect the combinatorics of a 4-simplex. 
\be
    \cS_{\cV} = 
    \int \dY^{30}\du^{20} \dX^{10}\dh^5 \, \cV \, \phi^{\times 5} \,.
\ee
We call the integration kernel $\cV$ of the \textit{vertex amplitude}.
As in the 1-GFT, the \textit{vertex amplitude} is the most geometrically meaningful object. We expect this interaction term to encode two types of flat holonomies.
\begin{itemize}
    \item A generalization of the flat 1-holonomies encountered in the 1-GFT. The generalization is due to the fact that we might have some non-trivial contribution coming from the t-map. These 1-holonomies are generated by the bulk gauge variables $h$'s and the boundary data $u$'s. This is illustrated in Fig. \ref{Fig_Vertex 0}. 
    \item A flat 2-holonomy generated by the bulk 2-gauge transformations $x$ and the boundary data $y$. 
\end{itemize}
These flat 1- and 2-holonomies are glued together through the propagator that identifies the boundary data $(\y,u)$'s.  As a result of the gluing, we will obtain  flat 1- and 2-holonomies which live in a 2-complex, dual to a 4d triangulation. To keep track of the combinatorics of the 4-simplex we are using, we take Fig. \ref{Fig_4-simplex} as a reference.

\begin{align}
    \cV
    = \,\, &
        \delta_{G_1}\big(
        t(\x_{2,1}) h_{1} u_{1;1} \, u_{2;4}\mone h_{2}\mone
        \big) \,
        \delta_{G_1}\big(
        t(\x_{3,1}) h_{1} u_{1;2} \, u_{3;3}\mone h_{3}\mone
        \big) \,
        \delta_{G_1}\big(
        t(\x_{4,1}) h_{1} u_{1;3} \, u_{4;2}\mone h_{4}\mone
        \big) \,
        \delta_{G_1}\big(
        t(\x_{5,1}) h_{1} u_{1;4} \, u_{5;1}\mone h_{5}\mone
        \big) \,
        \nonumber \\
        &
        \delta_{G_1}\big(
        t(\x_{3,2}) h_{2} u_{2;1} \, u_{3;4}\mone h_{3}\mone
        \big) \,
        \delta_{G_1}\big(
        t(\x_{4,2}) h_{2} u_{2;2} \, u_{4;3}\mone h_{4}\mone
        \big) \,
        \delta_{G_1}\big(
        t(\x_{5,2}) h_{2} u_{2;3} \, u_{5;2}\mone h_{5}\mone
        \big) \,
        \delta_{G_1}\big(
        t(\x_{4,3}) h_{3} u_{3;1} \, u_{4;4}\mone h_{4}\mone
        \big) \,
        \nonumber \\
        &
        \delta_{G_1}\big(
        t(\x_{5,3}) h_{3} u_{3;2} \, u_{5;3}\mone h_{5}\mone
        \big) \,
        \delta_{G_1}\big(
        t(\x_{5,4}) h_{4} u_{4;1} \, u_{5;4}\mone h_{5}\mone
        \big)
        \nonumber \\
        &
        \delta_{G_2}\big(
        (h_1 \rhd \y_{1;2,1}) \, \x_{1,2} \, 
        (h_2 \rhd \y_{2;4,1}) \,\x_{2,3} \,
        (h_3 \rhd \y_{3;4,3}) \, \x_{3,1}
        \big) \,
        \nonumber \\
        &
        \delta_{G_2}\big(
        (h_1 \rhd \y_{1;3,1}) \, \x_{1,2} \, 
        (h_2 \rhd \y_{2;4,2}) \,\x_{2,4} \,
        (h_4 \rhd \y_{4;3,2}) \, \x_{4,1}
        \big) \,
        \nonumber \\
        &
        \delta_{G_2}\big(
        (h_1 \rhd \y_{1;4,1}) \, \x_{1,2} \, 
        (h_2 \rhd \y_{2;4,3}) \,\x_{2,5} \,
        (h_5 \rhd \y_{5;2,1}) \, \x_{5,1}
        \big) \,
        \nonumber \\
        &
        \delta_{G_2}\big(
        (h_1 \rhd \y_{1;3,2}) \, \x_{1,3} \, 
        (h_3 \rhd \y_{3;3,1}) \,\x_{3,4} \,
        (h_4 \rhd \y_{4;4,2}) \, \x_{4,1}
        \big) \,
        \nonumber \\
        &
        \delta_{G_2}\big(
        (h_1 \rhd \y_{1;4,2}) \, \x_{1,3} \, 
        (h_3 \rhd \y_{3;3,2}) \,\x_{3,5} \,
        (h_5 \rhd \y_{5;3,1}) \, \x_{5,1}
        \big) \,
        \nonumber \\
        &
        \delta_{G_2}\big(
        (h_1 \rhd \y_{1;3,4}) \, \x_{1,4} \, 
        (h_4 \rhd \y_{4;2,1}) \,\x_{4,5} \,
        (h_5 \rhd \y_{5;4,1}) \, \x_{5,1}
        \big) \,
        \nonumber \\
        &
        \delta_{G_2}\big(
        (h_2 \rhd \y_{2;2,1}) \, \x_{2,3} \, 
        (h_3 \rhd \y_{3;4,1}) \,\x_{3,4} \,
        (h_4 \rhd \y_{4;4,3}) \, \x_{4,2}
        \big) \,
        \nonumber \\
        &
        \delta_{G_2}\big(
        (h_2 \rhd \y_{2;3,1}) \, \x_{2,3} \, 
        (h_3 \rhd \y_{3;4,2}) \,\x_{3,5} \,
        (h_5 \rhd \y_{5;3,2}) \, \x_{5,2}
        \big) \,
        \nonumber \\
        &
        \delta_{G_2}\big(
        (h_2 \rhd \y_{2;3,2}) \, \x_{2,4} \, 
        (h_4 \rhd \y_{4;3,1}) \,\x_{4,5} \,
        (h_5 \rhd \y_{5;4,2}) \, \x_{5,2}
        \big) \,
        \nonumber \\
        &
        \delta_{G_2}\big(
        (h_3 \rhd \y_{3;2,1}) \, \x_{3,4} \, 
        (h_4 \rhd \y_{4;4,1}) \,\x_{4,5} \,
        (h_5 \rhd \y_{5;4,3}) \, \x_{5,3}
        \big) \,.
    \label{VertexAmplitudeGauge}
\end{align}
The first ten deltas (on the group $G_1$) enforce the closure of the links dual to the ten identified 
faces while the other ten delta functions (on the group $G_2$) instead enforce the closure of the combination of the (boundary and bulk) wedges around the ten edges of the 4-simplex.

In order to express such closures in a compact way, we introduced the decorations of the fused external gauge wedges $X_{A,B}$ between tetrahedra $A,B$:
\be
    \begin{aligned}
        &
        \x_{1,2} = x_{1;1}\mone \, x_{2;4} \,,\\
        &
        \x_{1,3} = x_{1;2}\mone \, x_{3;3} \,,\\
        &
        \x_{1,4} = x_{1;3}\mone \, x_{4;2} \,,\\
        &
        \x_{1,5} = x_{1;4}\mone \, x_{5;1} \,,\\
        &
        \x_{2,3} = x_{2;1}\mone \, x_{3;4} \,,
    \end{aligned}
    \qquad \quad
    \begin{aligned}
        &
        \x_{2,4} = x_{2;2}\mone \, x_{4;3} \,,\\
        &
        \x_{2,5} = x_{2;3}\mone \, x_{5;2} \,,\\
        &
        \x_{3,4} = x_{3;1}\mone \, x_{4;4} \,,\\
        &
        \x_{3,5} = x_{3;2}\mone \, x_{5;3} \,,\\
        &
        \x_{4,5} = x_{4;1}\mone \, x_{5;4} \,,
    \end{aligned}
    \label{FusedExternalWedges}
\ee
with inverses $X_{A,B}\mone = X_{B,A}$.

\section{A topological model}\label{topoinv}

The general shape of the amplitude of the 2-GFT  which can be seen as the dual of a 4d triangulation $\cT$ (with no boundary) is given by
\be\label{StateSum_DualComplex}
    \cZ_\cT = \int [\rd X] [\rd h]\, \prod_{t} \delta_{G_1}( h_t t(X_t) ) \prod_{e}  \delta_{G_2}(X_e), 
\ee
where $X_t\in G_2$ decorates the face dual to the  triangle $t \in \cT$, while $h_t\in G_1$ decorates the boundary of this face.  $X_e\in G_2$ decorates the closed surface dual to an edge $e\in \cT$. 
We see that this is a natural generalization of the amplitude  obtained through a 1-GFT with a new contribution implementing the flatness of the 2-holonomy $X_e\in G_2$. 
With respect to the flatness of the 1-holonomy $h_t\in G_1$ that would appear in a regular 1-GFT, we have a possible contribution from the $t$-map, of the (open) 2-holonomy $X_t$ dual to a triangle $t$ in $\cT$.

\medskip

\begin{theorem}\label{prop:topoinv}
    Consider the GFT action given by the action \eqref{2GFT_Action} with kinetic term \eqref{PropagatorAmplitude_DualComplex} and  interaction term \eqref{VertexAmplitudeGauge}, then the associated Feynman diagrams are topological invariants. 
\end{theorem}
The proof of this theorem is given in Appendix \ref{appendix} as it is lengthy. 

To show that the constructed model is topological, one needs to check that the amplitudes are invariant under the the following Pachner moves. 
\begin{itemize}
    \item[$\mathrm{P_{1,5}}$:]
    \textit{relates the amplitude of one 4-simplex $\cV$ to the amplitude of the combination of five 4-simplices $\cV_5$};
    \item[$\mathrm{P_{2,4}}:$]
    \textit{relates the amplitude of the combination of two 4-simplices $\cV_2$ to the amplitude of the combination of four 4-simplices $\cV_4$};
    \item[$\mathrm{P_{3,3}}:$]
    \textit{relates the amplitude of the combination of three 4-simplices $\cV_3$ to the amplitude of the combination of three 4-simplices $\cV_3$}.
\end{itemize}

\paragraph{$P_{1,5}$ Pachner Move:}
First we show that the amplitudes of the two Feynman diagrams in Fig.\ \ref{Fig_Pachner(1,5)} are related to one another by an overall constant.
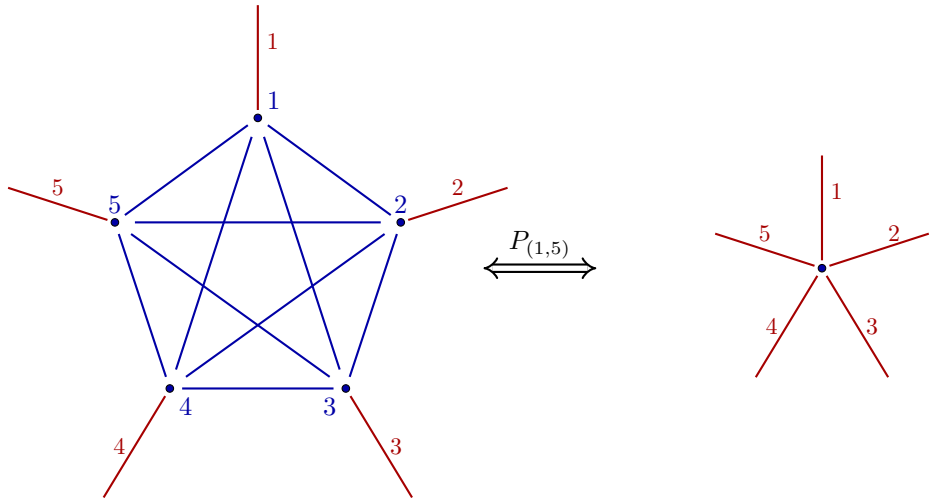
\begin{figure}[H]
    \centering
    \begin{tikzpicture}[scale=1]
        \coordinate (c) at (0,0);
        \coordinate (c1) at (0,1.5);
        \coordinate (c2) at (1.42,0.46);
        \coordinate (c3) at (0.88,-1.45);
        \coordinate (c4) at (-0.88,-1.45);
        \coordinate (c5) at (-1.42,0.46);
        
        \coordinate (incr1) at (0,2);
        \coordinate (x1) at ($ (c) + (incr1) $);
        \coordinate (x11) at ($ (c1) + (incr1) $);
        
        \coordinate (incr2) at (1.9,0.61);
        \coordinate (x2) at ($ (c) + (incr2) $);
        \coordinate (x21) at ($ (c2) + (incr2) $);
        
        \coordinate (incr3) at (1.17,-1.6);
        \coordinate (x3) at ($ (c) + (incr3) $);
        \coordinate (x31) at ($ (c3) + (incr3) $);
        
        \coordinate (incr4) at (-1.17,-1.6);
        \coordinate (x4) at ($ (c) + (incr4) $);
        \coordinate (x41) at ($ (c4) + (incr4) $);
        
        \coordinate (incr5) at (-1.9,0.61);
        \coordinate (x5) at ($ (c) + (incr5) $);
        \coordinate (x51) at ($ (c5) + (incr5) $);
        
        \coordinate (incr) at (7.5,0);
        \coordinate (c') at ($ (c) + (incr) $);
        \coordinate (c1') at ($ (c1) + (incr) $);
        \coordinate (c2') at ($ (c2) + (incr) $);
        \coordinate (c3') at ($ (c3) + (incr) $);
        \coordinate (c4') at ($ (c4) + (incr) $);
        \coordinate (c5') at ($ (c5) + (incr) $);
        
        \draw[fill=darkblue] (x1) circle [radius=0.05] node[above right , darkblue] {$1$};
        \draw[fill=darkblue] (x2) circle [radius=0.05] node[above , darkblue] {$2$};
        \draw[fill=darkblue] (x3) circle [radius=0.05] node[below left , darkblue] {$3$};
        \draw[fill=darkblue] (x4) circle [radius=0.05] node[below right , darkblue] {$4$};
        \draw[fill=darkblue] (x5) circle [radius=0.05] node[above , darkblue] {$5$};
        
        \draw[fill=darkblue] (c') circle [radius=0.05];
        
        \draw[- , thick , darkred] ($ 0.93*(x1) + 0.07*(x11) $) -- node[above right,scale=0.9] {$1$} (x11);
        
        \draw[- , thick , darkred] ($ 0.93*(x2) + 0.07*(x21) $) -- node[above,scale=0.9] {$2$} (x21);
        
        \draw[- , thick , darkred] ($ 0.93*(x3) + 0.07*(x31) $) -- node[right,scale=0.9] {$3$} (x31);
        
        \draw[- , thick , darkred] ($ 0.93*(x4) + 0.07*(x41) $) -- node[left,scale=0.9] {$4$} (x41);
        
        \draw[- , thick , darkred] ($ 0.93*(x5) + 0.07*(x51) $) -- node[above,scale=0.9] {$5$} (x51);
        
        \draw[- , thick , darkblue] ($ 0.93*(x1) + 0.07*(x2) $) --
        ($ 0.93*(x2) + 0.07*(x1) $);
        \draw[- , thick , darkblue] ($ 0.93*(x1) + 0.07*(x3) $) --
        ($ 0.93*(x3) + 0.07*(x1) $);
        \draw[- , thick , darkblue] ($ 0.93*(x1) + 0.07*(x4) $) --
        ($ 0.93*(x4) + 0.07*(x1) $);
        \draw[- , thick , darkblue] ($ 0.93*(x1) + 0.07*(x5) $) --
        ($ 0.93*(x5) + 0.07*(x1) $);
        
        \draw[- , thick , darkblue] ($ 0.93*(x2) + 0.07*(x3) $) --
        ($ 0.93*(x3) + 0.07*(x2) $);
        \draw[- , thick , darkblue] ($ 0.93*(x2) + 0.07*(x4) $) --
        ($ 0.93*(x4) + 0.07*(x2) $);
        \draw[- , thick , darkblue] ($ 0.93*(x2) + 0.07*(x5) $) --
        ($ 0.93*(x5) + 0.07*(x2) $);
        
        \draw[- , thick , darkblue] ($ 0.93*(x3) + 0.07*(x4) $) --
        ($ 0.93*(x4) + 0.07*(x3) $);
        \draw[- , thick , darkblue] ($ 0.93*(x3) + 0.07*(x5) $) --
        ($ 0.93*(x5) + 0.07*(x3) $);
        
        \draw[- , thick , darkblue] ($ 0.93*(x4) + 0.07*(x5) $) --
        ($ 0.93*(x5) + 0.07*(x4) $);

        \draw[{Implies[]}-{Implies[]} , thick , double distance=2] (3,0) -- node[above] {$\mathrel{P}_{(1,5)}$} (4.5,0);

        \draw[- , thick , darkred] ($ 0.93*(c') + 0.07*(c1') $) -- node[above right,scale=0.9] {$1$} (c1');
        \draw[- , thick , darkred] ($ 0.93*(c') + 0.07*(c2') $) -- node[above right,scale=0.9] {$2$} (c2');
        \draw[- , thick , darkred] ($ 0.93*(c') + 0.07*(c3') $) -- node[right,scale=0.9] {$3$} (c3');
        \draw[- , thick , darkred] ($ 0.93*(c') + 0.07*(c4') $) -- node[left,scale=0.9] {$4$} (c4');
        \draw[- , thick , darkred] ($ 0.93*(c') + 0.07*(c5') $) -- node[above,scale=0.9] {$5$} (c5');
    \end{tikzpicture}
    \caption{Pachner move $\mathrm{P}_{(1,5)}$.}
    \label{Fig_Pachner(1,5)}
\end{figure}
These diagrams correspond to a 4-simplex (on the right) and a graph obtained by adding an additional node (on the left). The diagram on the left is comprised of five nodes and ten bulk links, while the diagram on the right is a single node and no bulk links. The amplitude of these diagrams are 
\begin{align}
    \cA_{\cV_5} 
            & = 
            \int \dX^{50} \dh^{25} \du^{100} \dY^{150} \, 
            (\cK_{1,2} \, \cK_{1,3} \, \cK_{1,4} \, \cK_{1,5} \, \cK_{2,3} \, \cK_{2,4} \, \cK_{2,5} \, \cK_{3,4} \, \cK_{3,5} \, \cK_{4,5}) \,\, 
            \big(\cV_1 \, \cV_2 \, \cV_3 \, \cV_4 \, \cV_5\big)
\end{align}
and 
\begin{align}
\cA_{\cV} =
        \int \dX^{10} \dh^5 \du^{20} \dY^{30} \,\, \cV.
\end{align}
The subscripts refer to the labels on the diagram: for example, $\cK_{1,2}$ is the propagator between the first and second labelled vertex. The expressions are given explicitly in \eqref{Amplitude_4Simplex} and \eqref{Amplitude_Five4simplices} in the appendix.

Integrating out the bulk variables and using the invariance of the measures to define new variables, one can show that $\cA_{\cV_5}$ can be simplified and compared to $\cA_\cV$. The relevant change of variables are given in $\eqref{cv15}$ in the appendix. In the end, one finds that
\begin{align}
    \cA_{\cV_5} = V_{G_1}^4V_{G_2}^{37}\cA_{\cV}.
\end{align}
The $V_{G_i}$ stand for volumes of the group $G_i$ which may be infinite if the group is not compact.

\paragraph{$P_{2,4}$ Pachner move:}
Next, we consider the relation between the two amplitudes corresponding to the diagrams in Fig.\, \ref{Fig_Pachner(2,4)}.
\begin{figure}[H]
    \centering
    \begin{tikzpicture}[scale=1]
            \coordinate (c) at (0,0);
            \coordinate (c1) at (0,1.5);
            \coordinate (c2) at (1.42,0.46);
            \coordinate (c3) at (0.88,-1.45);
            \coordinate (c4) at (-0.88,-1.45);
            \coordinate (c5) at (-1.42,0.46);
            
            \coordinate (incr1) at (-1.25,1.25);
            \coordinate (x1) at ($ (c) + (incr1) $);
            \coordinate (x11) at ($ (c1) + (incr1) $);
            \coordinate (x15) at ($ (c5) + (incr1) $);
            
            \coordinate (incr2) at (1.25,1.25);
            \coordinate (x2) at ($ (c) + (incr2) $);
            \coordinate (x21) at ($ (c1) + (incr2) $);
            \coordinate (x24) at ($ (c2) + (incr2) $);
            
            \coordinate (incr3) at (1.25,-1.25);
            \coordinate (x3) at ($ (c) + (incr3) $);
            \coordinate (x31) at ($ (incr3) - (c1) $);
            \coordinate (x33) at ($ (incr3) - (c5) $);
            
            \coordinate (incr4) at (-1.25,-1.25);
            \coordinate (x4) at ($ (c) + (incr4) $);
            \coordinate (x41) at ($ (incr4) - (c1) $);
            \coordinate (x42) at ($ (incr4) - (c2) $);
            
            \coordinate (incr1') at (7,0);
            \coordinate (y1) at ($ (c) + (incr1') $);
            \coordinate (y11) at ($ (c1) + (incr1') $);
            \coordinate (y13) at ($ (c3) + (incr1') $);
            \coordinate (y14) at ($ (c4) + (incr1') $);
            \coordinate (y15) at ($ (c5) + (incr1') $);
            
            \coordinate (incr2') at (9.5,0);
            \coordinate (y2) at ($ (c) + (incr2') $);
            \coordinate (y21) at ($ (c1) + (incr2') $);
            \coordinate (y22) at ($ (c2) + (incr2') $);
            \coordinate (y23) at ($ (c3) + (incr2') $);
            \coordinate (y24) at ($ (c4) + (incr2') $);
            
            \draw[fill=darkblue] (x1) circle [radius=0.05] node[above right , darkblue] {$1$};
            \draw[fill=darkblue] (x2) circle [radius=0.05] node[above left , darkblue] {$2$};
            \draw[fill=darkblue] (x3) circle [radius=0.05] node[below left , darkblue] {$3$};
            \draw[fill=darkblue] (x4) circle [radius=0.05] node[below right , darkblue] {$4$};
            
            \draw[fill=darkblue] (y1) circle [radius=0.05] node[below , darkblue] {$1$};
            \draw[fill=darkblue] (y2) circle [radius=0.05] node[below , darkblue] {$2$};
            
            \draw[- , thick , darkred] ($ 0.93*(x1) + 0.07*(x11) $) -- node[above right,scale=0.9] {$1$} (x11);
            \draw[- , thick , darkred] ($ 0.93*(x1) + 0.07*(x15) $) -- node[above,scale=0.9] {$2$} (x15);
            
            \draw[- , thick , darkred] ($ 0.93*(x2) + 0.07*(x21) $) -- node[above right,scale=0.9] {$6$} (x21);
            \draw[- , thick , darkred] ($ 0.93*(x2) + 0.07*(x24) $) -- node[above,scale=0.9] {$3$} (x24);
            
            \draw[- , thick , darkred] ($ 0.93*(x3) + 0.07*(x31) $) -- node[below right,scale=0.9] {$4$} (x31);
            \draw[- , thick , darkred] ($ 0.93*(x3) + 0.07*(x33) $) -- node[below,scale=0.9] {$7$} (x33);
            
            \draw[- , thick , darkred] ($ 0.93*(x4) + 0.07*(x41) $) -- node[below right,scale=0.9] {$8$} (x41);
            \draw[- , thick , darkred] ($ 0.93*(x4) + 0.07*(x42) $) -- node[below,scale=0.9] {$5$} (x42);
            
            \draw[- , thick , darkblue] ($ 0.93*(x1) + 0.07*(x2) $) --
            ($ 0.93*(x2) + 0.07*(x1) $);
            \draw[- , thick , darkblue] ($ 0.93*(x1) + 0.07*(x3) $) --
            ($ 0.93*(x3) + 0.07*(x1) $);
            \draw[- , thick , darkblue] ($ 0.93*(x1) + 0.07*(x4) $) --
            ($ 0.93*(x4) + 0.07*(x1) $);
            \draw[- , thick , darkblue] ($ 0.93*(x2) + 0.07*(x3) $) --
            ($ 0.93*(x3) + 0.07*(x2) $);
            \draw[- , thick , darkblue] ($ 0.93*(x2) + 0.07*(x4) $) --
            ($ 0.93*(x4) + 0.07*(x2) $);
            \draw[- , thick , darkblue] ($ 0.93*(x3) + 0.07*(x4) $) --
            ($ 0.93*(x4) + 0.07*(x3) $);

            \draw[{Implies[]}-{Implies[]} , thick , double distance=2] (3.375,0) -- node[above] {$\mathrel{P}_{(2,4)}$} (4.875,0);

            \draw[- , thick , darkred] ($ 0.93*(y1) + 0.07*(y11) $) -- node[above right,scale=0.9] {$1$} (y11);
            \draw[- , thick , darkred] ($ 0.93*(y1) + 0.07*(y13) $) -- node[right,scale=0.9] {$6$} (y13);
            \draw[- , thick , darkred] ($ 0.93*(y1) + 0.07*(y14) $) -- node[left,scale=0.9] {$7$} (y14);
            \draw[- , thick , darkred] ($ 0.93*(y1) + 0.07*(y15) $) -- node[above,scale=0.9] {$8$} (y15);
            
            \draw[- , thick , darkred] ($ 0.93*(y2) + 0.07*(y21) $) -- node[above right,scale=0.9] {$2$} (y21);
            \draw[- , thick , darkred] ($ 0.93*(y2) + 0.07*(y22) $) -- node[above,scale=0.9] {$3$} (y22);
            \draw[- , thick , darkred] ($ 0.93*(y2) + 0.07*(y23) $) -- node[right,scale=0.9] {$4$} (y23);
            \draw[- , thick , darkred] ($ 0.93*(y2) + 0.07*(y24) $) -- node[left,scale=0.9] {$5$} (y24);
            
            \draw[- , thick , darkblue] ($ 0.93*(y1) + 0.07*(y2) $) --
            ($ 0.93*(y2) + 0.07*(y1) $);
        \end{tikzpicture}
    \caption{Pachner move $\mathrm{P}_{(2,4)}$.}
    \label{Fig_Pachner(2,4)}
\end{figure}
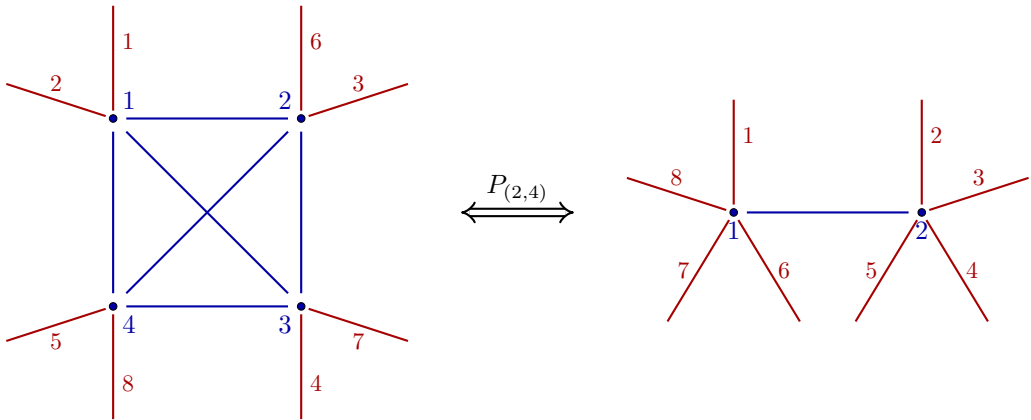
The amplitudes for these diagrams are 
    \begin{equation}
        \cA_{\cV_2}
        =
        \int \dX^{20} \dh^{10} \du^{40} \dY^{60} \,\, \cK_{1,2} \,\, \cV_1 \, \cV_2,
    \end{equation}
and 
\be
        \begin{aligned}
            \cA_{\cV_4} 
            & = 
            \int \dX^{40} \dh^{20} \du^{80} \dY^{120} \, 
            \cK_{1,2} \, \cK_{1,3} \, \cK_{1,4} \, \cK_{2,3} \, \cK_{2,4} \, \cK_{3,4} \,\, 
            \cV_1 \, \cV_2 \, \cV_3 \, \cV_4.
        \end{aligned}
    \ee
Once again the explicit expressions are in the appendix, see $\eqref{Amplitude_Two4simplices}$ and $\eqref{Amplitude_Four4simplices}$. The two amplitudes can once again be related to one another by integration and by renaming variables as shown in \eqref{cv24}. The result is that     
    \be
        \cA_{\cV_4} = (V_{G_1}^{2} V_{G_2}^{19}) \, \cA_{\cV_2} \,.
    \ee

\paragraph{$P_{3,3}$ Pachner Move:} Finally we investigate the diagrams in Fig.\, \ref{Fig_Pachner(3,3)}. The amplitudes of these diagrams are
    \be
        \cA_{\cV_3} = 
        \int \dX^{30} \dh^{15} \du^{60} \dY^{90} \, 
        (\cK_{1,2} \, \cK_{1,3} \, \cK_{2,3}) \,\, 
        \cV_1 \, \cV_2 \, \cV_3
    \ee
as well as the appropriate permutation of indices (see equation  \eqref{Amplitude_Three4simplices} in the appendix for details).
\begin{figure}
    \centering
    \begin{tikzpicture}[scale=1.]
            \coordinate (c) at (0,0);
            \coordinate (c1) at (0,1.5);
            \coordinate (c2) at (1.42,0.46);
            \coordinate (c3) at (0.88,-1.45);
            \coordinate (c4) at (-0.88,-1.45);
            \coordinate (c5) at (-1.42,0.46);
            
            \coordinate (incr1) at (-1.25,0.72);
            \coordinate (x1) at ($ (c) + (incr1) $);
            \coordinate (x11) at ($ (c1) + (incr1) $);
            \coordinate (x14) at ($ (c4) + (incr1) $);
            \coordinate (x15) at ($ (c5) + (incr1) $);
            
            \coordinate (incr2) at (1.25,0.72);
            \coordinate (x2) at ($ (c) + (incr2) $);
            \coordinate (x21) at ($ (c1) + (incr2) $);
            \coordinate (x23) at ($ (c2) + (incr2) $);
            \coordinate (x24) at ($ (c3) + (incr2) $);
            
            \coordinate (incr3) at (0,-1.5);
            \coordinate (x3) at ($ (c) + (incr3) $);
            \coordinate (x31) at ($ (incr3) - (c1) $);
            \coordinate (x32) at ($ (c5) + (incr3) $);
            \coordinate (x33) at ($ (c2) + (incr3) $);
            
            \coordinate (incr1') at (6.75,0.72);
            \coordinate (y1) at ($ (c) + (incr1') $);
            \coordinate (y11) at ($ (c1) + (incr1') $);
            \coordinate (y14) at ($ (c4) + (incr1') $);
            \coordinate (y15) at ($ (c5) + (incr1') $);
            
            \coordinate (incr2') at (9.25,0.72);
            \coordinate (y2) at ($ (c) + (incr2') $);
            \coordinate (y21) at ($ (c1) + (incr2') $);
            \coordinate (y23) at ($ (c2) + (incr2') $);
            \coordinate (y24) at ($ (c3) + (incr2') $);
            
            \coordinate (incr3') at (8,-1.5);
            \coordinate (y3) at ($ (c) + (incr3') $);
            \coordinate (y31) at ($ (incr3') - (c1) $);
            \coordinate (y32) at ($ (c5) + (incr3') $);
            \coordinate (y33) at ($ (c2) + (incr3') $);
            
            \draw[fill=darkblue] (x1) circle [radius=0.05] node[below , darkblue] {$1$};
            \draw[fill=darkblue] (x2) circle [radius=0.05] node[below , darkblue] {$2$};
            \draw[fill=darkblue] (x3) circle [radius=0.05] node[above , darkblue] {$3$};
            
            \draw[fill=darkblue] (y1) circle [radius=0.05] node[below , darkblue] {$1$};
            \draw[fill=darkblue] (y2) circle [radius=0.05] node[below , darkblue] {$2$};
            \draw[fill=darkblue] (y3) circle [radius=0.05] node[above , darkblue] {$3$};
            
            \draw[- , thick , darkred] ($ 0.93*(x1) + 0.07*(x11) $) --
            node[above right,scale=0.9] {$9$} (x11);
            \draw[- , thick , darkred] ($ 0.93*(x1) + 0.07*(x14) $) -- node[left,scale=0.9] {$4$} (x14);
            \draw[- , thick , darkred] ($ 0.93*(x1) + 0.07*(x15) $) -- node[above,scale=0.9] {$5$} (x15);
            
            \draw[- , thick , darkred] ($ 0.93*(x2) + 0.07*(x21) $) -- node[above right,scale=0.9] {$8$} (x21);
            \draw[- , thick , darkred] ($ 0.93*(x2) + 0.07*(x23) $) -- node[above,scale=0.9] {$3$} (x23);
            \draw[- , thick , darkred] ($ 0.93*(x2) + 0.07*(x24) $) -- node[left,scale=0.9] {$7$} (x24);
            
            \draw[- , thick , darkred] ($ 0.93*(x3) + 0.07*(x31) $) -- node[below right,scale=0.9] {$1$} (x31);
            \draw[- , thick , darkred] ($ 0.93*(x3) + 0.07*(x32) $) -- node[below,scale=0.9] {$2$} (x32);
            \draw[- , thick , darkred] ($ 0.93*(x3) + 0.07*(x33) $) -- node[below,scale=0.9] {$6$} (x33);
            
            \draw[- , thick , darkblue] ($ 0.93*(x1) + 0.07*(x2) $) --
            ($ 0.93*(x2) + 0.07*(x1) $);
            \draw[- , thick , darkblue] ($ 0.93*(x1) + 0.07*(x3) $) --
            ($ 0.93*(x3) + 0.07*(x1) $);
            \draw[- , thick , darkblue] ($ 0.93*(x2) + 0.07*(x3) $) --
            ($ 0.93*(x3) + 0.07*(x2) $);

            \draw[{Implies[]}-{Implies[]} , thick , double distance=2] (3.25,0) -- node[above] {$\mathrel{P}_{(3,3)}$} (4.75,0);

            \draw[- , thick , darkred] ($ 0.93*(y1) + 0.07*(y11) $) -- node[above right,scale=0.9] {$1$} (y11);
            \draw[- , thick , darkred] ($ 0.93*(y1) + 0.07*(y14) $) -- node[left,scale=0.9] {$8$} (y14);
            \draw[- , thick , darkred] ($ 0.93*(y1) + 0.07*(y15) $) -- node[above,scale=0.9] {$9$} (y15);
            
            \draw[- , thick , darkred] ($ 0.93*(y2) + 0.07*(y21) $) -- node[above right,scale=0.9] {$2$} (y21);
            \draw[- , thick , darkred] ($ 0.93*(y2) + 0.07*(y23) $) -- node[above,scale=0.9] {$3$} (y23);
            \draw[- , thick , darkred] ($ 0.93*(y2) + 0.07*(y24) $) -- node[left,scale=0.9] {$4$} (y24);
            
            \draw[- , thick , darkred] ($ 0.93*(y3) + 0.07*(y31) $) -- node[below right,scale=0.9] {$6$} (y31);
            \draw[- , thick , darkred] ($ 0.93*(y3) + 0.07*(y32) $) -- node[below,scale=0.9] {$7$} (y32);
            \draw[- , thick , darkred] ($ 0.93*(y3) + 0.07*(y33) $) -- node[below,scale=0.9] {$5$} (y33);
            
            \draw[- , thick , darkblue] ($ 0.93*(y1) + 0.07*(y2) $) --
            ($ 0.93*(y2) + 0.07*(y1) $);
            \draw[- , thick , darkblue] ($ 0.93*(y1) + 0.07*(y3) $) --
            ($ 0.93*(y3) + 0.07*(y1) $);
            \draw[- , thick , darkblue] ($ 0.93*(y2) + 0.07*(y3) $) --
            ($ 0.93*(y3) + 0.07*(y2) $);
        \end{tikzpicture}
    \caption{Pachner move $\mathrm{P}_{(3,3)}$.}
    \label{Fig_Pachner(3,3)}
\end{figure}
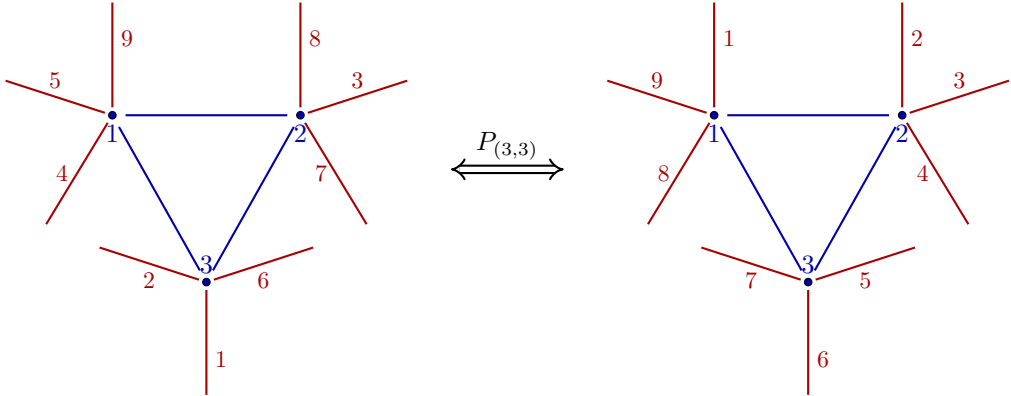
The two amplitudes are related to one another by the change of variables detailed in \eqref{cv33}.

\section{Outlook}
(Quantum) 2-groups are supposed to be the relevant structure to define quantum field theories for 4d manifolds and generate state-sums to encode the topological features of a 4d manifold. The Yetter-Mackaay model provides a generalization of the Ponzano-Regge/Turaev-Viro state sum, where 2-group structures are the basic symmetry structure. This model is built in the triangulation picture, just like the  Ponzano-Regge/Turaev-Viro models.  

Group field theories can be seen as a tool to generate state sum models as Feynman diagrams, on the dual complex of the triangulation. Since the initial work of Boulatov, GFTs have been extensively studied and used in the quantum gravity setting, both in 3d and in 4d. 

We have presented here a generalization of this construction using  strict 2-groups, or crossed modules. We have shown how the amplitude of different Feynman diagrams are related by Pachner moves so that the action proposed is indeed that of a topological model. Ultimately, we expect that the model we built is the dual version of  the Yetter-Mackaay model. To prove this we would need to define a notion of Fourier transform, which is not yet available to the best of our knowledge.  

We highlight a number of directions which we find interesting to develop in the future.  

\paragraph{\textbf{Fourier transform for 2-groups.}}
The first question we intend to explore is indeed the relation with the Yetter-Mackaay model. To relate our model to this model, we would need a generalization of the notion Fourier between the functions over a group and the non-commutative functions over the cotangent plane \cite{Guedes:2013vi} to the 2-group case. This is currently under investigation. An potential interesting direction to explore is the notion of co-commutative trialgebras and their dual analogue the co-commutative ones as discussed by Pfeiffer in \cite{Pfeiffer2007}. This is currently under investigation.


\paragraph{\textbf{Hidden 2-GFT's?}} Some models of GFT's have already been introduced which potentially could be re-interpreted as 2-GFT's. For example in 3d the notion of particle in the GFT has been discussed in \cite{Freidel:2005cg}. It involved the presence of extra decorations in the field which can be naturally interpreted as decorating the faces of the dual 2-complex, and be associated with the particle degrees of freedom.  In a different direction,  1-GFT's have been modified in order to account for a better probing of the "bubbles" \cite{Baratin:2014yra} which could also be interpreted as closed 2-holonomies. It would be interesting to see whether these two different  models could be re-interpreted as a 2-GFT's with a specific choice of 2-group in each case.   


\paragraph{\textbf{Divergences?}} A key-issue in any field theory deals with the divergences and their taming through renormalization. Heavy work was dedicated to 1-GFT's to assess their divergence structure according to different choices of groups or interactions see for example \cite{Freidel:2009hd, Rivasseau:2010cvl}. In the present work, we have not focused on this issue, but it should be explored accordingly.

\paragraph{\textbf{Large $N$ expansion.}}
In \cite{Carrozza:2012kt}, a large $N$ expansion of topological GFTs such as the Boulatov and Ooguri models, have been implemented (see also Chapter $4$ of S. Carrozza's PhD thesis \cite{Carrozza:2013oiy}). It was show that the geometric data encoded in the group variables can be exploited to prove that the first term in the large $N$ expansion is given by the celebrated melonic graphs, which are the dominant graphs in the large $N$ expansion of tensor models 
 \cite{Dartois:2013he, Gurau:2015tua, Bonzom:2021kjy, Dartois:2013sra}
or of the SYK model \cite{Witten-SYK}.

It would thus appear interesting to us to investigate whether or not the doubling of the group data represented by the $2-$group structure studied in our paper could lead, in this GFT context, to a different type of large $N$ limit, where other type of graphs that the melonic ones would be dominating.

\paragraph{\textbf{Application to quantum gravity?}}
2-group structures seem relevant to the construction of a quantum gravity theory. For example, one could demand to have information about the frame field and not only about the flux. Not only could this give a ``nicer" notion of operator for the notion of length \cite{Bianchi:2008es}, it could help to discriminate between degenerate non-degenerate manifolds in the semi-classical limit \cite{Crane:2003ep}. The flux is associated to faces and is therefore a natural candidate as a 2-connection. When there is no cosmological constant, the discrete (classical) flux is valued in an abelian group, but switching to a non-zero cosmological constant would typically lead to a non-abelian deformation and the appearance of a quantum group. Following the Eckmann-Hilton argument, one can only have a non-abelian group decoration on faces if, on top of that, edges are also decorated. 2-groups provide then a way to have  consistent non-abelian decorations on faces and edges.    These three arguments point to the relevance of 2-group to construct a more refined quantum gravity model. Once the relation between 2-groups and gravity has been identified, one could use then the GFT we proposed to constrain it to construct the quantum gravity amplitude, just like it was done for the Barrett-Crane model or the EPRL-FK model \cite{DePietri:1999bx, Baratin:2011tx, Baratin:2011hp}.  

\section*{Acknowledgement.} FG would like to thank Etera Livine for discussions regarding GFT's and 2-groups at the early stage of the project. ML, AT and PT have been partially supported by the ANR-20-CE48-0018 “3DMaps”
grant. A. T. has been partially supported by the PN 09370102 grant.

\appendix
\section{Feynman Diagrams and Amplitudes}\label{appendix}
We evaluate the relevant diagrams and then show explicitly how they are related through the Pachner moves.

\paragraph{Key Feynman diagrams}
We organize the amplitudes by factoring the delta functions according to  the group. For each amplitude, denote by $\delta_{G_1}^{\{\cV_N\}}$ and $\delta_{G_2}^{\{\cV_N\}}$ the set of deltas on links (group $G_1$) and wedges (group $G_2$) in the amplitude. The superscript $\cV_N$ specifies the number ($N$) of vertices in the diagram. 

We consider the specific Feynman diagrams which will be  dual respectively to a single 4-simplex, two, three, four, and five 4-simplices.
They can be derived by properly gluing a number of independent vertex amplitudes using the propagator. 

\begin{figure}
    \centering
    \begin{subfigure}[t]{0.35\textwidth}
        \centering
        \begin{tikzpicture}[scale=1.25]
            \coordinate (c) at (0,0);
            \coordinate (c1) at (0,1.5);
            \coordinate (c2) at (1.42,0.46);
            \coordinate (c3) at (0.88,-1.45);
            \coordinate (c4) at (-0.88,-1.45);
            \coordinate (c5) at (-1.42,0.46);
            
            \draw[fill=darkblue] (c) circle [radius=0.05];
            
            \draw[- , thick , darkred] ($ 0.93*(c) + 0.07*(c1) $) -- node[above right,scale=0.9] {$1$} (c1);
            \draw[- , thick , darkred] ($ 0.93*(c) + 0.07*(c2) $) -- node[above right,scale=0.9] {$2$} (c2);
            \draw[- , thick , darkred] ($ 0.93*(c) + 0.07*(c3) $) -- node[right,scale=0.9] {$3$} (c3);
            \draw[- , thick , darkred] ($ 0.93*(c) + 0.07*(c4) $) -- node[left,scale=0.9] {$4$} (c4);
            \draw[- , thick , darkred] ($ 0.93*(c) + 0.07*(c5) $) -- node[above,scale=0.9] {$5$} (c5);
        \end{tikzpicture}
        \caption{Simplest graph dual to a 4-simplex.}
        \label{SubFig_4Simplex}
    \end{subfigure}
    \hfill
    \begin{subfigure}[t]{0.4\textwidth}
        \centering
        \begin{tikzpicture}[scale=1.25]
            \coordinate (c) at (0,0);
            \coordinate (c1) at (0,1.5);
            \coordinate (c2) at (1.42,0.46);
            \coordinate (c3) at (0.88,-1.45);
            \coordinate (c4) at (-0.88,-1.45);
            \coordinate (c5) at (-1.42,0.46);
            
            \coordinate (incr1) at (-1.25,0);
            \coordinate (x1) at ($ (c) + (incr1) $);
            \coordinate (x11) at ($ (c1) + (incr1) $);
            \coordinate (x13) at ($ (c3) + (incr1) $);
            \coordinate (x14) at ($ (c4) + (incr1) $);
            \coordinate (x15) at ($ (c5) + (incr1) $);
            
            \coordinate (incr2) at (1.25,0);
            \coordinate (x2) at ($ (c) + (incr2) $);
            \coordinate (x21) at ($ (c1) + (incr2) $);
            \coordinate (x22) at ($ (c2) + (incr2) $);
            \coordinate (x23) at ($ (c3) + (incr2) $);
            \coordinate (x24) at ($ (c4) + (incr2) $);
            
            \draw[fill=darkblue] (x1) circle [radius=0.05] node[below , darkblue] {$1$};
            \draw[fill=darkblue] (x2) circle [radius=0.05] node[below , darkblue] {$2$};
            
            \draw[- , thick , darkred] ($ 0.93*(x1) + 0.07*(x11) $) -- node[above right,scale=0.9] {$1$} (x11);
            \draw[- , thick , darkred] ($ 0.93*(x1) + 0.07*(x13) $) -- node[right,scale=0.9] {$3$} (x13);
            \draw[- , thick , darkred] ($ 0.93*(x1) + 0.07*(x14) $) -- node[left,scale=0.9] {$4$} (x14);
            \draw[- , thick , darkred] ($ 0.93*(x1) + 0.07*(x15) $) -- node[above,scale=0.9] {$5$} (x15);
            
            \draw[- , thick , darkred] ($ 0.93*(x2) + 0.07*(x21) $) -- node[above right,scale=0.9] {$1$} (x21);
            \draw[- , thick , darkred] ($ 0.93*(x2) + 0.07*(x22) $) -- node[above,scale=0.9] {$2$} (x22);
            \draw[- , thick , darkred] ($ 0.93*(x2) + 0.07*(x23) $) -- node[right,scale=0.9] {$3$} (x23);
            \draw[- , thick , darkred] ($ 0.93*(x2) + 0.07*(x24) $) -- node[left,scale=0.9] {$4$} (x24);
            
            \draw[- , thick , darkblue] ($ 0.93*(x1) + 0.07*(x2) $) -- node(12)[above , darkred , scale=0.9] {$(2,5)$} node[above=1em of 12 , darkblue] {$1,2$} ($ 0.93*(x2) + 0.07*(x1) $);
        \end{tikzpicture}
        \caption{Graph dual to two 4-simplices sharing one tetrahedron, the one labelled 2 in the 4-simplex 1 and the one labelled 5 in the 4-simplex 2.}
        \label{SubFig_Two4simplices}
    \end{subfigure}
    \caption{The tetrahedron labelling is done in red, while in blue we noted the  4-simplex labelling.}
    \label{Fig_4Simplex(1-2)}
\end{figure}
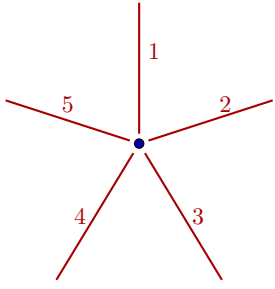
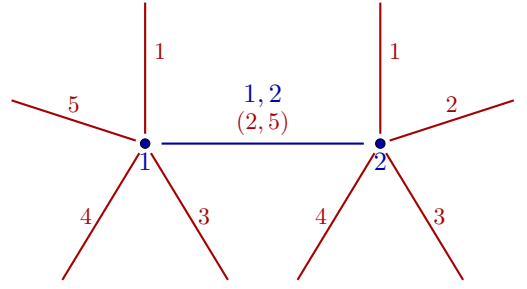

\begin{itemize}
    \item[i.] \textbf{The vertex term, a five-point Feynman diagram}, which is illustrated in Fig. \ref{SubFig_4Simplex} has amplitude
    \begin{equation}
        \cA_{\cV} =
        \int \dX^{10} \dh^5 \du^{20} \dY^{30} \,\, \cV_{gauge} =
        \int \dX^{10} \dh^5 \du^{20} \dY^{30} \,\, \delta_{G_1}^{\{\cV\}} \,,\delta_{G_2}^{\{\cV\}} \, 
        \label{Amplitude_4Simplex}
    \end{equation}
    where
    \begin{align}
        \delta_{G_1}^{\{\cV\}} = \,\,
        &
        \delta_{G_1}\big(
        t(\x_{2,1}) h_{1} u_{1;1} \, u_{2;4}\mone h_{2}\mone
        \big) \,
        \delta_{G_!}\big(
        t(\x_{3,1}) h_{1} u_{1;2} \, u_{3;3}\mone h_{3}\mone
        \big) \,
        \delta_{G_1}\big(
        t(\x_{4,1}) h_{1} u_{1;3} \, u_{4;2}\mone h_{4}\mone
        \big) \,
        \delta_{G_1}\big(
        t(\x_{5,1}) h_{1} u_{1;4} \, u_{5;1}\mone h_{5}\mone
        \big) \,
        \nonumber \\
        &
        \delta_{G_1}\big(
        t(\x_{3,2}) h_{2} u_{2;1} \, u_{3;4}\mone h_{3}\mone
        \big) \,
        \delta_{G_1}\big(
        t(\x_{4,2}) h_{2} u_{2;2} \, u_{4;3}\mone h_{4}\mone
        \big) \,
        \delta_{G_1}\big(
        t(\x_{5,2}) h_{2} u_{2;3} \, u_{5;2}\mone h_{5}\mone
        \big) \,
        \delta_{G_1}\big(
        t(\x_{4,3}) h_{3} u_{3;1} \, u_{4;4}\mone h_{4}\mone
        \big) \,
        \nonumber \\
        &
        \delta_{G_1}\big(
        t(\x_{5,3}) h_{3} u_{3;2} \, u_{5;3}\mone h_{5}\mone
        \big) \,
        \delta_{G_1}\big(
        t(\x_{5,4}) h_{4} u_{4;1} \, u_{5;4}\mone h_{5}\mone
        \big) \,,
        \label{One4Simplex_Links}
    \end{align}
    and
    \begin{align}
        \delta_{G_2}^{\{\cV\}} = \,\,
        &
        \delta_{G_2}\big(
        (h_1 \rhd \y_{1;2,1}) \, \x_{1,2} \, 
        (h_2 \rhd \y_{2;4,1}) \,\x_{2,3} \,
        (h_3 \rhd \y_{3;4,3}) \, \x_{3,1}
        \big) \,
        \nonumber \\
        &
        \delta_{{G_2}}\big(
        (h_1 \rhd \y_{1;3,1}) \, \x_{1,2} \, 
        (h_2 \rhd \y_{2;4,2}) \,\x_{2,4} \,
        (h_4 \rhd \y_{4;3,2}) \, \x_{4,1}
        \big) \,
        \nonumber \\
        &
        \delta_{{G_2}}\big(
        (h_1 \rhd \y_{1;4,1}) \, \x_{1,2} \, 
        (h_2 \rhd \y_{2;4,3}) \,\x_{2,5} \,
        (h_5 \rhd \y_{5;2,1}) \, \x_{5,1}
        \big) \,
        \nonumber \\
        &
        \delta_{{G_2}}\big(
        (h_1 \rhd \y_{1;3,2}) \, \x_{1,3} \, 
        (h_3 \rhd \y_{3;3,1}) \,\x_{3,4} \,
        (h_4 \rhd \y_{4;4,2}) \, \x_{4,1}
        \big) \,
        \nonumber \\
        &
        \delta_{{G_2}}\big(
        (h_1 \rhd \y_{1;4,2}) \, \x_{1,3} \, 
        (h_3 \rhd \y_{3;3,2}) \,\x_{3,5} \,
        (h_5 \rhd \y_{5;3,1}) \, \x_{5,1}
        \big) \,
        \nonumber \\
        &
        \delta_{{G_2}}\big(
        (h_1 \rhd \y_{1;3,4}) \, \x_{1,4} \, 
        (h_4 \rhd \y_{4;2,1}) \,\x_{4,5} \,
        (h_5 \rhd \y_{5;4,1}) \, \x_{5,1}
        \big) \,
        \nonumber \\
        &
        \delta_{{G_2}}\big(
        (h_2 \rhd \y_{2;2,1}) \, \x_{2,3} \, 
        (h_3 \rhd \y_{3;4,1}) \,\x_{3,4} \,
        (h_4 \rhd \y_{4;4,3}) \, \x_{4,2}
        \big) \,
        \nonumber \\
        &
        \delta_{{G_2}}\big(
        (h_2 \rhd \y_{2;3,1}) \, \x_{2,3} \, 
        (h_3 \rhd \y_{3;4,2}) \,\x_{3,5} \,
        (h_5 \rhd \y_{5;3,2}) \, \x_{5,2}
        \big) \,
        \nonumber \\
        &
        \delta_{{G_2}}\big(
        (h_2 \rhd \y_{2;3,2}) \, \x_{2,4} \, 
        (h_4 \rhd \y_{4;3,1}) \,\x_{4,5} \,
        (h_5 \rhd \y_{5;4,2}) \, \x_{5,2}
        \big) \,
        \nonumber \\
        &
        \delta_{{G_2}}\big(
        (h_3 \rhd \y_{3;2,1}) \, \x_{3,4} \, 
        (h_4 \rhd \y_{4;4,1}) \,\x_{4,5} \,
        (h_5 \rhd \y_{5;4,3}) \, \x_{5,3}
        \big) \,.
        \label{One4Simplex_Wedges}
    \end{align}
    
    \item[ii.] \textbf{The eight-point Feynman diagram,} illustrated in Fig. \ref{SubFig_Two4simplices} has amplitude,
    \begin{equation}
        \cA_{\cV_2}
        =
        \int \dX^{20} \dh^{10} \du^{40} \dY^{60} \,\, \cK_{1,2} \,\, \big(\cV_1 \, \cV_2\big)
        =
        \int \dX^{20} \dh^{9} \du^{32} \dY^{48} \,\,
        \delta_{G_1}^{\{\cV_2\}} \, \delta_{G_2}^{\{\cV_2\}} \,,
        \label{Amplitude_Two4simplices}
    \end{equation}
    where
    \begin{align}
        \delta_{G_1}^{\{\cV_2\}}
        = \Big(
        &
        \delta_{G_1}\big(
        t(\x_{1;3,2}) h_{1,2} t(\x_{2;5,2}) h_{2;2} u_{2;2;3} u_{1;3;4}\mone h_{1;3}\mone\big) \,
        \delta_{G_1}\big(
        t(\x_{1;4,2}) h_{1,2} t(\x_{2;5,3}) h_{2;3} u_{2;3;2} u_{1;4;3}\mone h_{1;4}\mone\big) \,
        \nonumber \\
        &
        \delta_{G_1}\big(
        t(\x_{1;5,2}) h_{1,2} t(\x_{2;5,4}) h_{2;4} u_{2;4;1} u_{1;5;2}\mone h_{1;5}\mone\big) \,
        \delta_{G_1}\big(
        t(\x_{1;2,1}) h_{1;1} u_{1;1;1} u_{2;1;4}\mone h_{2;1}\mone t(\x_{2;1,5}) h_{2,1}\big)
        \Big)
        \nonumber \\
        \Big(
        &
        \delta_{G_1}\big(
        t(\x_{1;3,1}) h_{1;1} u_{1;1;2} \, u_{1;3;3}\mone h_{1;3}\mone
        \big) \,
        \delta_{G_1}\big(
        t(\x_{1;4,1}) h_{1;1} u_{1;1;3} \, u_{1;4;2}\mone h_{1;4}\mone
        \big) \,
        \delta_{G_1}\big(
        t(\x_{1;5,1}) h_{1;1} u_{1;1;4} \, u_{1;5;1}\mone h_{1;5}\mone
        \big) \,
        \nonumber \\
        &
        \delta_{G_1}\big(
        t(\x_{1;4,3}) h_{1;3} u_{1;3;1} \, u_{1;4;4}\mone h_{1;4}\mone
        \big) \,
        \delta_{G_1}\big(
        t(\x_{1;5,3}) h_{1;3} u_{1;3;2} \, u_{1;5;3}\mone h_{1;5}\mone
        \big) \,
        \delta_{G_1}\big(
        t(\x_{1;5,4}) h_{1;4} u_{1;4;1} \, u_{1;5;4}\mone h_{1;5}\mone
        \big)
        \nonumber \\
        &
        \delta_{G_1}\big(
        t(\x_{2;2,1}) h_{2;1} u_{2;1;1} \, u_{2;2;4}\mone h_{2;2}\mone
        \big) \,
        \delta_{G_1}\big(
        t(\x_{2;3,1}) h_{2;1} u_{2;1;2} \, u_{2;3;3}\mone h_{2;3}\mone
        \big) \,
        \delta_{G_1}\big(
        t(\x_{2;4,1}) h_{2;1} u_{2;1;3} \, u_{2;4;2}\mone h_{2;4}\mone
        \big) \,
        \nonumber \\
        &
        \delta_{G_1}\big(
        t(\x_{2;3,2}) h_{2;2} u_{2;2;1} \, u_{2;3;4}\mone h_{2;3}\mone
        \big) \,
        \delta_{G_1}\big(
        t(\x_{2;4,2}) h_{2;2} u_{2;2;2} \, u_{2;4;3}\mone h_{2;4}\mone
        \big) \,
        \delta_{G_1}\big(
        t(\x_{2;4,3}) h_{2;3} u_{2;3;1} \, u_{2;4;4}\mone h_{2;4}\mone
        \big)
        \Big) \,,
        \label{Two4simplices_Links}
    \end{align} 
    and
    \begin{align}
        \delta_{G_2}^{\{\cV_2\}} 
        = \,\, &
        \delta_{G_2}\Big(
        h_{1,2} \rhd \big(\x_{2;5,3} \, (h_{2;3} \rhd \y_{2;3;2,4}) \, \x_{2;3,2} \, (h_{2;2} \rhd \y_{2;2;1,3}) \, X_{2;2,5}\big) \, \x_{1;2,3} \, (h_{1;3} \rhd \y_{1;3;4,1}) \, \x_{1;3,4} \, (h_{1;4} \rhd \y_{1;4;4,3}) \, \x_{1;4,2}
        \Big)
        \nonumber \\
        &
        \delta_{G_2}\Big(
        h_{1,2} \rhd \big(\x_{2;5,4} \, (h_{2;4} \rhd \y_{2;4;1,3}) \, \x_{2;4,2} \, (h_{2;2} \rhd \y_{2;2;2,3}) \, X_{2;2,5}\big) \, \x_{1;2,3} \, (h_{1;3} \rhd \y_{1;3;4,2}) \, \x_{1;3,5} \, (h_{1;5} \rhd \y_{1;5;3,2}) \, \x_{1;5,2}
        \Big)
        \nonumber \\
        &
        \delta_{G_2}\Big(
        h_{1,2} \rhd \big(\x_{2;5,1} \, (h_{2;1} \rhd \y_{2;1;4,1}) \, \x_{2;1,2} \, (h_{2;2} \rhd \y_{2;2;4,3}) \, \x_{2;2,5}\big) \, \x_{1;2,3} \, (h_{1;3} \rhd \y_{1;3;4,3}) \, \x_{1;3,1} \, (h_{1;1} \rhd \y_{1;1;2,1}) \, \x_{1;1,2}
        \Big)
        \nonumber \\
        &
        \delta_{G_2}\Big(
        h_{1,2} \rhd \big(\x_{2;5,4} \, (h_{2;4} \rhd \y_{2;4;1,4}) \, \x_{2;4,3} \, (h_{2;3} \rhd \y_{2;3;1,2}) \, \x_{2;3,5}\big) \, \x_{1;2,4} \, (h_{1;4} \rhd \y_{1;4;3,1}) \, \x_{1;4,5} \, (h_{1;5} \rhd \y_{1;5;4,2}) \, \x_{1;5,2} 
        \Big)
        \nonumber \\
        &
        \delta_{G_2}\Big(
        h_{1,2} \rhd \big(\x_{2;5,1} \, (h_{2;1} \rhd \y_{2;1;4,2}) \, \x_{2;1,3} \, (h_{2;3} \rhd \y_{2;3;3,2}) \, \x_{2;3,5}\big) \, \x_{1;2,4} \, (h_{1;4} \rhd \y_{1;4;3,2}) \, \x_{1;4,1} \, (h_{1;1} \rhd \y_{1;1;3,1}) \, \x_{1;1,2} \, 
        \Big)
        \nonumber \\
        &
        \delta_{G_2}\Big(
        h_{1,2} \rhd \big(\x_{2;5,1} \, (h_{2;1} \rhd \y_{2;1;3,4}) \, \x_{2;1,4} \, (h_{2;4} \rhd \y_{2;4;2,1}) \, \x_{2;4,5}\big) \, \x_{1;2,5} \, (h_{1;5} \rhd \y_{1;5;2,1}) \, \x_{1;5,1} \, (h_{1;1} \rhd \y_{1;1;4,1}) \, \x_{1;1,2} \, 
        \Big)
        \nonumber \\
        &
        \delta_{{G_2}}\big(
        (h_{1;1} \rhd \y_{1;1;3,2}) \, \x_{1;1,3} \, 
        (h_{1;3} \rhd \y_{1;3;3,1}) \,\x_{1;3,4} \,
        (h_{1;4} \rhd \y_{1;4;4,2}) \, \x_{1;4,1}
        \big) \,
        \nonumber \\
        &
        \delta_{{G_2}}\big(
        (h_{1;1} \rhd \y_{1;1;4,2}) \, \x_{1;1,3} \, 
        (h_{1;3} \rhd \y_{1;3;3,2}) \,\x_{1;3,5} \,
        (h_{1;5} \rhd \y_{1;5;3,1}) \, \x_{1;5,1}
        \big) \,
        \nonumber \\
        &
        \delta_{{G_2}}\big(
        (h_{1;1} \rhd \y_{1;1;3,4}) \, \x_{1;1,4} \, 
        (h_{1;4} \rhd \y_{1;4;2,1}) \,\x_{1;4,5} \,
        (h_{1;5} \rhd \y_{1;5;4,1}) \, \x_{1;5,1}
        \big) \,
        \nonumber \\
        &
        \delta_{{G_2}}\big(
        (h_{1;3} \rhd \y_{1;3;2,1}) \, \x_{1;3,4} \, 
        (h_{1;4} \rhd \y_{1;4;4,1}) \,\x_{1;4,5} \,
        (h_{1;5} \rhd \y_{1;5;4,3}) \, \x_{1;5,3}
        \big)
        \nonumber \\
        &
        \delta_{{G_2}}\big(
        (h_{2;1} \rhd \y_{2;1;2,1}) \, \x_{2;1,2} \, 
        (h_{2;2} \rhd \y_{2;2;4,1}) \,\x_{2;2,3} \,
        (h_{2;3} \rhd \y_{2;3;4,3}) \, \x_{2;3,1}
        \big) \,
        \nonumber \\
        &
        \delta_{{G_2}}\big(
        (h_{2;1} \rhd \y_{2;1;3,1}) \, \x_{2;1,2} \, 
        (h_{2;2} \rhd \y_{2;2;4,2}) \,\x_{2;2,4} \,
        (h_{2;4} \rhd \y_{2;4;3,2}) \, \x_{2;4,1}
        \big) \,
        \nonumber \\
        &
        \delta_{{G_2}}\big(
        (h_{2;1} \rhd \y_{2;1;3,2}) \, \x_{2;1,3} \, 
        (h_{2;3} \rhd \y_{2;3;3,1}) \,\x_{2;3,4} \,
        (h_{2;4} \rhd \y_{2;4;4,2}) \, \x_{2;4,1}
        \big) \,
        \nonumber \\
        &
        \delta_{{G_2}}\big(
        (h_{2;2} \rhd \y_{2;2;2,1}) \, \x_{2;2,3} \, 
        (h_{2;3} \rhd \y_{2;3;4,1}) \,\x_{2;3,4} \,
        (h_{2;4} \rhd \y_{2;4;4,3}) \, \x_{2;4,2}
        \big) \,,
        \label{Two4simplices_Wedges}
    \end{align}
   
    We defined the composed link $h_{1,2} \equiv h_{1;2} h_{2;5}\mone$ with inverse $h_{1,2}\mone = h_{2,1}$. 
    
    The first four deltas in \eqref{Two4simplices_Links} involve a combination of bulk and boundary links of the two 4-simplices, while the remaining twelve deltas involve bulk and boundary links of the first or second 4-simplices separately. 
    Similarly, the first six deltas in \eqref{Two4simplices_Wedges} involve a combination of bulk and boundary wedges of both the 4-simplices, while the remaining eight deltas involve bulk and boundary wedges of the first or the second 4-simplices separately.
    
    \item[iii.] \textbf{The  nine-point Feynman diagram,} illustrated in Fig. \ref{SubFig_Three4simplices} has amplitude,
    \be
        \cA_{\cV_3} = 
        \int \dX^{30} \dh^{15} \du^{60} \dY^{90} \, 
        (\cK_{1,2} \, \cK_{1,3} \, \cK_{2,3}) \,\, 
        \big(\cV_1 \, \cV_2 \, \cV_3\big)
        =
        \int \dX^{30} \dh^{12} \du^{36} \dY^{54} \,\,
        \delta_{G_1}^{\{\cV_3\}} \, \delta_{G_2}^{\{\cV_3\}} \,,
        \label{Amplitude_Three4simplices}
    \ee
    with
    \begin{align}
        \delta_{G_1}^{\{\cV_3\}} = \,\,
        &
        \delta_{G_1}\big(
        t(\x_{1;3,2}) h_{1,2} t(\x_{2;5,2}) h_{2,3} t(\x_{3;5,4}) h_{3,1}\big) \,
        \nonumber \\
        \Big(
        &
        \delta_{G_1}\big(
        t(\x_{1;4,2}) h_{1,2} t(\x_{2;5,3}) h_{2;3} u_{2;3;2} u_{1;4;3}\mone h_{1;4}\mone\big) \,
        \delta_{G_1}\big(
        t(\x_{1;5,2}) h_{1,2} t(\x_{2;5,4}) h_{2;4} u_{2;4;1} u_{1;5;2}\mone h_{1;5}\mone\big) \,
        \nonumber \\
        &
        \delta_{G_1}\big(
        t(\x_{1;2,1}) h_{1;1} u_{1;1;1} u_{2;1;4}\mone h_{2;1}\mone t(\x_{2;1,5}) h_{2,1}\big)
        \delta_{G_1}\big(
        t(\x_{1;3,1}) h_{1;1} u_{1;1;2} u_{3;1;3}\mone h_{3;1}\mone t(\x_{3;1,4}) h_{3,1}\big) \,
        \nonumber \\
        &
        \delta_{G_1}\big(
        t(\x_{1;4,3}) h_{1,3} t(\x_{3;4,2}) h_{3;2} u_{3;2;2} u_{1;4;4}\mone h_{1;4}\mone\big) \,
        \delta_{G_1}\big(
        t(\x_{1;5,3}) h_{1,3} t(\x_{3;4,3}) h_{3;3} u_{3;3;1} u_{1;5;3}\mone h_{1;5}\mone\big) \,
        \nonumber \\
        &
        \delta_{G_1}\big(
        t(\x_{2;2,1}) h_{2;1} u_{2;1;1} u_{3;1;4}\mone h_{3;1}\mone t(\x_{3;1,5}) h_{3,2}\big) \,
        \delta_{G_1}\big(
        t(\x_{2;3,2}) h_{2,3} t(\x_{3;5,2}) h_{3;2} u_{3;2;3} u_{2;3;4}\mone h_{2;3}\mone\big) \,
        \nonumber \\
        &
        \delta_{G_1}\big(
        t(\x_{2;4,2}) h_{2,3} t(\x_{3;5,3}) h_{3;3} u_{3;3;2} u_{2;4;3}\mone h_{2;4}\mone\big)
        \Big)
        \nonumber \\
        \Big(
        &
        \delta_{G_1}\big(
        t(\x_{1;5,4}) h_{1;1} u_{1;1;3} \, u_{1;4;2}\mone h_{1;4}\mone
        \big) \,
        \delta_{G_1}\big(
        t(\x_{1;5,1}) h_{1;1} u_{1;1;4} \, u_{1;5;1}\mone h_{1;5}\mone
        \big) \,
        \delta_{G_1}\big(
        t(\x_{1;4;1}) h_{1;4} u_{1;4;1} \, u_{1;5;4}\mone h_{1;5}\mone
        \big)
        \nonumber \\
        &
        \delta_{G_1}\big(
        t(\x_{2;3,1}) h_{2;1} u_{2;1;2} \, u_{2;3;3}\mone h_{2;3}\mone
        \big) \,
        \delta_{G_1}\big(
        t(\x_{2;4,1}) h_{2;1} u_{2;1;3} \, u_{2;4;2}\mone h_{2;4}\mone
        \big) \,
        \delta_{G_1}\big(
        t(\x_{2;4,3}) h_{2;3} u_{2;3;1} \, u_{2;4;4}\mone h_{2;4}\mone
        \big)
        \nonumber \\
        &
        \delta_{G_1}\big(
        t(\x_{3;2,1}) h_{3;1} u_{3;1;1} \, u_{3;2;4}\mone h_{3;2}\mone
        \big) \,
        \delta_{G_1}\big(
        t(\x_{3;3,1}) h_{3;1} u_{3;1;2} \, u_{3;3;3}\mone h_{3;3}\mone
        \big) \,
        \delta_{G_1}\big(
        t(\x_{3;3,2}) h_{3;2} u_{3;2;1} \, u_{3;3;4}\mone h_{3;3}\mone
        \big)
        \Big) \,,
        \label{Three4simplices_Links}
    \end{align}
    and
    \begin{align}
        \delta_{G_2}^{\{\cV_3\}} = \,\,
        &
        \delta_{G_2}\Big(
        h_{1,2} \rhd \big(\x_{2;5,3} \, (h_{2;3} \rhd \y_{2;3;2,4}) \, \x_{2;3,2} \, \x_{2;2,5}\big) \, \x_{1;2,3} \,  h_{1,3} \rhd \big(\x_{3;4,5} \, \x_{3;5,2} \, (h_{3;2} \rhd \y_{3;2;3,2}) \, \x_{3;2,4}\big) \,
        \nonumber \\
        & \quad
        \x_{1;3,4} \, (h_{1;4} \rhd \y_{1;4;4,3}) \, \x_{1;4,2}
        \Big)
        \nonumber \\
        &
        \delta_{G_2}\Big(
        h_{1,2} \rhd \big(\x_{2;5,4} \, (h_{2;4} \rhd \y_{2;4;1,3}) \, \x_{2;4,2} \, \x_{2;2,5}\big) \, \x_{1;2,3} \,  h_{1,3} \rhd \big(\x_{3;4,5} \, \x_{3;5,3} \, (h_{3;3} \rhd \y_{3;3;2,1}) \, \x_{3;3,4}\big) \,
        \nonumber \\
        & \quad
        \x_{1;3,5} \, (h_{1;5} \rhd \y_{1;5;3,2}) \, \x_{1;5,2}
        \Big)
        \nonumber \\
        &
        \delta_{G_2}\Big(
        h_{1,2} \rhd \big(\x_{2;5,1} \, (h_{2;1} \rhd \y_{2;1;4,1}) \, \x_{2;1,2} \, \x_{2;2,5}\big) \ \x_{1;2,3} \, h_{1,3} \rhd \big(\x_{3;4,5} \, \x_{3;5,1} \, (h_{3;1} \rhd \y_{3;1;3,4}) \, \x_{3;1,4}\big) \, 
        \nonumber \\
        & \quad
        \x_{1;3,1} \, (h_{1;1} \rhd \y_{1;1;2,1}) \, \x_{1;1,2}
        \Big)
        \nonumber \\
        &
        \delta_{G_2}\Big(
        h_{1,2} \rhd \big(\x_{2;5,4} \, (h_{2;4} \rhd \y_{2;4;1,4}) \, \x_{2;4,3} \, (h_{2;3} \rhd \y_{2;3;1,2}) \, \x_{2;3,5}\big) \, \x_{1;2,4} \, (h_{1;4} \rhd \y_{1;4;3,1}) \, \x_{1;4,5} \, (h_{1;5} \rhd \y_{1;5;4,2}) \, \x_{1;5,2} 
        \Big)
        \nonumber \\
        &
        \delta_{G_2}\Big(
        (h_{1;1} \rhd \y_{1;1;3,1}) \, \x_{1;1,2} \, h_{1,2} \rhd \big(\x_{2;5,1} \, (h_{2;1} \rhd \y_{2;1;4,2}) \, \x_{2;1,3} \, (h_{2;3} \rhd \y_{2;3;3,2}) \, \x_{2;3,5}\big) \, \x_{1;2,4} \, (h_{1;4} \rhd \y_{1;4;3,2}) \, \x_{1;4,1}
        \Big)
        \nonumber \\
        &
        \delta_{G_2}\Big(
        (h_{1;1} \rhd \y_{1;1;4,1}) \, \x_{1;1,2} \, h_{1,2} \rhd \big(\x_{2;5,1} \, (h_{2;1} \rhd \y_{2;1;3,4}) \, \x_{2;1,4} \, (h_{2;4} \rhd \y_{2;4;2,1}) \, \x_{2;4,5}\big) \, \x_{1;2,5} \, (h_{1;5} \rhd \y_{1;5;2,1}) \, \x_{1;5,1}
        \Big)
        \nonumber \\ 
        &
        \delta_{G_2}\Big(
        h_{1,3} \rhd \big(\x_{3;4,3} \, (h_{3;3} \rhd \y_{3;3;1,4}) \, \x_{3;3,2} \, (h_{3;2} \rhd \y_{3;2;1,2}) \, \x_{3;2,4}\big) \, \x_{1;3,4} \, (h_{1;4} \rhd \y_{1;4;4,1}) \, \x_{1;4,5} \, (h_{1;5} \rhd \y_{1;5;4,3}) \, \x_{1;5,3}
        \Big)
        \nonumber \\
        &
        \delta_{G_2}\Big(
        (h_{1;1} \rhd \y_{1;1;3,2}) \, \x_{1;1,3} \, 
        h_{1,3} \rhd \big(\x_{3;4,1} \, (h_{3;1} \rhd \y_{3;1;3,1}) \, \x_{3;1,2} \, (h_{3;2} \rhd \y_{3;2;4,2}) \, \x_{3;2,4}\big) \,\x_{1;3,4} \,
        (h_{1;4} \rhd \y_{1;4;4,2}) \, \x_{1;4,1}
        \Big)
        \nonumber \\
        &
        \delta_{G_2}\Big(
        (h_{1;1} \rhd \y_{1;1;4,2}) \, \x_{1;1,3} \, 
        h_{1,3} \rhd \big(\x_{3;4,1} \, (h_{3;1} \rhd \y_{3;1;3,2}) \, \x_{3;1,3} \, (h_{3;3} \rhd \y_{3;3;3,1}) \, \x_{3;3,4}\big) \,\x_{1;3,5} \,
        (h_{1;5} \rhd \y_{1;5;3,1}) \, \x_{1;5,1}
        \Big) \,
        \nonumber \\
        &
        \delta_{G_2}\Big(
        (h_{2;1} \rhd \y_{2;1;2,1}) \, \x_{2;1,2} \, 
        h_{2,3} \rhd \big(\x_{3;5,1} \, (h_{3;1} \rhd \y_{3;1;4,1}) \, \x_{3;1,2} \, (h_{3;2} \rhd \y_{3;2;4,3}) \, \x_{3;2,5}\big) \,\x_{2;2,3} \,
        (h_{2;3} \rhd \y_{2;3;4,3}) \, \x_{2;3,1}
        \Big) \,
        \nonumber \\
        &
        \delta_{G_2}\Big(
        (h_{2;1} \rhd \y_{2;1;3,1}) \, \x_{2;1,2} \, 
        h_{2,3} \rhd \big(\x_{3;5,1} \, (h_{3;1} \rhd \y_{3;1;4,2}) \, \x_{3;1,3} \, (h_{3;3} \rhd \y_{3;3;3,2}) \, \x_{3;3,5}\big) \,\x_{2;2,4} \,
        (h_{2;4} \rhd \y_{2;4;3,2}) \, \x_{2;4,1}
        \Big) \,
        \nonumber \\
        &
        \delta_{G_2}\Big(
        h_{2,3} \rhd \big(\x_{3;5,3} \, (h_{3;3} \rhd \y_{3;3;2,4}) \, \x_{3;3,2} \, (h_{3;2} \rhd \y_{3;2;1,3}) \, \x_{3;2,5}\big) \, \x_{2;2,3} \, 
        (h_{2;3} \rhd \y_{2;3;4,1}) \,\x_{2;3,4} \,
        (h_{2;4} \rhd \y_{2;4;4,3}) \, \x_{2;4,2}
        \Big)
        \nonumber \\
        &
        \delta_{G_2}\big(
        (h_{1;1} \rhd \y_{1;1;3,4}) \, \x_{1;1,4} \, 
        (h_{1;4} \rhd \y_{1;4;2,1}) \,\x_{1;4,5} \,
        (h_{1;5} \rhd \y_{1;5;4,1}) \, \x_{1;5,1}
        \big) \,
        \nonumber \\
        &
        \delta_{G_2}\big(
        (h_{2;1} \rhd \y_{2;1;3,2}) \, \x_{2;1,3} \, 
        (h_{2;3} \rhd \y_{2;3;3,1}) \,\x_{2;3,4} \,
        (h_{2;4} \rhd \y_{2;4;4,2}) \, \x_{2;4,1}
        \big) \,
        \nonumber \\
        &
        \delta_{G_2}\big(
        (h_{3;1} \rhd \y_{3;1;2,1}) \, \x_{3;1,2} \, 
        (h_{3;2} \rhd \y_{3;2;4,1}) \,\x_{3;2,3} \,
        (h_{3;3} \rhd \y_{3;3;4,3}) \, \x_{3;3,1}
        \big)  \,,
        \label{Three4simplices_Wedges}
    \end{align}
    
    We defined the composed links $h_{1,3} \equiv h_{1;3} h_{3;4}\mone$ and $h_{2,3} \equiv h_{2;2} h_{3;5}\mone$ with inverses $h_{1,3}\mone = h_{3,1}$ and $h_{2,3}\mone = h_{3,2}$. 
    
    The first delta in \eqref{Three4simplices_Links} enforces a closed path of only bulk links of the three 4-simplices; this is the loop of links dual to the single face shared by the three 4-simplices. The following nine delta functions involve a combination of bulk and boundary links of the 4-simplices $\{1,2\}$, $\{1,3\}$ and $\{2,3\}$, while the remaining nine deltas involve bulk and boundary links of the first, the second or the third 4-simplices separately.
    Similarly, the first three deltas in \eqref{Three4simplices_Wedges} involve a combination of bulk and boundary wedges of the three 4-simplices. The following nine deltas involve a combination of bulk and boundary wedges of the three 4-simplices pairwise, $\{1,2\}$, $\{1,3\}$ and $\{2,3\}$. The remaining three deltas involve bulk and boundary wedges of the first, second or third 4-simplices separately.
    
    \begin{figure}
    \centering
    \begin{subfigure}[t]{0.35\textwidth}
        \centering
        \begin{tikzpicture}[scale=1.25]
            \coordinate (c) at (0,0);
            \coordinate (c1) at (0,1.5);
            \coordinate (c2) at (1.42,0.46);
            \coordinate (c3) at (0.88,-1.45);
            \coordinate (c4) at (-0.88,-1.45);
            \coordinate (c5) at (-1.42,0.46);
            
            \coordinate (incr1) at (-1.25,0.72);
            \coordinate (x1) at ($ (c) + (incr1) $);
            \coordinate (x11) at ($ (c1) + (incr1) $);
            \coordinate (x14) at ($ (c4) + (incr1) $);
            \coordinate (x15) at ($ (c5) + (incr1) $);
            
            \coordinate (incr2) at (1.25,0.72);
            \coordinate (x2) at ($ (c) + (incr2) $);
            \coordinate (x21) at ($ (c1) + (incr2) $);
            \coordinate (x23) at ($ (c2) + (incr2) $);
            \coordinate (x24) at ($ (c3) + (incr2) $);
            
            \coordinate (incr3) at (0,-1.5);
            \coordinate (x3) at ($ (c) + (incr3) $);
            \coordinate (x31) at ($ (incr3) - (c1) $);
            \coordinate (x32) at ($ (c5) + (incr3) $);
            \coordinate (x33) at ($ (c2) + (incr3) $);
            
            \draw[fill=darkblue] (x1) circle [radius=0.05] node[below , darkblue] {$1$};
            \draw[fill=darkblue] (x2) circle [radius=0.05] node[below , darkblue] {$2$};
            \draw[fill=darkblue] (x3) circle [radius=0.05] node[above , darkblue] {$3$};
            
            \draw[- , thick , darkred] ($ 0.93*(x1) + 0.07*(x11) $) -- node[above right,scale=0.9] {$1$} (x11);
            \draw[- , thick , darkred] ($ 0.93*(x1) + 0.07*(x14) $) -- node[left,scale=0.9] {$4$} (x14);
            \draw[- , thick , darkred] ($ 0.93*(x1) + 0.07*(x15) $) -- node[above,scale=0.9] {$5$} (x15);
            
            \draw[- , thick , darkred] ($ 0.93*(x2) + 0.07*(x21) $) -- node[above right,scale=0.9] {$1$} (x21);
            \draw[- , thick , darkred] ($ 0.93*(x2) + 0.07*(x23) $) -- node[above,scale=0.9] {$3$} (x23);
            \draw[- , thick , darkred] ($ 0.93*(x2) + 0.07*(x24) $) -- node[left,scale=0.9] {$4$} (x24);
            
            \draw[- , thick , darkred] ($ 0.93*(x3) + 0.07*(x31) $) -- node[below right,scale=0.9] {$1$} (x31);
            \draw[- , thick , darkred] ($ 0.93*(x3) + 0.07*(x32) $) -- node[below,scale=0.9] {$2$} (x32);
            \draw[- , thick , darkred] ($ 0.93*(x3) + 0.07*(x33) $) -- node[below,scale=0.9] {$3$} (x33);
            
            \draw[- , thick , darkblue] ($ 0.93*(x1) + 0.07*(x2) $) -- node(12)[above , darkred , scale=0.9] {$(2,5)$} node[above=1em of 12 , darkblue] {$1,2$} ($ 0.93*(x2) + 0.07*(x1) $);
            \draw[- , thick , darkblue] ($ 0.93*(x1) + 0.07*(x3) $) -- node(13)[below , darkblue , rotate=-60] {$1,3$} node[below left=1.5em and 0.2em of 13 , darkred , scale=0.9 , rotate=-60] {$(3,4)$}  ($ 0.93*(x3) + 0.07*(x1) $);
            \draw[- , thick , darkblue] ($ 0.93*(x2) + 0.07*(x3) $) -- node(23)[below , darkblue , rotate=60] {$3,2$} node[below right=1.5em and 0.2em , darkred , scale=0.9 , rotate=60] {$(5,2)$}  ($ 0.93*(x3) + 0.07*(x2) $);
        \end{tikzpicture}
        \caption{Graph dual to three 4-simplices.}
        \label{SubFig_Three4simplices}
    \end{subfigure}
    \hfill
    \begin{subfigure}[t]{0.4\textwidth}
        \centering
        \begin{tikzpicture}[scale=1.25]
            \coordinate (c) at (0,0);
            \coordinate (c1) at (0,1.5);
            \coordinate (c2) at (1.42,0.46);
            \coordinate (c3) at (0.88,-1.45);
            \coordinate (c4) at (-0.88,-1.45);
            \coordinate (c5) at (-1.42,0.46);
            
            \coordinate (incr1) at (-1.25,1.25);
            \coordinate (x1) at ($ (c) + (incr1) $);
            \coordinate (x11) at ($ (c1) + (incr1) $);
            \coordinate (x15) at ($ (c5) + (incr1) $);
            
            \coordinate (incr2) at (1.25,1.25);
            \coordinate (x2) at ($ (c) + (incr2) $);
            \coordinate (x21) at ($ (c1) + (incr2) $);
            \coordinate (x24) at ($ (c2) + (incr2) $);
            
            \coordinate (incr3) at (1.25,-1.25);
            \coordinate (x3) at ($ (c) + (incr3) $);
            \coordinate (x31) at ($ (incr3) - (c1) $);
            \coordinate (x33) at ($ (incr3) - (c5) $);
            
            \coordinate (incr4) at (-1.25,-1.25);
            \coordinate (x4) at ($ (c) + (incr4) $);
            \coordinate (x41) at ($ (incr4) - (c1) $);
            \coordinate (x42) at ($ (incr4) - (c2) $);
            
            \draw[fill=darkblue] (x1) circle [radius=0.05] node[above right , darkblue] {$1$};
            \draw[fill=darkblue] (x2) circle [radius=0.05] node[above left , darkblue] {$2$};
            \draw[fill=darkblue] (x3) circle [radius=0.05] node[below left , darkblue] {$3$};
            \draw[fill=darkblue] (x4) circle [radius=0.05] node[below right , darkblue] {$4$};
            
            \draw[- , thick , darkred] ($ 0.93*(x1) + 0.07*(x11) $) -- node[above right,scale=0.9] {$1$} (x11);
            \draw[- , thick , darkred] ($ 0.93*(x1) + 0.07*(x15) $) -- node[above,scale=0.9] {$5$} (x15);
            
            \draw[- , thick , darkred] ($ 0.93*(x2) + 0.07*(x21) $) -- node[above right,scale=0.9] {$1$} (x21);
            \draw[- , thick , darkred] ($ 0.93*(x2) + 0.07*(x24) $) -- node[above,scale=0.9] {$4$} (x24);
            
            \draw[- , thick , darkred] ($ 0.93*(x3) + 0.07*(x31) $) -- node[below right,scale=0.9] {$1$} (x31);
            \draw[- , thick , darkred] ($ 0.93*(x3) + 0.07*(x33) $) -- node[below,scale=0.9] {$3$} (x33);
            
            \draw[- , thick , darkred] ($ 0.93*(x4) + 0.07*(x41) $) -- node[below right,scale=0.9] {$1$} (x41);
            \draw[- , thick , darkred] ($ 0.93*(x4) + 0.07*(x42) $) -- node[below,scale=0.9] {$2$} (x42);
            
            \draw[- , thick , darkblue] ($ 0.93*(x1) + 0.07*(x2) $) -- node[above , darkred , scale=0.9] {$(2,5)$} node[above=1em , darkblue] {$1,2$} ($ 0.93*(x2) + 0.07*(x1) $);
            \draw[- , thick , darkblue] ($ 0.93*(x1) + 0.07*(x3) $) -- node[below left , darkblue , rotate=-45] {$1,3$} node[below left=0.8em and 0.5em , darkred , scale=0.9 , rotate=-45] {$(3,4)$}  ($ 0.93*(x3) + 0.07*(x1) $);
            \draw[- , thick , darkblue] ($ 0.93*(x1) + 0.07*(x4) $) -- node[left=1em , darkblue , rotate=-90] {$1,4$} node[below left=0.3em and 1.25em , darkred , scale=0.9 , rotate=-90] {$(4,3)$}  ($ 0.93*(x4) + 0.07*(x1) $);
            \draw[- , thick , darkblue] ($ 0.93*(x2) + 0.07*(x3) $) -- node[right=1em , darkblue , rotate=90] {$3,2$} node[below right=0.3em and 1.25em , darkred , scale=0.9 , rotate=90] {$(5,2)$}  ($ 0.93*(x3) + 0.07*(x2) $);
            \draw[- , thick , darkblue] ($ 0.93*(x2) + 0.07*(x4) $) -- node[below right , darkblue , rotate=45] {$4,2$} node[below right=0.8em and 0.5em , darkred , scale=0.9 , rotate=45] {$(4,3)$}  ($ 0.93*(x4) + 0.07*(x2) $);
            \draw[- , thick , darkblue] ($ 0.93*(x3) + 0.07*(x4) $) -- node[below , darkblue] {$4,3$} node[below=1em , darkred , scale=0.9] {$(5,2)$} ($ 0.93*(x4) + 0.07*(x3) $);
        \end{tikzpicture}
        \caption{Graph dual to four 4-simplices.}
        \label{SubFig_Four4simplices}
    \end{subfigure}
    \caption{We labelled the  tetrahedra in red, and in blue the 4-simplices.}
    \label{Fig_4Simplex(3-4)}
\end{figure}
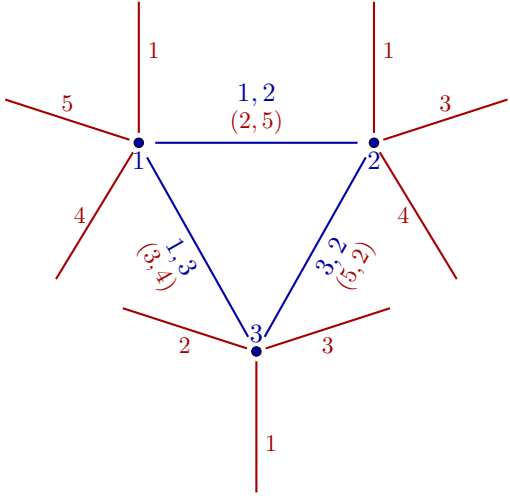
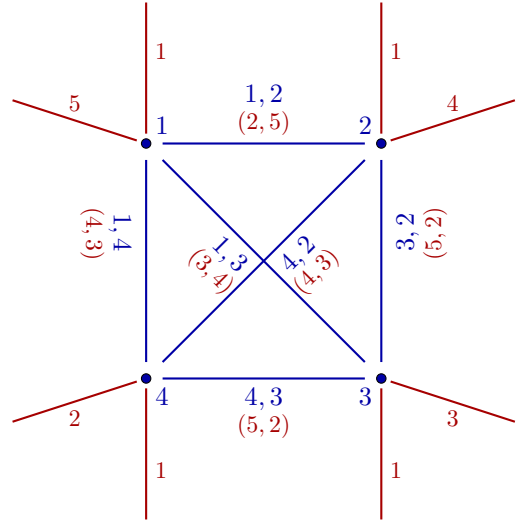

    \item[iv.] \textbf{The  eight-point Feynman diagram,} illustrated in \ref{SubFig_Four4simplices},
    \be
        \begin{aligned}
            \cA_{\cV_4} 
            & = 
            \int \dX^{40} \dh^{20} \du^{80} \dY^{120} \, 
            (\cK_{1,2} \, \cK_{1,3} \, \cK_{1,4} \, \cK_{2,3} \, \cK_{2,4} \, \cK_{3,4}) \,\, 
            \big(\cV_1 \, \cV_2 \, \cV_3 \, \cV_4\big)
            \\
            & =
            \int \dX^{40} \dh^{14} \du^{32} \dY^{48} \,\, 
            \delta_{G_1}^{\{\cV_4\}} \, \delta_{G_2}^{\{\cV_4\}} \,.
            \label{Amplitude_Four4simplices}
        \end{aligned}
    \ee
    We denoted
    \begin{align}
        \delta_{G_1}^{\{\cV_4\}} = \,\,
        &
        \delta_{G_1}\big(
        t(\x_{1;3,2}) h_{1,2} t(\x_{2;5,2}) h_{2,3} t(\x_{3;5,4}) h_{3,1}\big) \,
        \delta_{G_1}\big(
        t(\x_{1;4,2}) h_{1,2} t(\x_{2;5,3}) h_{2,4} t(\x_{4;4,3}) h_{4,1}\big) \,
        \nonumber \\
        &
        \delta_{G_1}\big(
        t(\x_{1;4,3}) h_{1,3} t(\x_{3;4,2}) h_{3,4} t(\x_{4;5,3}) h_{4,1}\big) \,
        \delta_{G_1}\big(
        t(\x_{2;3,2}) h_{2,3} t(\x_{3;5,2}) h_{3,4} t(\x_{4;5,4}) h_{4,2}\big)
        \Big)
        \nonumber \\
        \Big(
        &
        \delta_{G_1}\big(
        t(\x_{1;5,2}) h_{1,2} t(\x_{2;5,4}) h_{2;4} u_{2;4;1} u_{1;5;2}\mone h_{1;5}\mone\big) \,
        \delta_{G_1}\big(
        t(\x_{1;2,1}) h_{1;1} u_{1;1;1} u_{2;1;4}\mone h_{2;1}\mone t(\x_{2;1,5}) h_{2,1}\big)
        \nonumber \\
        &
        \delta_{G_1}\big(
        t(\x_{1;3,1}) h_{1;1} u_{1;1;2} u_{3;1;3}\mone h_{3;1}\mone t(\x_{3;1,4}) h_{3,1}\big) \,
        \delta_{G_1}\big(
        t(\x_{1;5,3}) h_{1,3} t(\x_{3;4,3}) h_{3;3} u_{3;3;1} u_{1;5;3}\mone h_{1;5}\mone\big) \,
        \nonumber \\
        &
        \delta_{G_1}\big(
        t(\x_{2;2,1}) h_{2;1} u_{2;1;1} u_{3;1;4}\mone h_{3;1}\mone t(\x_{3;1,5}) h_{3,2}\big) \,
        \delta_{G_1}\big(
        t(\x_{2;4,2}) h_{2,3} t(\x_{3;5,3}) h_{3;3} u_{3;3;2} u_{2;4;3}\mone h_{2;4}\mone\big) \,
        \nonumber \\
        &
        \delta_{G_1}\big(
        t(\x_{1;5,4}) h_{1;1} u_{1;1;3} \, u_{4;1;2}\mone h_{4;1}\mone t(\x_{4;1,3}) h_{4,1}
        \big) \,
        \delta_{G_1}\big(
        t(\x_{1;4;1}) h_{1,4} t(\x_{4;3,2}) h_{4;2} u_{4;2;1} \, u_{1;5;4}\mone h_{1;5}\mone
        \big)
        \nonumber \\
        &
        \delta_{G_1}\big(
        t(\x_{2;3,1}) h_{2;1} u_{2;1;2} \, u_{4;1;3}\mone h_{4;1}\mone t(\x_{4;1,4}) h_{4,2}
        \big) \,
        \delta_{G_1}\big(
        t(\x_{2;4,3}) h_{2,4} t(\x_{4;4,2}) h_{4;2} u_{4;2;2} \, u_{2;4;4}\mone h_{2;4}\mone
        \big)
        \nonumber \\
        &
        \delta_{G_1}\big(
        t(\x_{3;2,1}) h_{3;1} u_{3;1;1} \, u_{4;1;4}\mone h_{4;1}\mone t(\x_{4;1,5})  h_{4,3}
        \big) \,
        \delta_{G_1}\big(
        t(\x_{3;3,2}) h_{3,4} t(\x_{4;5,2}) h_{4;2} u_{4;2;3} \, u_{3;3;4}\mone h_{3;3}\mone
        \big)
        \Big)
        \nonumber \\
        \Big(
        &
        \delta_{G_1}\big(
        t(\x_{1;5,1}) h_{1;1} u_{1;1;4} \, u_{1;5;1}\mone h_{1;5}\mone
        \big) \,
        \delta_{G_1}\big(
        t(\x_{2;4,1}) h_{2;1} u_{2;1;3} \, u_{2;4;2}\mone h_{2;4}\mone
        \big) \,
        \delta_{G_1}\big(
        t(\x_{3;3,1}) h_{3;1} u_{3;1;2} \, u_{3;3;3}\mone h_{3;3}\mone
        \big)
        \nonumber \\
        &
        \delta_{G_1}\big(
        t(\x_{4;2,1}) h_{4;1} u_{4;1;1} \, u_{4;2;4}\mone h_{4;2}\mone
        \big) 
        \Big) \,,
        \label{Four4simplices_Links}
    \end{align}
    the set of delta functions on the group $G_1$ (on the links) and
    \begin{align}
        \delta_{G_2}^{\{\cV_4\}} = \,\,
        &
        \delta_{G_2}\Big(\x_{1;4,2} \,
        h_{1,2} \rhd \big(\x_{2;5,3} \x_{2;3,2} \x_{2;2,5}\big) \, \x_{1;2,3} \,
        h_{1,3} \rhd \big(\x_{3;4,5} \x_{3;5,2} \x_{3;2,4}\big) \, \x_{1;3,4} \, 
        h_{1,4} \rhd \big(\x_{4;3,5} \x_{4;5,4} \x_{4;4,3}\big)
        \Big)
        \nonumber \\
        &
        \delta_{G_2}\Big(
        h_{1,2} \rhd \big(\x_{2;5,4} \, (h_{2;4} \rhd \y_{2;4;1,3}) \, \x_{2;4,2} \, \x_{2;2,5}\big) \, \x_{1;2,3} \, h_{1,3} \rhd \big(\x_{3;4,5} \, \x_{3;5,3} \, (h_{3;3} \rhd \y_{3;3;2,1}) \, \x_{3;3,4}\big) \,
        \nonumber \\
        & \quad
        \x_{1;3,5} \, (h_{1;5} \rhd \y_{1;5;3,2}) \, \x_{1;5,2}
        \Big)
        \nonumber \\
        &
        \delta_{G_2}\Big(
        h_{1,2} \rhd \big(\x_{2;5,1} \, (h_{2;1} \rhd \y_{2;1;4,1}) \, \x_{2;1,2} \, \x_{2;2,5}\big) \, \x_{1;2,3} \, h_{1,3} \rhd \big(\x_{3;4,5} \, \x_{3;5,1} \, (h_{3;1} \rhd \y_{3;1;3,4}) \, \x_{3;1,4}\big) \, 
        \nonumber \\
        & \quad
        \x_{1;3,1} \, (h_{1;1} \rhd \y_{1;1;2,1}) \, \x_{1;1,2}
        \Big)
        \nonumber \\
        &
        \delta_{G_2}\Big(
        h_{1,2} \rhd \big(\x_{2;5,4} \, (h_{2;4} \rhd \y_{2;4;1,4}) \, \x_{2;4,3} \, \x_{2;3,5}\big) \, \x_{1;2,4} \, h_{1,4} \rhd \big(\x_{4;3,4} \, \x_{4;4,2} \, (h_{4;2} \rhd \y_{4;2;2,1}) \, \x_{4;2,3}\big) \, 
        \nonumber \\
        & \quad
        \x_{1;4,5} \, (h_{1;5} \rhd \y_{1;5;4,2}) \, \x_{1;5,2} 
        \Big)
        \nonumber \\
        &
        \delta_{G_2}\Big(
        h_{1,2} \rhd \big(\x_{2;5,1} \, (h_{2;1} \rhd \y_{2;1;4,2}) \, \x_{2;1,3} \, \x_{2;3,5}\big) \, \x_{1;2,4} \, h_{1,4} \rhd \big(\x_{3;3,4} \, \x_{4;4,1} \, (h_{4;1} \rhd \y_{4;1;3,2}) \, \x_{4;1,3}\big) \, 
        \nonumber \\
        & \quad
        \x_{1;4,1} \, (h_{1;1} \rhd \y_{1;1;3,1}) \, \x_{1;1,2}
        \Big)
        \nonumber \\
        &
        \delta_{G_2}\Big(
        h_{1,3} \rhd \big(\x_{3;4,3} \, (h_{3;3} \rhd \y_{3;3;1,4}) \, \x_{3;3,2} \, \x_{3;2,4}\big) \, \x_{1;3,4} \, h_{1,4} \rhd \big(\x_{4;3,5} \, \x_{4;5,2} \, (h_{4;2} \rhd \y_{4;2;3,1}) \, \x_{4;2,3}\big)\, 
        \nonumber \\
        & \quad
        \x_{1;4,5} \, (h_{1;5} \rhd \y_{1;5;4,3}) \, \x_{1;5,3}
        \Big)
        \nonumber \\
        &
        \delta_{G_2}\Big(
        h_{1,3} \rhd \big(\x_{3;4,1} \, (h_{3;1} \rhd \y_{3;1;3,1}) \, \x_{3;1,2} \, \x_{3,2,4}\big) \, \x_{1;3,4} \, h_{1,4} \rhd \big(\x_{4;3,5} \, \x_{4;5,1} (h_{4;1} \rhd \y_{4;1;4,2}) \, \x_{4;1,3}\big) \, 
        \nonumber \\
        & \quad
        \x_{1;4,1} \, (h_{1;1} \rhd \y_{1;1;3,2}) \, \x_{1;1,3} \, 
        \Big)
        \nonumber \\
        &
        \delta_{G_2}\Big(
        h_{2,3} \rhd \big(\x_{3;5,1} \, (h_{3;1} \rhd \y_{3;1;4,1}) \, \x_{3;1,2} \, \x_{3;2,5}\big) \, \x_{2;2,3} \, h_{2,4} \rhd \big(\x_{4;4,5}\x_{4;5,1} \, (h_{4;1} \rhd \y_{4;1;3,4}) \x_{4;1,4}\big) \,
        \nonumber \\
        & \quad
        \x_{2;3,1} \, (h_{2;1} \rhd \y_{2;1;2,1}) \, \x_{2;1,2}
        \Big) \,
        \nonumber \\
        &
        \delta_{G_2}\big(
        h_{2,3} \rhd \big(\x_{3;5,3} \, (h_{3;3} \rhd \y_{3;3;2,4}) \, \x_{3;3,2} \, \x_{3;2,5}\big) \, \x_{2;2,3} \, h_{2,4} \rhd \big(\x_{4;4,5} \, \x_{4;5,2} \, (h_{4;2} \rhd \y_{4;2;3,2}) \, \x_{4;2,4}\big) \, 
        \nonumber \\
        & \quad
        \x_{2;3,4} \, (h_{2;4} \rhd \y_{2;4;4,3}) \, \x_{2;4,2}
        \Big)
        \nonumber \\
        &
        \delta_{G_2}\Big(
        h_{1,2} \rhd \big(\x_{2;5,1} \, (h_{2;1} \rhd \y_{2;1;3,4}) \, \x_{2;1,4} \, (h_{2;4} \rhd \y_{2;4;2,1}) \, \x_{2;4,5}\big) \, \x_{1;2,5} \, (h_{1;5} \rhd \y_{1;5;2,1}) \, \x_{1;5,1} \, (h_{1;1} \rhd \y_{1;1;4,1}) \, \x_{1;1,2}
        \Big)
        \nonumber \\
        &
        \delta_{G_2}\Big(
        h_{1,3} \rhd \big(\x_{3;4,1} \, (h_{3;1} \rhd \y_{3;1;3,2}) \, \x_{3;1,3} \, (h_{3;3} \rhd \y_{3;3;3,1}) \, \x_{3;3,4}\big) \,\x_{1;3,5} \, (h_{1;5} \rhd \y_{1;5;3,1}) \, \x_{1;5,1} \, (h_{1;1} \rhd \y_{1;1;4,2}) \, \x_{1;1,3}
        \Big) \,
        \nonumber \\
        &
        \delta_{G_2}\Big(
        h_{2,3} \rhd \big(\x_{3;5,1} \, (h_{3;1} \rhd \y_{3;1;4,2}) \, \x_{3;1,3} \, (h_{3;3} \rhd \y_{3;3;3,2}) \, \x_{3;3,5}\big) \,\x_{2;2,4} \, (h_{2;4} \rhd \y_{2;4;3,2}) \, \x_{2;4,1} \, (h_{2;1} \rhd \y_{2;1;3,1}) \, \x_{2;1,2}
        \Big) \,
        \nonumber \\
        &
        \delta_{G_2}\Big(
        h_{1,4} \rhd \big(\x_{4;3,1} \, (h_{4;1} \rhd \y_{4;1;2,1}) \, \x_{4;1,2} \, (h_{4;2} \rhd \y_{4;2;4,1}) \, \x_{4;2,3}\big) \, \x_{1;4,5} \, (h_{1;5} \rhd \y_{1;5;4,1}) \, \x_{1;5,1} \, (h_{1;1} \rhd \y_{1;1;3,4}) \, \x_{1;1,4}
        \Big) \,
        \nonumber \\
        &
        \delta_{G_2}\Big(
        h_{2,4} \rhd \big(\x_{4;4,1} \, (h_{4;1} \rhd \y_{4;1;3,1}) \, \x_{4;1,2} \, (h_{4;2} \rhd \y_{4;2;4,2}) \, \x_{4;2,4}\big) \,\x_{2;3,4} \, (h_{2;4} \rhd \y_{2;4;4,2}) \, \x_{2;4,1} \, (h_{2;1} \rhd \y_{2;1;3,2}) \, \x_{2;1,3}
        \Big) \,
        \nonumber \\
        &
        \delta_{G_2}\Big(
        h_{3,4} \rhd \big(\x_{4;5,1} \, (h_{4;1} \rhd \y_{4;1;4,1}) \, \x_{4;1,2}  \, (h_{4;2;} \rhd \y_{4;2;4,3}) \, \x_{4;2,5}\big) \,\x_{3;2,3} \, (h_{3;3} \rhd \y_{3;3;4,3}) \, \x_{3;3,1} \, (h_{3;1} \rhd \y_{3;1;2,1}) \, \x_{3;1,2}
        \Big)  \,,
        \label{Four4simplices_Wedges}
    \end{align}
    the set of delta functions on the group $G_2$ (on the wedges).
    We defined the composed links
    $h_{1,4} \equiv h_{1;4} h_{4;3}\mone$, 
    $h_{2,4} \equiv h_{2;3} h_{4;4}\mone$ and
    $h_{3,4} \equiv h_{3;2} h_{4;5}\mone$, with inverses
    $h_{1,4}\mone = h_{4,1}$, $h_{2,4}\mone = h_{4,2}$ and $h_{3,4}\mone = h_{4,3}$. \\
    The first four delta functions in \eqref{Four4simplices_Links} enforces a closed path of only bulk links of the three 4-simplices pairwise. These are the loops of links dual to the four faces shared by triplets of the four 4-simplices, $\{1,2,3\}$, $\{1,2,4\}$, $\{1,3,4\}$ and $\{2,3,4\}$. The following twelve delta functions involve a combination of bulk and boundary links of the 4-simplices pairwise, $\{A,B\}$, with $A,B = 1,2,3,4$ and $B > A$; while the remaining four deltas involve bulk and boundary links of the four 4-simplices separately.
    Similarly, the first delta functions in \eqref{Four4simplices_Wedges} is a closed 2-path of only bulk wedges shared by the four 4-simplices; this is the closed surface in the dual complex dual to the single edge shared by the four 4-simplices. The following eight delta functions involve a combination of bulk and boundary wedges of triplets of 4-simplices, $\{A,B,C\}$ for $A,B,C = 1,2,3,4$ and $C>B>A$; the last six delta functions involve a combination of bulk and boundary wedges of the 4-simplices pairwise.
    
    \smallskip 
    

\begin{figure}
    \centering
    \begin{tikzpicture}[scale=1.25]
        \coordinate (c) at (0,0);
        \coordinate (c1) at (0,1.5);
        \coordinate (c2) at (1.42,0.46);
        \coordinate (c3) at (0.88,-1.45);
        \coordinate (c4) at (-0.88,-1.45);
        \coordinate (c5) at (-1.42,0.46);
        
        \coordinate (incr1) at (0,2);
        \coordinate (x1) at ($ (c) + (incr1) $);
        \coordinate (x11) at ($ (c1) + (incr1) $);
        
        \coordinate (incr2) at (1.9,0.61);
        \coordinate (x2) at ($ (c) + (incr2) $);
        \coordinate (x21) at ($ (c2) + (incr2) $);
        
        \coordinate (incr3) at (1.17,-1.6);
        \coordinate (x3) at ($ (c) + (incr3) $);
        \coordinate (x31) at ($ (c3) + (incr3) $);
        
        \coordinate (incr4) at (-1.17,-1.6);
        \coordinate (x4) at ($ (c) + (incr4) $);
        \coordinate (x41) at ($ (c4) + (incr4) $);
        
        \coordinate (incr5) at (-1.9,0.61);
        \coordinate (x5) at ($ (c) + (incr5) $);
        \coordinate (x51) at ($ (c5) + (incr5) $);
        
        \draw[fill=darkblue] (x1) circle [radius=0.05] node[above right , darkblue] {$1$};
        \draw[fill=darkblue] (x2) circle [radius=0.05] node[above , darkblue] {$2$};
        \draw[fill=darkblue] (x3) circle [radius=0.05] node[below left , darkblue] {$3$};
        \draw[fill=darkblue] (x4) circle [radius=0.05] node[below right , darkblue] {$4$};
        \draw[fill=darkblue] (x5) circle [radius=0.05] node[above , darkblue] {$5$};
        
        \draw[- , thick , darkred] ($ 0.93*(x1) + 0.07*(x11) $) -- node[above right,scale=0.9] {$1$} (x11);
        
        \draw[- , thick , darkred] ($ 0.93*(x2) + 0.07*(x21) $) -- node[above,scale=0.9] {$1$} (x21);
        
        \draw[- , thick , darkred] ($ 0.93*(x3) + 0.07*(x31) $) -- node[right,scale=0.9] {$1$} (x31);
        
        \draw[- , thick , darkred] ($ 0.93*(x4) + 0.07*(x41) $) -- node[left,scale=0.9] {$1$} (x41);
        
        \draw[- , thick , darkred] ($ 0.93*(x5) + 0.07*(x51) $) -- node[above,scale=0.9] {$1$} (x51);
        
        \draw[- , thick , darkblue] ($ 0.93*(x1) + 0.07*(x2) $) -- node[above , darkred , scale=0.9 , rotate=-35] {$(2,5)$} node[above right=1.5em and -0.4em , darkblue , rotate=-35] {$1,2$} ($ 0.93*(x2) + 0.07*(x1) $);
        \draw[- , thick , darkblue] ($ 0.93*(x1) + 0.07*(x3) $) -- node[above right=1.1em and 0.0em , darkred , scale=0.9  , rotate=-75] {$(3,4)$} node[above right=1.25em and 0.9em , darkblue, rotate=-75] {$1,3$}  ($ 0.93*(x3) + 0.07*(x1) $);
        \draw[- , thick , darkblue] ($ 0.93*(x1) + 0.07*(x4) $) -- node[above left=1.1em and 0.0em , darkred , scale=0.9  , rotate=75] {$(4,3)$} node[above left=1.25em and 0.9em , darkblue, rotate=75] {$4,1$}  ($ 0.93*(x4) + 0.07*(x1) $);
        \draw[- , thick , darkblue] ($ 0.93*(x1) + 0.07*(x5) $) -- node[above , darkred , scale=0.9 , rotate=35] {$(2,5)$} node[above left=1.5em and -0.4em , darkblue , rotate=35] {$5,1$} ($ 0.93*(x5) + 0.07*(x1) $);
        
        \draw[- , thick , darkblue] ($ 0.93*(x2) + 0.07*(x3) $) -- node[below right , darkblue , rotate=75] {$3,2$} node[below right=0.4em and 1em , darkred , scale=0.9 , rotate=75] {$(5,2)$}  ($ 0.93*(x3) + 0.07*(x2) $);
        \draw[- , thick , darkblue] ($ 0.93*(x2) + 0.07*(x4) $) -- node[below , darkblue , rotate=40] {$4,2$} node[below right=1.6em and -0.5em, darkred , scale=0.9 , rotate=40] {$(4,3)$}  ($ 0.93*(x4) + 0.07*(x2) $);
        \draw[- , thick , darkblue] ($ 0.93*(x2) + 0.07*(x5) $) -- node[above , darkred , scale=0.9] {$(3,4)$} node[above=1em , darkblue] {$5,2$} ($ 0.93*(x5) + 0.07*(x2) $);
        
        \draw[- , thick , darkblue] ($ 0.93*(x3) + 0.07*(x4) $) -- node[below , darkblue] {$4,3$} node[below=1em , darkred , scale=0.9] {$(5,2)$} ($ 0.93*(x4) + 0.07*(x3) $);
        \draw[- , thick , darkblue] ($ 0.93*(x3) + 0.07*(x5) $) -- node[below , darkblue , rotate=-40] {$5,3$} node[below left=1.6em and -0.5em, darkred , scale=0.9 , rotate=-40] {$(4,3)$}  ($ 0.93*(x5) + 0.07*(x3) $);
        
        \draw[- , thick , darkblue] ($ 0.93*(x4) + 0.07*(x5) $) -- node[below left , darkblue , rotate=-75] {$5,4$} node[below left=0.4em and 1em , darkred , scale=0.9 , rotate=-75] {$(5,2)$}  ($ 0.93*(x5) + 0.07*(x4) $);
    \end{tikzpicture}
    \caption{Graphs dual to five 4-simplices. We labelled the  tetrahedra in red, and in blue the 4-simplices.}
    \label{Fig_Five4simplices}
\end{figure}
    
    \item[v.] \textbf{The 5-point Feynman diagram,} illustrated in Fig. \ref{Fig_Five4simplices}, 
    \be
        \begin{aligned}
            \cA_{\cV_5} 
            & = 
            \int \dX^{50} \dh^{25} \du^{100} \dY^{150} \, 
            (\cK_{1,2} \, \cK_{1,3} \, \cK_{1,4} \, \cK_{1,5} \, \cK_{2,3} \, \cK_{2,4} \, \cK_{2,5} \, \cK_{3,4} \, \cK_{3,5} \, \cK_{4,5}) \,\, 
            \big(\cV_1 \, \cV_2 \, \cV_3 \, \cV_4 \, \cV_5\big)
            \\
            & =
            \int \dX^{50} \dh^{15} \du^{20} \dY^{30} \,\,
            \delta_{G_1}^{\{\cV_5\}} \, \delta_{G_2}^{\{\cV_5\}} \,
            \label{Amplitude_Five4simplices}
        \end{aligned}
    \ee
    We denoted
    \begin{align}
        \delta_{G_1}^{\{\cV_5\}} =
        \Big(
        &
        \delta_{G_1}\big(
        t(\x_{1;3,2}) h_{1,2} t(\x_{2;5,2}) h_{2,3} t(\x_{3;5,4}) h_{3,1}\big) \,
        \delta_{G_1}\big(
        t(\x_{1;4,2}) h_{1,2} t(\x_{2;5,3}) h_{2,4} t(\x_{4;4,3}) h_{4,1}\big) \,
        \nonumber \\
        &
        \delta_{G_1}\big(
        t(\x_{1;4,3}) h_{1,3} t(\x_{3;4,2}) h_{3,4} t(\x_{4;5,3}) h_{4,1}\big) \,
        \delta_{G_1}\big(
        t(\x_{2;3,2}) h_{2,3} t(\x_{3;5,2}) h_{3,4} t(\x_{4;5,4}) h_{4,2}\big) \,
        \nonumber \\
        &
        \delta_{G_1}\big(
        t(\x_{1;5,2}) h_{1,2} t(\x_{2;5,4}) h_{2,5} t(\x_{5;3,2}) h_{5,1}\big) \,
        \delta_{G_1}\big(
        t(\x_{1;5,3}) h_{1,3} t(\x_{3;4,3}) h_{3,5} t(\x_{5;4,2}) h_{5,1}\big) \,
        \nonumber \\
        &
        \delta_{G_1}\big(
        t(\x_{2;4,2}) h_{2,3} t(\x_{3;5,3}) h_{3,5} t(\x_{5;4,3}) h_{5,2}\big) \,
        \delta_{G_1}\big(
        t(\x_{1;5,4}) h_{1,4} t(\x_{4;3,2}) h_{4,5} t(\x_{5;5,2}) h_{5,1}\big) \,
        \nonumber \\
        &
        \delta_{G_1}\big(
        t(\x_{2;4,3}) h_{2,4} t(\x_{4;4,2}) h_{4,5} t(\x_{5;5,3}) h_{5,2}\big) \,
        \delta_{G_1}\big(
        t(\x_{3;3,2}) h_{3,4} t(\x_{4;5,2}) h_{4,5} t(\x_{5;5,4}) h_{5,3}\big)
        \Big) \,
        \nonumber \\
        \Big(
        &
        \delta_{G_1}\big(
        t(\x_{1;2,1}) h_{1;1} u_{1;1;1} u_{2;1;4}\mone h_{2;1}\mone t(\x_{2;1,5}) h_{2,1}\big) \,
        \delta_{G_1}\big(
        t(\x_{1;3,1}) h_{1;1} u_{1;1;2} u_{3;1;3}\mone h_{3;1}\mone t(\x_{3;1,4}) h_{3,1}\big) \,
        \nonumber \\
        &
        \delta_{G_1}\big(
        t(\x_{1;4,1}) h_{1;1} u_{1;1;3} u_{4;1;2}\mone h_{4;1}\mone t(\x_{4;1,3}) h_{4,1}\big) \,
        \delta_{G_1}\big(
        t(\x_{2;2,1}) h_{2;1} u_{2;1;1} u_{3;1;4}\mone h_{3;1}\mone t(\x_{3;1,5}) h_{3,2}\big) \,
        \nonumber \\
        &
        \delta_{G_1}\big(
        t(\x_{2;3,1}) h_{2;1} u_{2;1;2} u_{4;1;3}\mone h_{4;1}\mone t(\x_{4;1,4}) h_{4,2}\big) \,
        \delta_{G_1}\big(
        t(\x_{3;2,1}) h_{3;1} u_{3;1;1} u_{4;1;4}\mone h_{4;1}\mone t(\x_{4;1,5}) h_{4,3}\big)
        \nonumber \\
        &
        \delta_{G_1}\big(
        t(\x_{1;5,1}) h_{1;1} u_{1;1;4} u_{5;1;1}\mone h_{5;1}\mone t(\x_{5;1,2}) h_{5,1}\big) \,
        \delta_{G_1}\big(
        t(\x_{2;4,1}) h_{2;1} u_{2;1;3} u_{5;1;2}\mone h_{5;1}\mone t(\x_{5;1,3}) h_{5,2}\big) \,
        \nonumber \\
        &
        \delta_{G_1}\big(
        t(\x_{3;3,1}) h_{3;1} u_{3;1;2} u_{5;1;3}\mone h_{5;1}\mone t(\x_{5;1,4}) h_{5,3}\big) \,
        \delta_{G_1}\big(
        t(\x_{4;2,1}) h_{4;1} u_{4;1;1} u_{5;1;4}\mone h_{5;1}\mone t(\x_{5;1,5}) h_{5,4}\big) 
        \Big) \,,
        \label{Five4simplices_Links}
    \end{align}
    the set of delta functions on the group $G_1$ (on the links) and
    \begin{align}
        \delta_{G_2}^{\{\cV_5\}} = \,\,
        &
        \delta_{G_2}\Big(\x_{1;4,2} \,
        h_{1,2} \rhd \big(\x_{2;5,3} \x_{2;3,2} \x_{2;2,5}\big) \, \x_{1;2,3} \,
        h_{1,3} \rhd \big(\x_{3;4,5} \x_{3;5,2} \x_{3;2,4}\big) \, \x_{1;3,4} \, 
        h_{1,4} \rhd \big(\x_{4;3,5} \x_{4;5,4} \x_{4;4,3}\big)
        \Big)
        \nonumber \\
        &
        \delta_{G_2}\Big(\x_{1;5,2} \,
        h_{1,2} \rhd \big(\x_{2;5,4} \x_{2;4,2} \x_{2;2,5}\big) \, \x_{1;2,3} \,
        h_{1,3} \rhd \big(\x_{3;4,5} \x_{3;5,3} \x_{3;3,4}\big) \, \x_{1;3,5} \, 
        h_{1,5} \rhd \big(\x_{5;2,4} \x_{5;4,3} \x_{5;3,2}\big)
        \Big)
        \nonumber \\
        &
        \delta_{G_2}\Big(\x_{1;5,2} \,
        h_{1,2} \rhd \big(\x_{2;5,4} \x_{2;4,3} \x_{2;3,5}\big) \, \x_{1;2,4} \,
        h_{1,4} \rhd \big(\x_{4;4,3} \x_{4;3,2} \x_{4;2,4}\big) \, \x_{1;4,5} \, 
        h_{1,5} \rhd \big(\x_{5;2,5} \x_{5;5,3} \x_{5;3,2}\big)
        \Big)
        \nonumber \\
        &
        \delta_{G_2}\Big(\x_{1;5,3} \,
        h_{1,3} \rhd \big(\x_{3;4,3} \x_{3;3,2} \x_{3;2,4}\big) \, \x_{1;3,4} \,
        h_{1,4} \rhd \big(\x_{4;3,5} \x_{4;5,2} \x_{4;2,3}\big) \, \x_{1;4,5} \, 
        h_{1,5} \rhd \big(\x_{5;2,5} \x_{5;5,4} \x_{5;4,2}\big)
        \Big)
        \nonumber \\
        &
        \delta_{G_2}\Big(\x_{2;4,2} \,
        h_{2,3} \rhd \big(\x_{3;5,3} \x_{3;3,2} \x_{3;2,5}\big) \, \x_{2;2,3} \,
        h_{2,4} \rhd \big(\x_{4;4,5} \x_{4;5,2} \x_{4;2,4}\big) \, \x_{2;3,4} \, 
        h_{2,5} \rhd \big(\x_{5;3,5} \x_{5;5,4} \x_{5;4,3}\big)
        \Big)
        \nonumber \\
        &
        \delta_{G_2}\Big(
        h_{1,2} \rhd \big(\x_{2;5,1} \, (h_{2;1} \rhd \y_{2;1;4,1}) \, \x_{2;1,2} \, \x_{2;2,5}\big) \, \x_{1;2,3} \, h_{1,3} \rhd \big(\x_{3;4,5} \, \x_{3;5,1} \, (h_{3;1} \rhd \y_{3;1;3,4}) \, \x_{3;1,4}\big) \, 
        \nonumber \\
        & \quad
        \x_{1;3,1} \, (h_{1;1} \rhd \y_{1;1;2,1}) \, \x_{1;1,2}
        \Big)
        \nonumber \\
        &
        \delta_{G_2}\Big(
        h_{1,2} \rhd \big(\x_{2;5,1} \, (h_{2;1} \rhd \y_{2;1;4,2}) \, \x_{2;1,3} \, \x_{2;3,5}\big) \, \x_{1;2,4} \, h_{1,4} \rhd \big(\x_{3;3,4} \, \x_{4;4,1} \, (h_{4;1} \rhd \y_{4;1;3,2}) \, \x_{4;1,3}\big) \, 
        \nonumber \\
        & \quad
        \x_{1;4,1} \, (h_{1;1} \rhd \y_{1;1;3,1}) \, \x_{1;1,2}
        \Big)
        \nonumber \\
        &
        \delta_{G_2}\Big(
        h_{1,3} \rhd \big(\x_{3;4,1} \, (h_{3;1} \rhd \y_{3;1;3,1}) \, \x_{3;1,2} \, \x_{3,2,4}\big) \, \x_{1;3,4} \, h_{1,4} \rhd \big(\x_{4;3,5} \, \x_{4;5,1} (h_{4;1} \rhd \y_{4;1;4,2}) \, \x_{4;1,3}\big) \, 
        \nonumber \\
        & \quad
        \x_{1;4,1} \, (h_{1;1} \rhd \y_{1;1;3,2}) \, \x_{1;1,3} \, 
        \Big)
        \nonumber \\
        &
        \delta_{G_2}\Big(
        h_{2,3} \rhd \big(\x_{3;5,1} \, (h_{3;1} \rhd \y_{3;1;4,1}) \, \x_{3;1,2} \, \x_{3;2,5}\big) \, \x_{2;2,3} \, h_{2,4} \rhd \big(\x_{4;4,5}\x_{4;5,1} \, (h_{4;1} \rhd \y_{4;1;3,4}) \x_{4;1,4}\big) \,
        \nonumber \\
        & \quad
        \x_{2;3,1} \, (h_{2;1} \rhd \y_{2;1;2,1}) \, \x_{2;1,2}
        \Big) \,
        \nonumber \\
        &
        \delta_{G_2}\Big(
        h_{1,2} \rhd \big(\x_{2;5,1} \, (h_{2;1} \rhd \y_{2;1;3,4}) \, \x_{2;1,4} \, \x_{2;4,5}\big) \, \x_{1;2,5} \, h_{1,5} \rhd \big(\x_{5;2,3} \, \x_{5;3,1} \, (h_{5;1} \rhd \y_{5;1;2,1}) \, \x_{5;1,2}\big) \,
        \nonumber \\
        & \quad
        \x_{1;5,1} \, (h_{1;1} \rhd \y_{1;1;4,1}) \, \x_{1;1,2}
        \Big)
        \nonumber \\
        &
        \delta_{G_2}\Big(
        h_{1,3} \rhd \big(\x_{3;4,1} \, (h_{3;1} \rhd \y_{3;1;3,2}) \, \x_{3;1,3} \, \x_{3;3,4}\big) \, \x_{1;3,5} \, h_{1,5} \rhd \big(\x_{5;4,3} \, \x_{5;3,1} \, (h_{5;1} \rhd \y_{5;1;2,3}) \, \x_{5;1,4}\big) \,
        \nonumber \\
        & \quad
        \x_{1;5,1} \, (h_{1;1} \rhd \y_{1;1;4,2}) \, \x_{1;1,3}
        \Big) \,
        \nonumber \\
        &
        \delta_{G_2}\Big(
        h_{2,3} \rhd \big(\x_{3;5,1} \, (h_{3;1} \rhd \y_{3;1;4,2}) \, \x_{3;1,3} \, \x_{3;3,5}\big) \, \x_{2;2,4} \, h_{2,5} \rhd \big(\x_{5;5,3} \, \x_{5;3,1} \, (h_{5;1} \rhd \y_{5;1;2,4}) \, \x_{5;1,5}\big) \,
        \nonumber \\
        & \quad
        \x_{2;4,1} \, (h_{2;1} \rhd \y_{2;1;3,1}) \, \x_{2;1,2}
        \Big) \,
        \nonumber \\
        &
        \delta_{G_2}\Big(
        h_{1,4} \rhd \big(\x_{4;3,1} \, (h_{4;1} \rhd \y_{4;1;2,1}) \, \x_{4;1,2} \, \x_{4;2,3}\big) \, \x_{1;4,5} \, h_{1,5} \rhd \big(\x_{5;2;5} \, \x_{5;5,1} \, (h_{5;1} \rhd \y_{5;1;4,1}) \, \x_{5;1,2}\big) \, 
        \nonumber \\
        & \quad
        \x_{1;5,1} \, (h_{1;1} \rhd \y_{1;1;3,4}) \, \x_{1;1,4}
        \Big) \,
        \nonumber \\
        &
        \delta_{G_2}\Big(
        h_{2,4} \rhd \big(\x_{4;4,1} \, (h_{4;1} \rhd \y_{4;1;3,1}) \, \x_{4;1,2} \, \x_{4;2,4}\big) \, \x_{2;3,4} \, h_{2,5} \rhd \big(\x_{5;3;5} \, x_{5;5,1} \, (h_{5;1} \rhd \y_{5;1;4,2}) \, \x_{5;1,3}\big) \, 
        \nonumber \\
        & \quad
        \x_{2;4,1} \, (h_{2;1} \rhd \y_{2;1;3,2}) \, \x_{2;1,3}
        \Big) \,
        \nonumber \\
        &
        \delta_{G_2}\Big(
        h_{3,4} \rhd \big(\x_{4;5,1} \, (h_{4;1} \rhd \y_{4;1;4,1}) \, \x_{4;1,2} \, \x_{4;2,5}\big) \, \x_{3;2,3} \, h_{3,5} \rhd \big(\x_{5;4;5} \, \x_{5;5,1} \, (h_{5;1} \rhd \y_{5;1;3,4}) \, \x_{5;1,4}\big) \, 
        \nonumber \\
        & \quad
        \x_{3;3,1} \, (h_{3;1} \rhd \y_{3;1;2,1}) \, \x_{3;1,2}
        \Big)  \,,
        \label{Five4simplices_Wedges}
    \end{align}
    the set of delta functions on the group ${G_2}$ (on the wedges).
    We defined the composed links
    $h_{1,5} \equiv h_{1;5} h_{5;2}\mone$, 
    $h_{2,5} \equiv h_{2;4} h_{5;3}\mone$,
    $h_{3,5} \equiv h_{2;3} h_{5;4}\mone$ and
    $h_{4,5} \equiv h_{3;2} h_{5;5}\mone$, with inverses
    $h_{1,5}\mone = h_{5,1}$, $h_{2,5}\mone = h_{5,2}$, $h_{3,5}\mone = h_{5,3}$ and $h_{4,5}\mone = h_{5,4}$. \\
    {The first six delta functions in \eqref{Five4simplices_Links} enforces a closed path of only bulk links of the 4-simplices pairwise; these are the loops of links dual to the six faces shared by triplets of four 4-simplices. The remaining ten delta functions involve a combination of bulk and boundary links of the 4-simplices pairwise.
    Similarly, the first five delta functions in \eqref{Five4simplices_Wedges} enforce the closure of five 2-path of only bulk wedges shared by the 4-simplices four by four; these are closed surfaces in the dual complex dual to the edges shared by the four 4-simplices.
    The remaining ten delta functions involve a combination of bulk and boundary wedges of triplets of 4-simplices.}
\end{itemize}

\begin{itemize}
    \item[$\ast$]
    \textbf{Pachner move $\mathrm{P_{1,5}}$}, illustrated in Fig. \ref{Fig_Pachner(1,5)}.  We consider the amplitude of five 4-simplices \eqref{Amplitude_Five4simplices}. We integrate over six bulk links ($h_{1,2},h_{1,3},h_{1,4},h_{2,3},h_{2,4},h_{3,4}$) and over three bulk wedges ($\x_{4;3,4},\x_{4;3,5},\x_{4;4,5}$). Upon integration, four delta functions on the group $G_1$ and one on the group $G_2$ are automatically satisfied and give the factors $\delta_{G_1}(1)^{\times 4}$ and $\delta_{G_2}(1)$.
    In the Lie group case these factors are divergent, but can be removed by properly normalizing the amplitudes, just as in the ordinary GFT case; for finite groups, these factors are equal to $1$ and thus automatically regularized.

    Consider the map from the variables that decorate the dual of the combination of five 4-simplices to the variables that decorate the dual of a single 4-simplex.
    \be
        \begin{aligned}
            &
            \begin{aligned}
                &
                \textbf{\textit{Boundary links}} 
                \\
                &
                u_{A;1;i} \,\,\to\,\, u_{A;i} 
                \quad \text{ for } \quad 
                \begin{aligned}
                    A & = 1,2,3,4,5 \\ 
                    i & = 1,2,3,4
                \end{aligned}
            \end{aligned}
            \\
            &
            \begin{aligned}
                &
                \textbf{\textit{Boundary wedges}} 
                \\
                &
                \y_{A;1;i,j} \,\,\to\,\, \y_{A;i,j}
                \quad \text{ for } \quad 
                \begin{aligned}
                    A & = 1,2,3,4,5 \\ 
                    i,j & = 1,2,3,4
                \end{aligned}
            \end{aligned}
            \\
            &
            \begin{aligned}
                &
                \textbf{\textit{Bulk links}}
                \\
                &
                h_{5,A} h_{A;1} \,\,\to\,\, h_A
                \quad \text{ for } \,\,\, A = 1,2,3,4
                \\
                &
                h_{5;1} \,\,\to\,\, h_5
            \end{aligned}
        \end{aligned}
        \qquad \qquad
        \begin{aligned}
            &
            \textbf{\textit{Bulk wedges}}
            \\
            &
            h_{5,2} \rhd (\x_{2;1,5} \x_{2;5,4}) \x_{5;3,2} h_{5,1} \rhd (\x_{1;5,2} \x_{1;2,1})
            \,\,\to\,\, \x_{2,1} \\
            &
            h_{5,3} \rhd (\x_{3;1,4} \x_{3;4,3}) \x_{5;4,2} h_{5,1} \rhd (\x_{1;5,3} \x_{1;3,1})
            \,\,\to\,\, \x_{3,1} \\
            &
            h_{5,4} \rhd (\x_{4;1,3} \x_{4;3,2}) \x_{5;5,2} h_{5,1} \rhd (\x_{1;5,4} \x_{1;4,1})
            \,\,\to\,\, \x_{4,1} \\
            &
            h_{5,3} \rhd (\x_{3;1,5} \x_{3;5,3}) \x_{5;4,3} h_{5,2} \rhd (\x_{2;4,2} \x_{2;2,1})
            \,\,\to\,\, \x_{3,2} \\
            &
            h_{5,4} \rhd (\x_{4;1,4} \x_{4;4,2}) \x_{5;5,3} h_{5,2} \rhd (\x_{2;4,3} \x_{2;3,1})
            \,\,\to\,\, \x_{4,2} \\
            &
            h_{5,4} \rhd (\x_{4;1,5} \x_{4;5,2}) \x_{5;5,4} h_{5,3} \rhd (\x_{3;3,2} \x_{3;2,1})
            \,\,\to\,\, \x_{4,3} \\
            &
            \x_{5;1,2} h_{5,1} \rhd (\x_{1;5,1})
            \,\,\to\,\, \x_{5,1} \\
            &
            \x_{5;1,3} h_{5,2} \rhd (\x_{2;4,1})
            \,\,\to\,\, \x_{5,2} \\
            &
            \x_{5;1,4} h_{5,3} \rhd (\x_{3;3,1})
            \,\,\to\,\, \x_{5,3} \\
            &
            \x_{5;1,5} h_{5,4} \rhd (\x_{4;2,1})
            \,\,\to\,\, \x_{5,4}
        \end{aligned}\label{cv15}
    \ee
    Under such change of variables, the five point Feynman diagram \eqref{Amplitude_Five4simplices} turns out to be proportional to that of a vertex diagram \eqref{Amplitude_4Simplex}:
    \be
        \cA_{\cV_5} = (V_{G_1}^{4} V_{G_2}^{37}) \, \cA_{\cV_1} \,.
    \ee

    \item[$\ast$]
    \textbf{Pachner move $\mathrm{P_{2,4}}$}, illustrated in Fig. \ref{Fig_Pachner(2,4)}.  Consider the amplitude of four 4-simplices \eqref{Amplitude_Four4simplices}. We integrate over three bulk links ($h_{1,2},h_{1,3},h_{2,3}$). Upon integration, the last delta function on the bulk links and the only delta function on the bulk wedges have the shape 
    \be
        \delta_{G_1}(t(\x)) \, \delta_{G_2}(\x) \,,    
    \ee
    with $\x$ being a combination of twelve wedge decorations in $G_2$. 
    We introduce then the map from the variables that decorate the combination of four 4-simplices to the variables that decorate two 4-simplices.
    \be                         
        \begin{aligned}
            &
            \begin{aligned}
                &
                \textbf{\textit{Boundary links}} 
                \\
                &
                u_{1;1;1} \,\,\to\,\, u_{1;1;2} \\
                &
                u_{1;1;2} \,\,\to\,\, u_{1;1;3} \\
                &
                u_{1;1;3} \,\,\to\,\, u_{1;1;4} \\
                &
                u_{1;1;4} \,\,\to\,\, u_{1;1;1} \\
                &
                u_{1;5;1} \,\,\to\,\, u_{2;1;4} \\
                &
                u_{1;5;2} \,\,\to\,\, u_{2;1;1} \\
                &
                u_{1;5;3} \,\,\to\,\, u_{2;1;2} \\
                &
                u_{1;5;4} \,\,\to\,\, u_{2;1;3} 
            \end{aligned}
            \quad
            \begin{aligned}
                &
                \\
                &
                u_{2;1;1} \,\,\to\,\, u_{1;3;1} \\
                &
                u_{2;1;2} \,\,\to\,\, u_{1;3;2} \\
                &
                u_{2;1;3} \,\,\to\,\, u_{1;3;4} \\
                &
                u_{2;1;4} \,\,\to\,\, u_{1;3;3} \\
                &
                u_{2;4;1} \,\,\to\,\, u_{2;2;4} \\
                &
                u_{2;4;2} \,\,\to\,\, u_{2;2;3} \\
                &
                u_{2;4;3} \,\,\to\,\, u_{2;2;1} \\
                &
                u_{2;4;4} \,\,\to\,\, u_{2;2;2} 
            \end{aligned}
            \quad
            \begin{aligned}
                &
                \\
                &
                u_{3;1;1} \,\,\to\,\, u_{1;4;1} \\
                &
                u_{3;1;2} \,\,\to\,\, u_{1;4;3} \\
                &
                u_{3;1;3} \,\,\to\,\, u_{1;4;2} \\
                &
                u_{3;1;4} \,\,\to\,\, u_{1;4;4} \\
                &
                u_{3;3;1} \,\,\to\,\, u_{2;3;3} \\
                &
                u_{3;3;2} \,\,\to\,\, u_{2;3;4} \\
                &
                u_{3;3;3} \,\,\to\,\, u_{2;3;2} \\
                &
                u_{3;3;4} \,\,\to\,\, u_{2;3;1} 
            \end{aligned}
            \quad
            \begin{aligned}
                &
                \\
                &
                u_{4;1;1} \,\,\to\,\, u_{1;5;2} \\
                &
                u_{4;1;2} \,\,\to\,\, u_{1;5;1} \\
                &
                u_{4;1;3} \,\,\to\,\, u_{1;5;3} \\
                &
                u_{4;1;4} \,\,\to\,\, u_{1;5;4} \\
                &
                u_{4;2;1} \,\,\to\,\, u_{2;4;2} \\
                &
                u_{4;2;2} \,\,\to\,\, u_{2;4;3} \\
                &
                u_{4;2;3} \,\,\to\,\, u_{2;4;4} \\
                &
                u_{4;2;4} \,\,\to\,\, u_{2;4;1} 
            \end{aligned}
            \qquad
            \begin{aligned}
                &
                \textbf{\textit{Bulk links}}
                \\
                &
                h_{4,1} h_{1;1} \,\,\to\,\, h_{1;1} \\
                &
                h_{4,2} h_{2;1} \,\,\to\,\, h_{1;3} \\
                &
                h_{4,3} h_{3;1} \,\,\to\,\, h_{1;4} \\
                &
                h_{4;1} \,\,\to\,\, h_{1;5} \\
                &
                h_{4,1} h_{1;5} \,\,\to\,\, h_{1,2} h_{2;1} \\
                &
                h_{4,2} h_{2;4} \,\,\to\,\, h_{1,2} h_{2;2} \\
                &
                h_{4,3} h_{3;3} \,\,\to\,\, h_{1,2} h_{2;3} \\
                &
                h_{4;2} \,\,\to\,\, h_{1,2} h_{2;4}  
            \end{aligned}
        \end{aligned}
    \ee
    \be
        \begin{aligned}
            &
            \begin{aligned}
                &
                \textbf{\textit{Boundary wedges}} 
                \\
                &
                h_{4,1} \rhd (h_{1;1} \rhd \y_{1;1;2,1})
                \,\,\to\,\,
                h_{1;1} \rhd \y_{1;1;3,2} \\
                &
                h_{4,1} \rhd (h_{1;1} \rhd \y_{1;1;3,1})
                \,\,\to\,\,
                h_{1;1} \rhd \y_{1;1;4,2} \\
                &
                h_{4,1} \rhd (h_{1;1} \rhd \y_{1;1;3,2})
                \,\,\to\,\,
                h_{1;1} \rhd \y_{1;1;3,4} \\
                &
                h_{4,1} \rhd (h_{1;5} \rhd \y_{1;5;3,2})
                \,\,\to\,\,
                (h_{1,2} h_{2;1}) \rhd \y_{2;1;2,1} \\
                &
                h_{4,1} \rhd (h_{5;1} \rhd \y_{5;1;4,2})
                \,\,\to\,\,
                (h_{1,2} h_{2;1}) \rhd \y_{2;1;3,1} \\
                &
                h_{4,1} \rhd (h_{1;5} \rhd \y_{1;5;4,3})
                \,\,\to\,\,
                (h_{1,2} h_{2;1}) \rhd \y_{2;1;3,2} \\
                &
                h_{4,1} \rhd (h_{1;1} \rhd \y_{1;1;4,1})
                \,\,\to\,\,
                h_{1;1} \rhd \y_{1;1;1,2} \\
                &
                h_{4,1} \rhd (h_{1;5} \rhd \y_{1;5;2,1})
                \,\,\to\,\,
                (h_{1,2} h_{2;1}) \rhd \y_{2;1;1,4} \\
                &
                h_{4,1} \rhd (h_{1;5} \rhd \y_{1;5;3,1})
                \,\,\to\,\,
                (h_{1,2} h_{2;1}) \rhd \y_{2;1;2,4} \\
                &
                h_{4,1} \rhd (h_{1;1} \rhd \y_{1;1;4,2})
                \,\,\to\,\,
                h_{1;1} \rhd \y_{1;1;1,3} \\
                &
                h_{4,1} \rhd (h_{1;1} \rhd \y_{1;1;3,4})
                \,\,\to\,\,
                h_{1;1} \rhd \y_{1;1;1,4} \\
                &
                h_{4,1} \rhd (h_{1;5} \rhd \y_{1;5;4,1})
                \,\,\to\,\,
                (h_{1,2} h_{2;1}) \rhd \y_{2;1;4,3} \\
                &
                h_{4,3} \rhd (h_{3;1} \rhd \y_{3;1;3,4})
                \,\,\to\,\,
                h_{1;4} \rhd \y_{1;4;4,2} \\
                &
                h_{4,3} \rhd (h_{3;1} \rhd \y_{3;1;3,1})
                \,\,\to\,\,
                h_{1;4} \rhd \y_{1;4;2,1} \\
                &
                h_{4,3} \rhd (h_{3;1} \rhd \y_{3;1;4,1})
                \,\,\to\,\,
                h_{1;4} \rhd \y_{1;4;4,1} \\
                &
                h_{4,3} \rhd (h_{3;3} \rhd \y_{3;3;2,1})
                \,\,\to\,\,
                (h_{1,2} h_{2;3}) \rhd \y_{2;3;4,3} \\
                &
                h_{4,3} \rhd (h_{3;3} \rhd \y_{3;3;1,4})
                \,\,\to\,\,
                (h_{1,2} h_{2;3}) \rhd \y_{2;3;3,1} \\
                &
                h_{4,3} \rhd (h_{3;3} \rhd \y_{3;3;2,4})
                \,\,\to\,\,
                (h_{1,2} h_{2;3}) \rhd \y_{2;3;4,1} \\
                &
                h_{4,3} \rhd (h_{3;1} \rhd \y_{3;1;3,2})
                \,\,\to\,\,
                h_{1;4} \rhd \y_{1;4;2,3} \\
                &
                h_{4,3} \rhd (h_{3;3} \rhd \y_{3;3;3,1})
                \,\,\to\,\,
                (h_{1,2} h_{2;3}) \rhd \y_{2;3;2,3} \\
                &
                h_{4,3} \rhd (h_{3;1} \rhd \y_{3;1;4,2})
                \,\,\to\,\,
                h_{1;4} \rhd \y_{1;4;4,3} \\
                &
                h_{4,3} \rhd (h_{3;3} \rhd \y_{3;3;3,2})
                \,\,\to\,\,
                (h_{1,2} h_{2;3}) \rhd \y_{2;3;2,4} \\
                &
                h_{4,3} \rhd (h_{3;1} \rhd \y_{3;1;2,1})
                \,\,\to\,\,
                h_{1;4} \rhd \y_{1;4;3,1} \\
                &
                h_{4,3} \rhd (h_{3;3} \rhd \y_{3;3;4,3})
                \,\,\to\,\,
                (h_{1,2} h_{2;3}) \rhd \y_{2;3;1,2}
            \end{aligned}
            \qquad
            \begin{aligned}
                & \\
                &
                h_{4,2} \rhd (h_{2;1} \rhd \y_{2;1;4,1})
                \,\,\to\,\,
                h_{1;3} \rhd \y_{1;3;3,1} \\
                &
                h_{4,2} \rhd (h_{2;1} \rhd \y_{2;1;4,2})
                \,\,\to\,\,
                h_{1;3} \rhd \y_{1;3;3,2} \\
                &
                h_{4,2} \rhd (h_{2;1} \rhd \y_{2;1;2,1})
                \,\,\to\,\,
                h_{1;3} \rhd \y_{1;3;2,1} \\
                &
                h_{4,2} \rhd (h_{2;4} \rhd \y_{2;4;1,3})
                \,\,\to\,\,
                (h_{1,2} h_{2;2}) \rhd \y_{2;2;4,1} \\
                &
                h_{4,2} \rhd (h_{2;4} \rhd \y_{2;4;1,4})
                \,\,\to\,\,
                (h_{1,2} h_{2;2}) \rhd \y_{2;2;4,2} \\
                &
                h_{4,2} \rhd (h_{2;4} \rhd \y_{2;4;4,3})
                \,\,\to\,\,
                (h_{1,2} h_{2;2}) \rhd \y_{2;2;2,1} \\
                &
                h_{4,2} \rhd (h_{2;1} \rhd \y_{2;1;3,4})
                \,\,\to\,\,
                h_{1;3} \rhd \y_{1;3;3,4} \\
                &
                h_{4,2} \rhd (h_{2;4} \rhd \y_{2;4;4,2})
                \,\,\to\,\,
                (h_{1,2} h_{2;2}) \rhd \y_{2;2;3,4} \\
                &
                h_{4,2} \rhd (h_{2;1} \rhd \y_{2;1;3,1})
                \,\,\to\,\,
                h_{1;3} \rhd \y_{1;3;4,1} \\
                &
                h_{4,2} \rhd (h_{2;4} \rhd \y_{2;4;3,2})
                \,\,\to\,\,
                (h_{1,2} h_{2;2}) \rhd \y_{2;2;1,3} \\
                &
                h_{4;2} \rhd \y_{4;2;4,1}
                \,\,\to\,\,
                (h_{1,2} h_{2;4}) \rhd \y_{2;4;1,2} \\
                &
                h_{4,2} \rhd (h_{2;1} \rhd \y_{2;1;3,2})
                \,\,\to\,\,
                h_{1;3} \rhd \y_{1;3;4,2} \\
                &
                h_{4,2} \rhd (h_{2;4} \rhd \y_{2;4;4,2})
                \,\,\to\,\,
                (h_{1,2} h_{2;2}) \rhd \y_{2;2;2,3} \\
                &
                h_{4;1} \rhd \y_{4;1;4,2}
                \,\,\to\,\,
                h_{1;5} \rhd \y_{1;5;4,1} \\
                &
                h_{4;1} \rhd \y_{4;1;3,4}
                \,\,\to\,\,
                h_{1;5} \rhd \y_{1;5;4,3} \\
                &
                h_{4;1} \rhd \y_{4;1;3,2}
                \,\,\to\,\,
                h_{1;5} \rhd \y_{1;5;3,1} \\
                &
                h_{4;2} \rhd \y_{4;2;2,1}
                \,\,\to\,\,
                (h_{1,2} h_{2;4}) \rhd \y_{2;4;3,2} \\
                &
                h_{4;2} \rhd \y_{4;2;3,1}
                \,\,\to\,\,
                (h_{1,2} h_{2;4}) \rhd \y_{2;4;4,2} \\
                &
                h_{4;2} \rhd \y_{4;2;3,2})
                \,\,\to\,\,
                (h_{1,2} h_{2;4}) \rhd \y_{2;4;4,3} \\
                &
                h_{4;1} \rhd \y_{4;1;2,1}
                \,\,\to\,\,
                h_{1;5} \rhd \y_{1;5;1,2} \\
                &
                h_{4;1} \rhd \y_{4;1;3,1}
                \,\,\to\,\,
                h_{1;5} \rhd \y_{1;5;3,2} \\
                &
                h_{4;2} \rhd \y_{4;2;4,2}
                \,\,\to\,\,
                (h_{1,2} h_{2;4}) \rhd \y_{2;4;1,3} \\
                &
                h_{4;1} \rhd \y_{4;1;4,1}
                \,\,\to\,\,
                h_{1;5} \rhd \y_{1;5;4,2} \\
                &
                h_{4;2} \rhd \y_{4;2;4,3}
                \,\,\to\,\,
                (h_{1,2} h_{2;4}) \rhd \y_{2;4;1,4}
            \end{aligned}
            \\
            &
            \begin{aligned}
                &
                \textbf{\textit{Bulk wedges}}
                \\
                &
                h_{4,1} \rhd \x_{1;1,5} \,\,\to\,\, \x_{1;1,2} h_{1,2} \rhd \x_{2;5,1} \\
                &
                h_{4,3} \rhd \x_{3;1,3} \,\,\to\,\, \x_{1;4,2} h_{1,2} \rhd \x_{2;5,3} \\
                &
                h_{4,1} \rhd (\x_{1;5,2} \x_{1;2,4}) \x_{4;3,4}
                h_{4,2} \rhd (\x_{2;3,5} \x_{2;5,4})
                \,\,\to\,\,
                h_{1,2} \rhd \x_{2;1,2} \\
                &
                h_{4,1} \rhd (\x_{1;1,2} \x_{1;2,4}) \x_{4;3,4}
                h_{4,2} \rhd (\x_{2;3,5} \x_{2;5,1})
                \,\,\to\,\,
                \x_{1;1,3} \\
                &
                h_{4,1} \rhd (\x_{1;1,3} \x_{1;3,4}) \x_{4;3,5} 
                h_{4,3} \rhd (\x_{3;2,4} \x_{3;4,1})
                \,\,\to\,\,
                \x_{1;1,4} \\
                &
                h_{4,1} \rhd (\x_{1;5,3} \x_{1;3,4}) \x_{4;3,5}
                h_{4,3} \rhd (\x_{3;2,4} \x_{3;4,3})
                \,\,\to\,\,
                h_{1,2} \rhd \x_{2;1,3} \\
                &
                h_{4,2} \rhd (\x_{2;4,2} \x_{2;2,3}) \x_{4;4,5}
                h_{4,3} \rhd (\x_{3;2,5} \x_{3;5,3})
                \,\,\to\,\,
                h_{1,2} \rhd \x_{2;2,3} \\
                &
                h_{4,2} \rhd (\x_{2;1,2} \x_{2;2,3}) \x_{4;4,5} 
                h_{4,3} \rhd (\x_{3;2,5} \x_{3;5,1})
                \,\,\to\,\,
                \x_{1;3,4}
            \end{aligned}
                \qquad
            \begin{aligned}
                & \\
                &
                h_{4,2} \rhd \x_{2;1,4} \,\,\to\,\, \x_{1;3,2} h_{1,2} \rhd \x_{2;5,2} \\
                &
                \x_{4;1,2} \,\,\to\,\, \x_{1;5,2} h_{1,2} \rhd \x_{2;5,4} \\
                &
                \x_{4;1,3} h_{4,1} \rhd \x_{1;4,1}
                \,\,\to\,\,
                \x_{1;5,1} \\
                &
                \x_{4;2,3} h_{4,1} \rhd \x_{1;4,5}
                \,\,\to\,\,
                h_{1,2} \rhd \x_{2;4,1} \\
                &
                \x_{4;1,4} h_{4,2} \rhd \x_{2;3,1}
                \,\,\to\,\,
                \x_{1;5,3} \\
                &
                \x_{4;2,4} h_{4,2} \rhd \x_{2;3,4}
                \,\,\to\,\,
                h_{1,2} \rhd \x_{2;4,2} \\
                &
                \x_{4;1,5} h_{4,3} \rhd \x_{3;2,1}
                \,\,\to\,\,
                \x_{1;5,4} \\
                &
                \x_{4;2,5} h_{4,3} \rhd \x_{3;2,3}
                \,\,\to\,\,
                h_{1,2} \rhd \x_{2;4,3}
            \end{aligned}\label{cv24}
        \end{aligned}
    \ee
    Under such change of variables, the four point Feynman diagram  \eqref{Amplitude_Four4simplices} turns out to be proportional to that of two point Feynman diagram \eqref{Amplitude_Two4simplices}:
    \be
        \cA_{\cV_4} = (V_{G_1}^{2} V_{G_2}^{19}) \, \cA_{\cV_2} \,.
    \ee

    \item[$\ast$]
    \textbf{Pachner move $\mathrm{P_{3,3}}$}: consider the amplitude of three point Feynman diagram  \eqref{Amplitude_Three4simplices}.
    We consider the map from the variables that decorate the combination of three point Feynman diagram  to the variables that decorate the different three point Feynman diagram.
    \be                         
        \begin{aligned}
            &
            \begin{aligned}
                &
                \textbf{\textit{Boundary links}} 
                \\
                &
                u_{1;1;1} \,\,\to\,\, u_{1;5;4} \\
                &
                u_{1;1;2} \,\,\to\,\, u_{1;5;1} \\
                &
                u_{1;1;3} \,\,\to\,\, u_{1;5;2} \\
                &
                u_{1;1;4} \,\,\to\,\, u_{1;5;3} \\
                &
                u_{1;4;1} \,\,\to\,\, u_{2;4;3} \\
                &
                u_{1;4;2} \,\,\to\,\, u_{2;4;1} \\
                &
                u_{1;4;3} \,\,\to\,\, u_{2;4;4} \\
                &
                u_{1;4;4} \,\,\to\,\, u_{2;4;2} \\
                &
                u_{1;5;1} \,\,\to\,\, u_{3;3;1} \\
                &
                u_{1;5;2} \,\,\to\,\, u_{3;3;4} \\
                &
                u_{1;5;3} \,\,\to\,\, u_{3;3;3} \\
                &
                u_{1;5;4} \,\,\to\,\, u_{3;3;2} 
            \end{aligned}
            \quad
            \begin{aligned}
                &
                \\
                &
                u_{2;1;1} \,\,\to\,\, u_{1;4;2} \\
                &
                u_{2;1;2} \,\,\to\,\, u_{1;4;3} \\
                &
                u_{2;1;3} \,\,\to\,\, u_{1;4;4} \\
                &
                u_{2;1;4} \,\,\to\,\, u_{1;4;1} \\
                &
                u_{2;3;1} \,\,\to\,\, u_{2;3;4} \\
                &
                u_{2;3;2} \,\,\to\,\, u_{2;3;1} \\
                &
                u_{2;3;3} \,\,\to\,\, u_{2;3;2} \\
                &
                u_{2;3;4} \,\,\to\,\, u_{2;3;3} \\
                &
                u_{2;4;1} \,\,\to\,\, u_{3;2;1} \\
                &
                u_{2;4;2} \,\,\to\,\, u_{3;2;2} \\
                &
                u_{2;4;3} \,\,\to\,\, u_{3;2;4} \\
                &
                u_{2;4;4} \,\,\to\,\, u_{3;2;3} 
            \end{aligned}
            \qquad
            \begin{aligned}
                &
                \\
                &
                u_{3;1;1} \,\,\to\,\, u_{1;1;1} \\
                &
                u_{3;1;2} \,\,\to\,\, u_{1;1;2} \\
                &
                u_{3;1;3} \,\,\to\,\, u_{1;1;4} \\
                &
                u_{3;1;4} \,\,\to\,\, u_{1;1;3} \\
                &
                u_{3;2;1} \,\,\to\,\, u_{2;1;1} \\
                &
                u_{3;2;2} \,\,\to\,\, u_{2;1;3} \\
                &
                u_{3;2;3} \,\,\to\,\, u_{2;1;2} \\
                &
                u_{3;2;4} \,\,\to\,\, u_{2;1;4} \\
                &
                u_{3;3;1} \,\,\to\,\, u_{3;1;2} \\
                &
                u_{3;3;2} \,\,\to\,\, u_{3;1;1} \\
                &
                u_{3;3;3} \,\,\to\,\, u_{3;1;3} \\
                &
                u_{3;3;4} \,\,\to\,\, u_{3;1;4} 
            \end{aligned}
            \qquad
            \begin{aligned}
                &
                \textbf{\textit{Bulk links}}
                \\
                &
                h_{1;1} \,\,\to\,\, h_{1;5} \\
                &
                h_{1;4} \,\,\to\,\, h_{1,2} h_{2;4} \\
                &
                h_{1;5} \,\,\to\,\, h_{1,3} h_{3;3} \\
                %
                &
                h_{1,2} h_{2;1} \,\,\to\,\, h_{1;4} \\
                &
                h_{1,2} h_{2;3} \,\,\to\,\, h_{1,2} h_{2;3} \\
                &
                h_{1,2} h_{2;4} \,\,\to\,\, h_{1,3} h_{3;2} \\
                %
                &
                h_{1,3} h_{3;1} \,\,\to\,\, h_{1;1} \\
                &
                h_{1,3} h_{3;2} \,\,\to\,\, h_{1,2} h_{2;1} \\
                &
                h_{1,3} h_{3;3} \,\,\to\,\, h_{1,3} h_{3,1} 
            \end{aligned}
            \\
            &
            \begin{aligned}
                &
                \textbf{\textit{Bulk wedges}} 
                \\
                &
                \x_{1;1,2} h_{1,2} \rhd \x_{2;5,1} \,\,\to\,\, \x_{1;5,4} \\
                &
                \x_{1;1,3} h_{1,3} \rhd \x_{3;4,1} \,\,\to\,\, \x_{1;5,1} \\
                &
                \x_{1;1,4} \,\,\to\,\, \x_{1;5,2} h_{1,2} \rhd \x_{2;5,4} \\
                &
                \x_{1;1,5} \,\,\to\,\, \x_{1;5,3} h_{1,3} \rhd \x_{3;4,3} \\
                &
                \x_{1;4,2} h_{1,2} \rhd \x_{2;5,3} \,\,\to\,\, h_{1,2} \rhd \x_{2;4,3} \\
                &
                \x_{1;4,3} h_{1,3} \rhd \x_{3;4,2} \,\,\to\,\, h_{1,2} \rhd \x_{2;4,1} \\
                &
                \x_{1;4,5} \,\,\to\,\, h_{1,2} \rhd (\x_{2;4,2} \x_{2;2,5}) \x_{1;2,3} h_{1,3} \rhd (\x_{3;4,5} \x_{3;5,3}) \\
                &
                \x_{1;5,2} h_{1,2} \rhd \x_{2;5,4} \,\,\to\,\, h_{1,3} \rhd \x_{3;3,2} \\
                &
                \x_{1;5,3} h_{1,3} \rhd \x_{3;4,3} \,\,\to\,\, h_{1,2} \rhd \x_{3;3,1} \\
                &
                h_{1,2} \rhd \x_{2;1,3} \,\,\to\,\, \x_{1;4,2} h_{1,2} \rhd \x_{2;5,3} \\
                &
                h_{1,2} \rhd \x_{2;1,4} \,\,\to\,\, \x_{1;4,3} h_{1,3} \rhd \x_{3;4,2} \\
                &
                h_{1,2} \rhd \x_{2;3,4} \,\,\to\,\, h_{1,2} \rhd (\x_{2;3,2} \x_{2;2,5}) \x_{1;2,3} h_{1,3} \rhd (\x_{3;4,5} \x_{3;5,2}) \\
                &
                h_{1,3} \rhd \x_{3;1,2} \,\,\to\,\, \x_{1;1,2} h_{1,2} \rhd \x_{2;5,1} \\
                &
                h_{1,3} \rhd \x_{3;1,3} \,\,\to\,\, \x_{1;1,3} h_{1,3} \rhd \x_{3;4,1} \\
                &
                h_{1,3} \rhd \x_{3;2,3} \,\,\to\,\, h_{1,2} \rhd (\x_{2;1,2} \x_{2;2,5}) \x_{1;2,3} h_{1,3} \rhd (\x_{3;4,5} \x_{3;5,1}) \\
                &
                h_{1,2} \rhd (\x_{2;1,2} \x_{2;2,5}) \x_{1;2,3} h_{1,3} \rhd (\x_{3;4,5} \x_{3;5,1}) \,\,\to\,\, \x_{1;4,1} \\
                &
                h_{1,2} \rhd (\x_{2;3,2} \x_{2;2,5}) \x_{1;2,3} h_{1,3} \rhd (\x_{3;4,5} \x_{3;5,2}) \,\,\to\,\, h_{1,2} \rhd \x_{2;3,1} \\
                &
                h_{1,2} \rhd (\x_{2;4,2} \x_{2;2,5}) \x_{1;2,3} h_{1,3} \rhd (\x_{3;4,5} \x_{3;5,3}) \,\,\to\,\, h_{1,3} \rhd \x_{3;2,1} 
            \end{aligned} 
        \end{aligned}
    \ee

    \be
        \begin{aligned}
            &
            \begin{aligned}
                &
                \textbf{\textit{Boundary wedges}} 
                \\
                &
                (h_{1;1} \rhd \y_{1;1;1,2}) \,\,\to\,\, (h_{1;5} \rhd \y_{1;5;4,1}) \\ 
                &
                (h_{1;1} \rhd \y_{1;1;1,3}) \,\,\to\,\, (h_{1;5} \rhd \y_{1;5;4,2}) \\ 
                &
                (h_{1;1} \rhd \y_{1;1;1,4}) \,\,\to\,\, (h_{1;5} \rhd \y_{1;5;4,3}) \\ 
                &
                (h_{1;1} \rhd \y_{1;1;2,3}) \,\,\to\,\, (h_{1;5} \rhd \y_{1;5;1,2}) \\ 
                &
                (h_{1;1} \rhd \y_{1;1;2,4}) \,\,\to\,\, (h_{1;5} \rhd \y_{1;5;1,3}) \\ 
                &
                (h_{1;1} \rhd \y_{1;1;4,3}) \,\,\to\,\, (h_{1;5} \rhd \y_{1;5;2,3}) \\ 
                &
                (h_{1;4} \rhd \y_{1;4;1,2}) \,\,\to\,\, h_{1,2} \rhd (h_{2;4} \rhd \y_{2;4;3,1}) \\
                &
                (h_{1;4} \rhd \y_{1;4;1,3}) \,\,\to\,\, h_{1,2} \rhd (h_{2;4} \rhd \y_{2;4;3,4}) \\
                &
                (h_{1;4} \rhd \y_{1;4;1,4}) \,\,\to\,\, h_{1,2} \rhd (h_{2;4} \rhd \y_{2;4;3,2}) \\
                &
                (h_{1;4} \rhd \y_{1;4;2,3}) \,\,\to\,\, h_{1,2} \rhd (h_{2;4} \rhd \y_{2;4;1,4}) \\
                &
                (h_{1;4} \rhd \y_{1;4;2,4}) \,\,\to\,\, h_{1,2} \rhd (h_{2;4} \rhd \y_{2;4;1,2}) \\ 
                &
                (h_{1;4} \rhd \y_{1;4;3,4}) \,\,\to\,\, h_{1,2} \rhd (h_{2;4} \rhd \y_{2;4;4,2}) \\ 
                &
                (h_{1;5} \rhd \y_{1;5;1,2}) \,\,\to\,\, h_{1,3} \rhd (h_{3;3} \rhd \y_{3;3;1,4}) \\ 
                &
                (h_{1;5} \rhd \y_{1;5;1,3}) \,\,\to\,\, h_{1,3} \rhd (h_{3;3} \rhd \y_{3;3;1,3}) \\ 
                &
                (h_{1;5} \rhd \y_{1;5;1,4}) \,\,\to\,\, h_{1,3} \rhd (h_{3;3} \rhd \y_{3;3;1,2}) \\ 
                &
                (h_{1;5} \rhd \y_{1;5;2,3}) \,\,\to\,\, h_{1,3} \rhd (h_{3;3} \rhd \y_{3;3;4,3}) \\ 
                &
                (h_{1;5} \rhd \y_{1;5;2,4}) \,\,\to\,\, h_{1,3} \rhd (h_{3;3} \rhd \y_{3;3;4,2}) \\ 
                &
                (h_{1;5} \rhd \y_{1;5;3,4}) \,\,\to\,\, h_{1,3} \rhd (h_{3;3} \rhd \y_{3;3;3,2}) \\ 
            \end{aligned}
            \qquad \quad \,\,\,
            \begin{aligned}
                & \\
                &
                h_{1,2} \rhd (h_{2;1} \rhd \y_{2;1;1,2}) \,\,\to\,\, (h_{1;4} \rhd \y_{1;4;2,3}) \\
                &
                h_{1,2} \rhd (h_{2;1} \rhd \y_{2;1;1,3}) \,\,\to\,\, (h_{1;4} \rhd \y_{1;4;2,4}) \\
                &
                h_{1,2} \rhd (h_{2;1} \rhd \y_{2;1;1,4}) \,\,\to\,\, (h_{1;4} \rhd \y_{1;4;2,1}) \\
                &
                h_{1,2} \rhd (h_{2;1} \rhd \y_{2;1;2,4}) \,\,\to\,\, (h_{1;4} \rhd \y_{1;4;3,1}) \\
                &
                h_{1,2} \rhd (h_{2;1} \rhd \y_{2;1;2,3}) \,\,\to\,\, (h_{1;4} \rhd \y_{1;4;3,4}) \\
                &
                h_{1,2} \rhd (h_{2;1} \rhd \y_{2;1;4,3}) \,\,\to\,\, (h_{1;4} \rhd \y_{1;4;4,1}) \\
                &
                h_{1,2} \rhd (h_{2;3} \rhd \y_{2;3;2,1}) \,\,\to\,\, h_{1,2} \rhd (h_{2;3} \rhd \y_{2;3;1,4}) \\
                &
                h_{1,2} \rhd (h_{2;3} \rhd \y_{2;3;1,3}) \,\,\to\,\, h_{1,2} \rhd (h_{2;3} \rhd \y_{2;3;4,2}) \\
                &
                h_{1,2} \rhd (h_{2;3} \rhd \y_{2;3;1,4}) \,\,\to\,\, h_{1,2} \rhd (h_{2;3} \rhd \y_{2;3;4,3}) \\
                &
                h_{1,2} \rhd (h_{2;3} \rhd \y_{2;3;2,3}) \,\,\to\,\, h_{1,2} \rhd (h_{2;3} \rhd \y_{2;3;1,2}) \\
                &
                h_{1,2} \rhd (h_{2;3} \rhd \y_{2;3;3,4}) \,\,\to\,\, h_{1,2} \rhd (h_{2;3} \rhd \y_{2;3;2,3}) \\
                &
                h_{1,2} \rhd (h_{2;3} \rhd \y_{2;3;4,2}) \,\,\to\,\, h_{1,2} \rhd (h_{2;3} \rhd \y_{2;3;3,1}) \\
                &
                h_{1,2} \rhd (h_{2;4} \rhd \y_{2;4;1,2}) \,\,\to\,\, h_{1,3} \rhd (h_{3;2} \rhd \y_{3;2;1,2}) \\
                &
                h_{1,2} \rhd (h_{2;4} \rhd \y_{2;4;3,1}) \,\,\to\,\, h_{1,3} \rhd (h_{3;2} \rhd \y_{3;2;4,1}) \\
                &
                h_{1,2} \rhd (h_{2;4} \rhd \y_{2;4;2,3}) \,\,\to\,\, h_{1,3} \rhd (h_{3;2} \rhd \y_{3;2;2,4}) \\
                &
                h_{1,2} \rhd (h_{2;4} \rhd \y_{2;4;2,4}) \,\,\to\,\, h_{1,3} \rhd (h_{3;2} \rhd \y_{3;2;2,3}) \\
                &
                h_{1,2} \rhd (h_{2;4} \rhd \y_{2;4;4,1}) \,\,\to\,\, h_{1,3} \rhd (h_{3;2} \rhd \y_{3;2;3,1}) \\
                &
                h_{1,2} \rhd (h_{2;4} \rhd \y_{2;4;3,4}) \,\,\to\,\, h_{1,3} \rhd (h_{3;2} \rhd \y_{3;2;4,3}) 
            \end{aligned}
            \\
            & \\
            &
            \begin{aligned}
                &
                h_{1,3} \rhd (h_{3;1} \rhd \y_{3;1;1,2}) \,\,\to\,\, (h_{1;1} \rhd \y_{1;1;1,2}) \\
                &
                h_{1,3} \rhd (h_{3;1} \rhd \y_{3;1;1,3}) \,\,\to\,\, (h_{1;1} \rhd \y_{1;1;1,4}) \\
                &
                h_{1,3} \rhd (h_{3;1} \rhd \y_{3;1;1,4}) \,\,\to\,\, (h_{1;1} \rhd \y_{1;1;1,3}) \\
                &
                h_{1,3} \rhd (h_{3;1} \rhd \y_{3;1;2,3}) \,\,\to\,\, (h_{1;1} \rhd \y_{1;1;2,4}) \\
                &
                h_{1,3} \rhd (h_{3;1} \rhd \y_{3;1;2,4}) \,\,\to\,\, (h_{1;1} \rhd \y_{1;1;2,3}) \\
                &
                h_{1,3} \rhd (h_{3;1} \rhd \y_{3;1;4,3}) \,\,\to\,\, (h_{1;1} \rhd \y_{1;1;3,4}) \\
                &
                h_{1,3} \rhd (h_{3;2} \rhd \y_{3;2;1,2}) \,\,\to\,\, h_{1,2} \rhd (h_{2;1} \rhd \y_{2;1;1,3}) \\
                &
                h_{1,3} \rhd (h_{3;2} \rhd \y_{3;2;1,4}) \,\,\to\,\, h_{1,2} \rhd (h_{2;1} \rhd \y_{2;1;1,4}) \\
                &
                h_{1,3} \rhd (h_{3;2} \rhd \y_{3;2;2,3}) \,\,\to\,\, h_{1,2} \rhd (h_{2;1} \rhd \y_{2;1;3,2}) 
            \end{aligned}
            \quad
            \begin{aligned}
                &
                h_{1,3} \rhd (h_{3;2} \rhd \y_{3;2;2,4}) \,\,\to\,\, h_{1,2} \rhd (h_{2;1} \rhd \y_{2;1;4,3}) \\
                &
                h_{1,3} \rhd (h_{3;2} \rhd \y_{3;2;3,1}) \,\,\to\,\, h_{1,2} \rhd (h_{2;1} \rhd \y_{2;1;2,1}) \\
                &
                h_{1,3} \rhd (h_{3;2} \rhd \y_{3;2;3,4}) \,\,\to\,\, h_{1,2} \rhd (h_{2;1} \rhd \y_{2;1;2,4}) \\
                &
                h_{1,3} \rhd (h_{3;3} \rhd \y_{3;3;1,2}) \,\,\to\,\, h_{1,3} \rhd (h_{3;1} \rhd \y_{3;1;2,1}) \\
                &
                h_{1,3} \rhd (h_{3;3} \rhd \y_{3;3;1,3}) \,\,\to\,\, h_{1,3} \rhd (h_{3;1} \rhd \y_{3;1;2,3}) \\
                &
                h_{1,3} \rhd (h_{3;3} \rhd \y_{3;3;4,1}) \,\,\to\,\, h_{1,3} \rhd (h_{3;1} \rhd \y_{3;1;4,2}) \\
                &
                h_{1,3} \rhd (h_{3;3} \rhd \y_{3;3;2,3}) \,\,\to\,\, h_{1,3} \rhd (h_{3;1} \rhd \y_{3;1;1,3}) \\
                &
                h_{1,3} \rhd (h_{3;3} \rhd \y_{3;3;4,2}) \,\,\to\,\, h_{1,3} \rhd (h_{3;1} \rhd \y_{3;1;4,1}) \\
                &
                h_{1,3} \rhd (h_{3;3} \rhd \y_{3;3;3,4}) \,\,\to\,\, h_{1,3} \rhd (h_{3;1} \rhd \y_{3;1;4,3}) 
            \end{aligned}\label{cv33}
        \end{aligned}
    \ee
    Under such change of variables the generating function of the three point Feynman diagram \eqref{Amplitude_Three4simplices} turns out to be \textit{equal} to that of the three point Feynman diagram  \eqref{Amplitude_Three4simplices} with a different combinatorics:
    \be
        \cA_{\cV_3} = \cA_{\cV_3}' \,.
    \ee
\end{itemize}

\clearpage

\bibliographystyle{Biblio}
\bibliography{biblio}

\end{document}